\documentclass[useAMS,usenatbib]{mn2e}
\usepackage{url}
\usepackage{stmaryrd}

\usepackage{graphics}
\usepackage[pdftex]{graphicx}
  \usepackage[draft]{hyperref}

\def\aj{AJ}%
\def\apj{ApJ}%
\def\apjl{ApJ}%
\def\apjs{ApJS}%
\def\aap{A\&A}%
\def\aapr{A\&A~Rev.}%
\def\aaps{A\&AS}%
\def\mnras{MNRAS}%
\def\nar{New A Rev.}%
\def\hi{H\,{\sc i}}

\def\m20{$M_{20}$}
\def\gm{$G_{M}$}

\def\whisp{{\em WHISP}}
\def\things{{\em THINGS}}
\def\fuv{{FUV}}
\def\nuv{{NUV}}

\def\xuv{{\sc xuv}}
\def\xuvc{{\sc xuv$_c$}}
\def\xhi{{\sc xhi}}

\def\uv{{UV}}

\def\galex{{\sc galex}}

\begin{document}

\title[Extended Disks in \hi\ and \uv ]{Quantified \hi\ Morphology VII:\\ 
The Morphology of Extended Disks in \uv\ and \hi}

\author[Holwerda et al.]{B. W. Holwerda$^{1}$\thanks{E-mail: benne.holwerda@esa.int}, N. Pirzkal,$^{2}$ and J. S. Heiner$^{3}$ \\
$^{1}$ European Space Agency, ESTEC, Keplerlaan 1, 2200 AG, Noordwijk, the Netherlands\\
$^{2}$ Space Telescope Science Institute, Baltimore, MD 21218, USA\\
$^{3}$ Centro de Radiastronom\'{i}a y Astrof\'{i}sica, Universidad Nacional Aut\'{o}noma de M\'{e}xico, 58190 Morelia, Michoac\'an, Mexico\\
}
\date{Accepted 1988 December 15. Received 1988 December 14; in original form 1988 October 11}

\pagerange{\pageref{firstpage}--\pageref{lastpage}} \pubyear{2002}

\maketitle

\label{firstpage}

\begin{abstract}

Extended UltraViolet (\xuv) disks have been found in a substantial fraction of late-type --S0, spiral and irregular-- galaxies. Similarly, most late-type spirals have an extended gas disk, observable in the 21cm radio line (\hi). 
The morphology of galaxies can be quantified well using a series of scale-invariant parameters; Concentration-Asymmetry-Smoothness (CAS), Gini, \m20, and $G_M$ parameters. In this series of papers, we apply these to \hi\ column density maps to identify mergers and interactions, lopsidedness and now \xuv\ disks. 

In this paper, we compare the quantified morphology and effective radius ($R_{50}$) of the Westerbork observations of neutral Hydrogen in Irregular and Spiral galaxies Project (\whisp) \hi\ maps to those of far-and near-ultraviolet images obtained with \galex, to explore how close the morphology and scales of \hi\ and \uv\ in these disks correlate.
%
%
We find that \xuv\ disks do not stand out by their effective radii in \uv\ or \hi. However, the concentration index in \fuv\ appears to select some \xuv\ disks.
And known \xuv\ disks can be identified via a criterion using Asymmetry and \m20; 80\% of \xuv\ disks are included but with 55\% contamination. This translates 
into 61 candidate \xuv\ disk out of our 266 galaxies, --23\%-- consistent with previous findings.
Otherwise, the \uv\ and \hi\ morphology parameters do not appear closely related.

Our motivation is to identify \xuv\ disks and their origin. We consider three scenarios; tidal features from major mergers, the typical extended \hi\ disk is a photo-dissociation product of the \xuv\ regions and 
both \hi\ and \uv\ features originate in cold flows fueling the main galaxy.

We define extended \hi\ and \uv\ disks based on their concentration ($C_{HI} > 5$ and $C_{FUV} > 4$ respectively), but that these two subsamples never overlap in the \whisp\ sample. 
This appears to discount a simple photo-dissociation origin of the outer \hi\ disk. 

Previously, we identified the morphology space occupied by ongoing major mergers. Known \xuv\ disks rarely reside in the merger dominated part of \hi\ morphology space but those that do are Type 1.
Exceptions, \xuv\ disks in ongoing mergers, are the previously identified UGC 4862 and UGC 7081, 7651, and 7853.
This suggests cold flows as the origin for the \xuv\ complexes and their surrounding \hi\ structures.

\end{abstract}


\section{\label{s:intro}Introduction}

Interest in the outskirts of spiral galaxy disks has increased over recent years as these regions are the site of the most recent acquisition of gas for these systems \citep[e.g.][]{Sancisi08}, as well as low-level star-formation \cite[e.g.,][for recent results]{Dong08,Bigiel10b, Alberts11}, making these faint outskirts the interface between the island universes --the galaxies themselves-- and the cosmic web of primordial gas.

The low-level star-formation was first discovered in H$\alpha$ emission  by \cite{Ferguson98a} and \cite{Lelievre00}. After the launch of the Galaxy Evolution EXplorer \citep[{\sc galex},][]{galex}, initial anecdotal evidence pointed to ultraviolet disks of spiral galaxies extending much beyond their optical radius \citep{Thilker05a,Thilker05b, Gil-de-Paz05, Gil-de-Paz07, Zaritsky07}. 
Subsequent structural searches for these Extended Ultraviolet (\xuv) Disks by \cite{Thilker07b} and \cite{Lemonias11} find that some 20--30\% of spirals posses an \xuv\  disk and 40\% of S0s \citep{Moffett11}, making this type of disk common but not typical for spiral and S0 galaxies. 
These \xuv\  disk complexes are generally $\sim100$ Myr old, explaining why most lack H$\alpha$ \citep{Alberts11}, as opposed to a top-light IMF \citep[as proposed by][]{Meurer09}, and sub-solar but not excessively low metallicities \citep[][$0.1-1 ~ Z_\odot$, based on emission lines]{Gil-de-Paz07,Bresolin09a, Werk10}.
Additionally, it has been known for some time now that atomic hydrogen (\hi) as observed by the 21cm fine structure line also extends well beyond the optical disk of spiral galaxies \citep[e.g.,][]{Begeman89, Meurer96, Meurer98, Swaters02, Noordermeer05, Walter08, Boomsma08, Elson11, Heald11a, Heald11b, Zschaechner11b}.
In those few cases where both high-quality \hi\ and deep {\sc galex} data are available, a close relation in their respective morphology was remarked upon \citep{Bigiel10b}. While we have to wait for the all-sky surveys in \hi\ to catch up to the coverage of the {\sc galex} surveys (e.g., the wide survey with WSRT/APERTIF or the WALLABY survey with ASKAP), we can compare the morphology for those galaxies for which uniform \hi\ information is available. 

The canonical view of the origin of the \xuv\  disks, is that the Kennicutt-Schmidt law \citep{Kennicutt98} needs to be extended to low global surface densities of gas and the formation of individual O-stars in the very outskirts of disks \citep{Cuillandre01,Bigiel10c}.
The recent accretion of cold gas flows \citep[][]{Keres05} into the \hi\ disk is the origin for the young stars \citep[the fueling rate implied by XUV disks is explored in][]{Lemonias11} and \hi\ warps\citep[][]{Roskar10a}. The fraction of spirals that have an \xuv\  disk ($\sim20-30$\%) supports this scenario as the remaining spirals may simply have no current cool gas inflow. In this case, one would expect \uv\ and \hi\ morphology to follow each other reasonably closely for the \xuv\ disks but not for many of the others, as the star-formation in the inner disk is more closely related to the molecular phase \citep{Bigiel08}.
However, the existence of \xuv\  disks pose an intriguing alternate possibility for the origin of the atomic hydrogen disk. 
Instead of primordial gas accreting onto the disk, the \hi\ disk could also be the byproduct of photodissociation of molecular hydrogen on the `skins'  of molecular clouds by the ultraviolet flux of the young stars in the \xuv\  disk \citep[see][]{Allen97, Allen02, Allen04}. This explanation has been explored in the stellar disks of several nearby galaxies \citep{Heiner08a, Heiner08b, Heiner09,Heiner10}. \cite{Gil-de-Paz07} calculate the time-scales (molecular gas dissociation and re-formation) involved but these are inconclusive regarding the origin of the \xuv\ disk.
In this scenario, one would expect the \uv\ and \hi\ morphologies to follow each other closely in all cases; in the outer disk, the \uv\ flux from a few young stars would reach out to large areas of the low-column density gas to dissociate enough hydrogen to form the outer \hi\ disk. Thus, the low-flux and low \hi\ column density morphology --those defining the limits and extent of the \xuv\ and \hi\ disks-- should show a close relation in parameters such as concentration, Gini, \m20\ and the effective radius.
A third explanation is in terms of recent tidal interaction. A major merger often pulls gas out of the planes of galaxies and triggers star-forming events. Some anecdotal evidence \citep[e.g., UGC 04862 is a late-stage major merger][]{Torres-Flores12} does point to this possible origin. In this scenario, one would expect that most \hi\ disks hosting an \xuv\ disk would be seriously tidally disrupted.

In this series of papers, we have explored the quantified morphology of available \hi\ maps with the common parameters for visible morphology; concentration-asymmetry-smoothness, Gini and \m20 and $G_M$.
In \cite{Holwerda11a}, we compare the \hi\  morphology to other wavelengths, noting that the \hi\ and ultraviolet morphologies are closely related. In subsequent papers of the series, we use the \hi\ morphology to identify mergers \citep{Holwerda11b}, their visibility time \citep{Holwerda11c} and subsequently infer a merger rate from \whisp\ \citep{Holwerda11d} as well as identify phenomena unique to cluster members \citep{Holwerda11e}.

In this paper, we explore the morphological link between the \hi\ and \xuv\  disks in the Westerbork \hi\ Survey Project \citep[\whisp,][]{whisp,whisp2}, a survey of several hundred \hi\ observations of nearby galaxies. We complement this data with {\sc galex} images to explore the morphology and typical scales of these maps. A direct and quantified comparison between the gas and ongoing star-formation morphology could help answer open questions regarding the origin and nature of \xuv\  disks: how do their respective sizes relate? Are their morphologies closely related in every case? Do their respective morphologies point to a dominant formation mechanism; gas accretion, photodissociation or tidal? Are \xuv\ disks in morphologically distinct or typical \hi\ disks? Are \xuv\  disks embedded in \hi\ disks that appear to be in an active interaction? What is the relation between \uv\ flux and \hi\ column density in the \xuv\  disks, especially the outer disk?

The paper is organized as follows; \S \ref{s:morph} gives the definitions of the quantified morphology parameters we employ, and \S \ref{s:data} describes the origin the data. 
We describe the application of the morphological parameters and the results in \S \ref{s:app} and \ref{s:analysis}, and we discuss them in \S \ref{s:disc}.
We list our conclusions and discuss possible future work in \S \ref{s:concl}.

\section{Morphological Parameters}
\label{s:morph}

In this series of papers, we use the Concentration-Asymmetry-Smoothness parameters \citep[CAS,][]{CAS}, combined with the Gini-$M_{20}$ parameters from \cite{Lotz04} and one addition of our own $G_M$. We have discussed the definitions of these parameters in the previous papers, as well as how we estimate uncertainties for each. 
Here, we will give a brief overview but for details we refer the reader to \cite{Holwerda11a,Holwerda11b}.

We select pixels in an image as belonging to the galaxy based on the outer \hi\ contour and adopt the position from the 2MASS catalog \citep{2MASS} as the central position of the galaxy.
Given a set of $n$ pixels in each object, iterating over pixel $i$ with value $I_i$, position $x_i,y_i$ with the centre of the object at $x_c,y_c$ these parameters are defined as:

\begin{equation}
C = 5 ~ \log (r_{80} /  r_{20}),
\label{eq:c}
\end{equation}
\noindent with $r_{f}$ as the radial aperture, centered on $x_c,y_c$ containing percentage $f$ of the light of the galaxy \citep[see definitions of $r_f$ in][]{se,seman}.
\footnote{We must note that the earlier version of our code contained an error, artificially inflating the concentration values. A check revealed this to be $C_{new} = 0.38 C_{old}$, and we adopt the new, correct values in this paper.}. We include the $r_{50}$, or ``effective radius'' in our catalog as well.
\begin{equation}
A = {\Sigma_{i} | I_i - I_{180} |  \over \Sigma_{i} | I(i) |  },
\label{eq:a}
\end{equation}
\noindent where $I_{180}$ is the pixel at position $i$ in the galaxy's image, after it was rotated $180^\circ$ around the centre of the galaxy.
\begin{equation}
S = {\Sigma_{i,j} | I(i,j) - I_{S}(i,j) | \over \Sigma_{i,j} | I(i,j) | },
\label{eq:s}
\end{equation}
\noindent where $I_{S}$ is pixel $i$ in a smoothed image. The type of smoothing (e.g., boxcar or Gaussian) has changed over the years. 
We chose a fixed 5" Gaussian smoothing kernel for simplicity. 

The Gini coefficient is defined as:
\begin{equation}
G = {1\over \bar{I} n (n-1)} \Sigma_i (2i - n - 1) I_i ,
\label{eq:g}
\end{equation}
\noindent where the list of $n$ pixels was first ordered according to value and $\bar{I}$ is the mean pixel value in the image. 
\begin{equation}
M_{20} = \ log \left( {\Sigma_i M_i  \over  M_{tot}}\right), ~ {\rm for} ~ \Sigma_i I_i < 0.2 I_{tot}, \\
\label{eq:m20}
\end{equation}
\noindent where $M_i$ is the second order moment of pixel $i$; $M_i = I_i \times [(x-x_c)^2 + (y-y_c)^2 ]$. $M_{tot}$ is the second order moment summed over all pixels in the object and $M_{20}$ is the relative contribution of the brightest 20\% of the pixels in the object. 
Instead of using the intensity of  pixel $i$, the Gini parameter can be defined using the second order moment:
\begin{equation}
G_M = {1\over \bar{M} n (n-1)} \Sigma_i (2i - n - 1) M_i ,
\label{eq:gm}
\end{equation}

These parameters trace different structural characteristics of a galaxy's image but these do not span an orthogonal parameter space \citep[see also the discussion in][]{Scarlata07}.


\section{Data}
\label{s:data}

We use three data sets for this paper; \hi\ radio observations from \whisp\ or \things, and ultraviolet observations from {\sc galex} space telescope observations.



\subsection{\whisp\ Sample}
\label{s:whisp}

The starting dataset here is the 266 observations done as part of the Westerbork observations of neutral Hydrogen in Irregular and SPiral galaxies \citep[\whisp,][; 339 individual galaxies]{whisp,whisp2} that also have reasonable quality \fuv\ and \nuv\ data.
The \whisp\ observation targets were selected from the Uppsala General Catalogue of Galaxies \citep{Nilson73}, with blue major axis diameters $> ~ 2\farcm0$, declination (B1950) $>$ 20 degrees and flux densities at 21-cm larger than 100 mJy, later lowered to 20 mJy. Observation times were typically 12 hours of integration. The galaxies satisfying these selection criteria generally have redshifts less than 20000 km/s ($z<0.07$).
\whisp\ has mapped the distribution and velocity structure of \hi\ in several hundreds of nearby galaxies, increasing the number of \hi\ observations of galaxies by an order of magnitude. The \whisp\ project provides a uniform database of data-cubes, zeroth-order and velocity maps. Its focus has been on the structure of the dark matter halo as a function of Hubble type, the Tully-Fisher relation and the dark matter content of dwarf galaxies \citep{WHISPI, WHISPII, WHISPIII,Zwaan05b}. Until the large all-sky surveys with new instruments are completed, \whisp\ is the largest, publicly available data-set of resolved \hi\ observations\footnote{Available at ``Westerbork On the Web'' (WOW) project at ASTRON (\url{http://www.astron.nl/wow/}).}.
We compiled a catalogue of basic data (position, radial velocity etc.) \citep[see for details][]{Holwerda11d}.

\subsubsection{\hi\ Data}

We use the highest available resolution zero-moment maps (beam size of $\sim$12" x 12"/sin($\delta$)), from the WOW website and converted these to $M_\odot / pc^2$ column density maps with J2000 coordinates and of the same size as the \galex\ postage stamp \citep[see for details][]{Holwerda11d}.

\subsubsection{{\sc galex} data}
\label{s:galex}

To complement the \hi~ column density maps, we retrieved {\sc galex} \citep{galex} postage stamps from 
\url{http://skyview.gsfc.nasa.gov}, near- and far-ultraviolet (1350--1750 and 1750--2750 \AA). The majority of 
observations were taken as part of the AIS, MIS, NGS, etc. imaging surveys and hence vary in depth and signal-to-noise. 
Spatial resolution is 4--6", depending on position in the field-of-view.

Of the 339 galaxies in \whisp, 266 had reliable \fuv\ information and this cross-section is what we use for the remainder of the paper. The \fuv\ images with \hi\ contours overlaid are shown in Figures \ref{f:xuvt1}, \ref{f:xuvt2}, and Figures \ref{f:xuv} \ref{f:xhi} and all galaxies in appendix C ({\em electronic edition}).

\subsubsection{\xuv\  classifications}
\label{s:xuv}

In the \whisp\ sample, \cite{Thilker07b} classified 22 galaxies for the presence and type of \xuv\ disks.
The Type 1 \xuv\ disks are characterized by discrete regions out to high radii and Type 2 by an anomalously large, UV-bright outer zone of low optical surface brightness.
Table \ref{t:thilker} lists the \xuv\ classifications for the \whisp\ sample; 22 from \cite{Thilker07b} and a single one from \cite{Lemonias11}. The latter study's lack of overlap is unsurprising as they focus on a more distant sample in complement to the Thilker et al.'s nearby sample ($D<40 $Mpc).
We designate an \xuv\  disk with both a Type 1 and Type 2 as Type 1/2 to discriminate in the plots below and mark those galaxies in WHISP surveyed by Thilker et al. that do not have an \xuv\ disk as Type 0.

\begin{figure*}
\begin{center}
\includegraphics[width=0.49\textwidth]{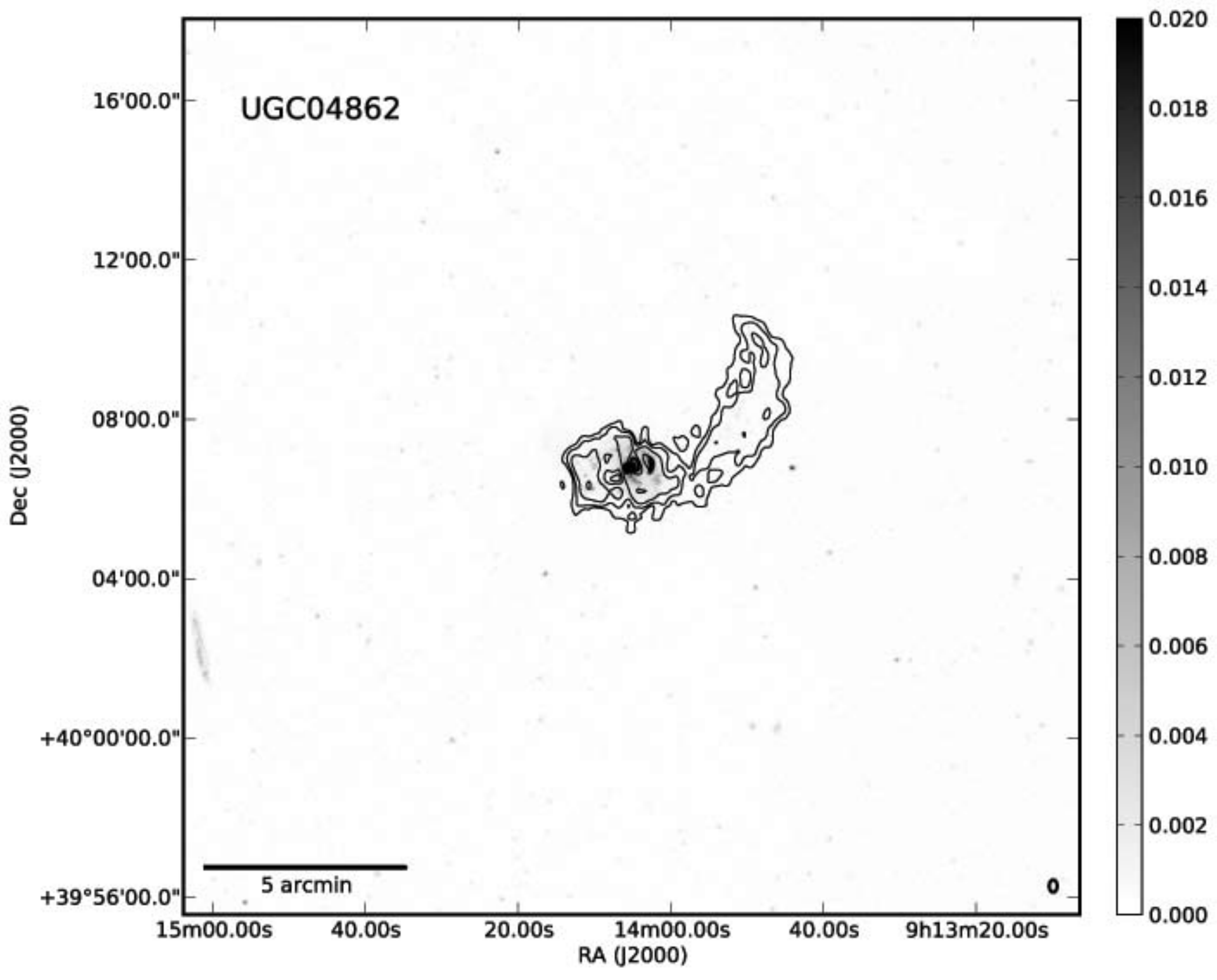}
\includegraphics[width=0.49\textwidth]{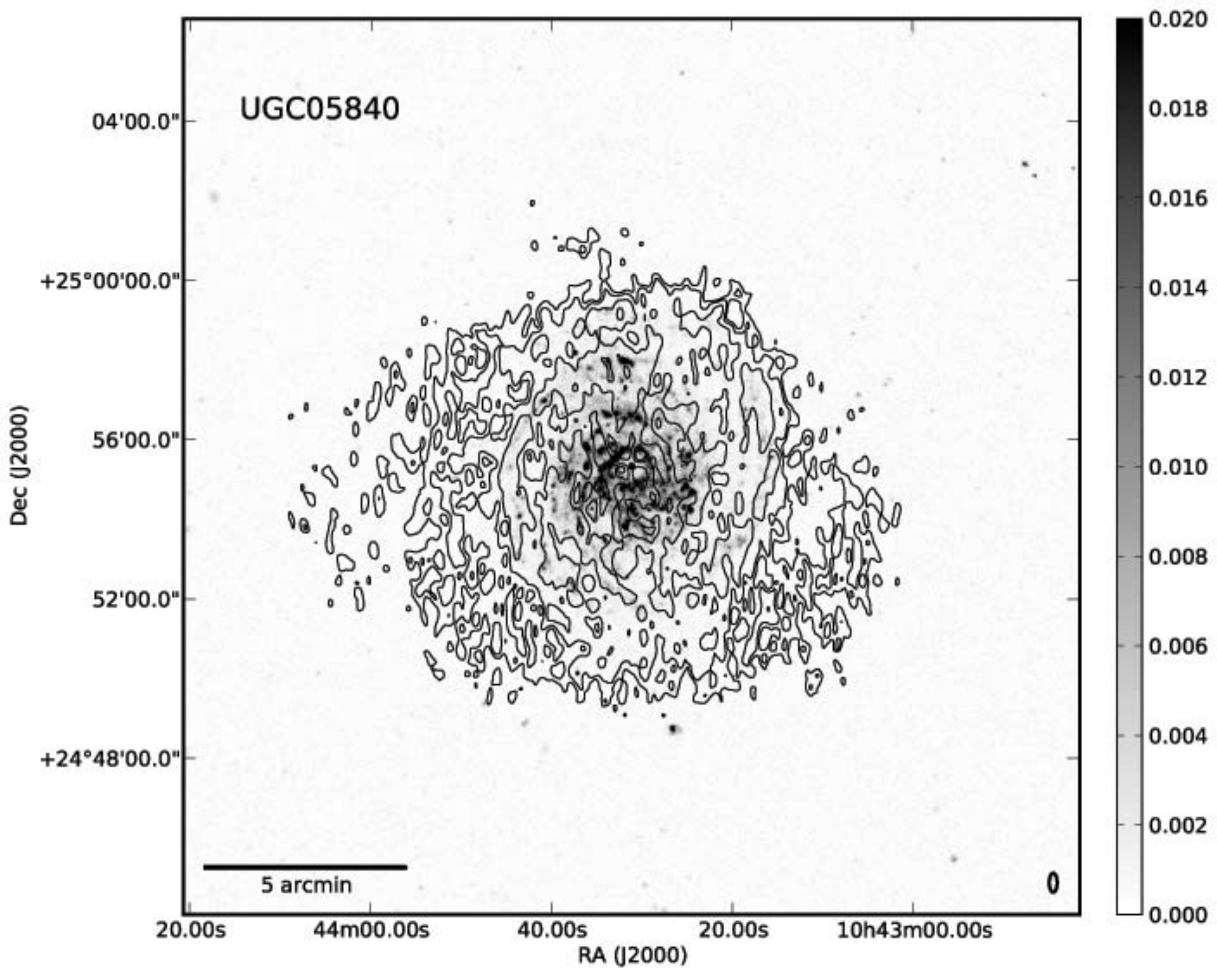}
\includegraphics[width=0.49\textwidth]{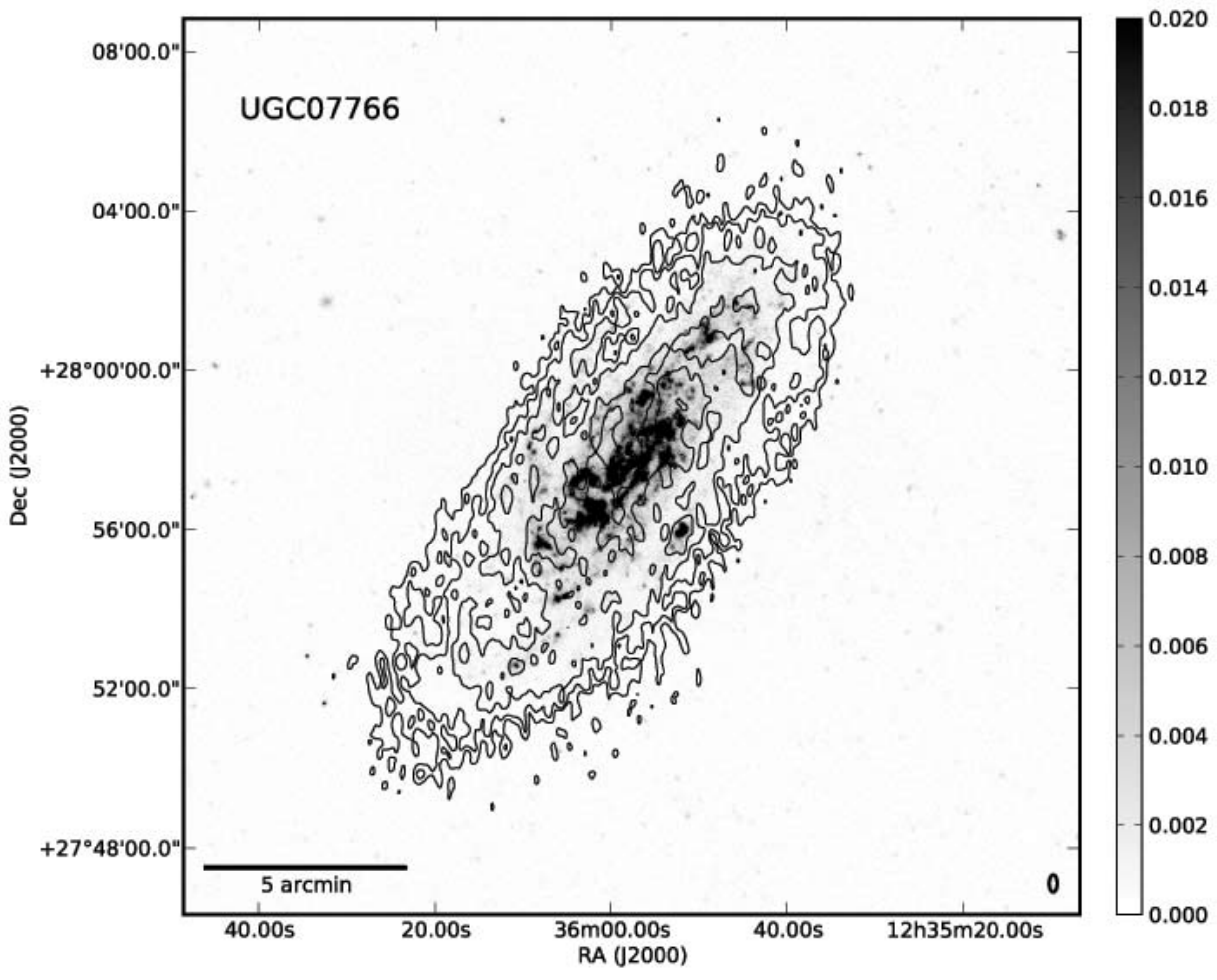}
\includegraphics[width=0.49\textwidth]{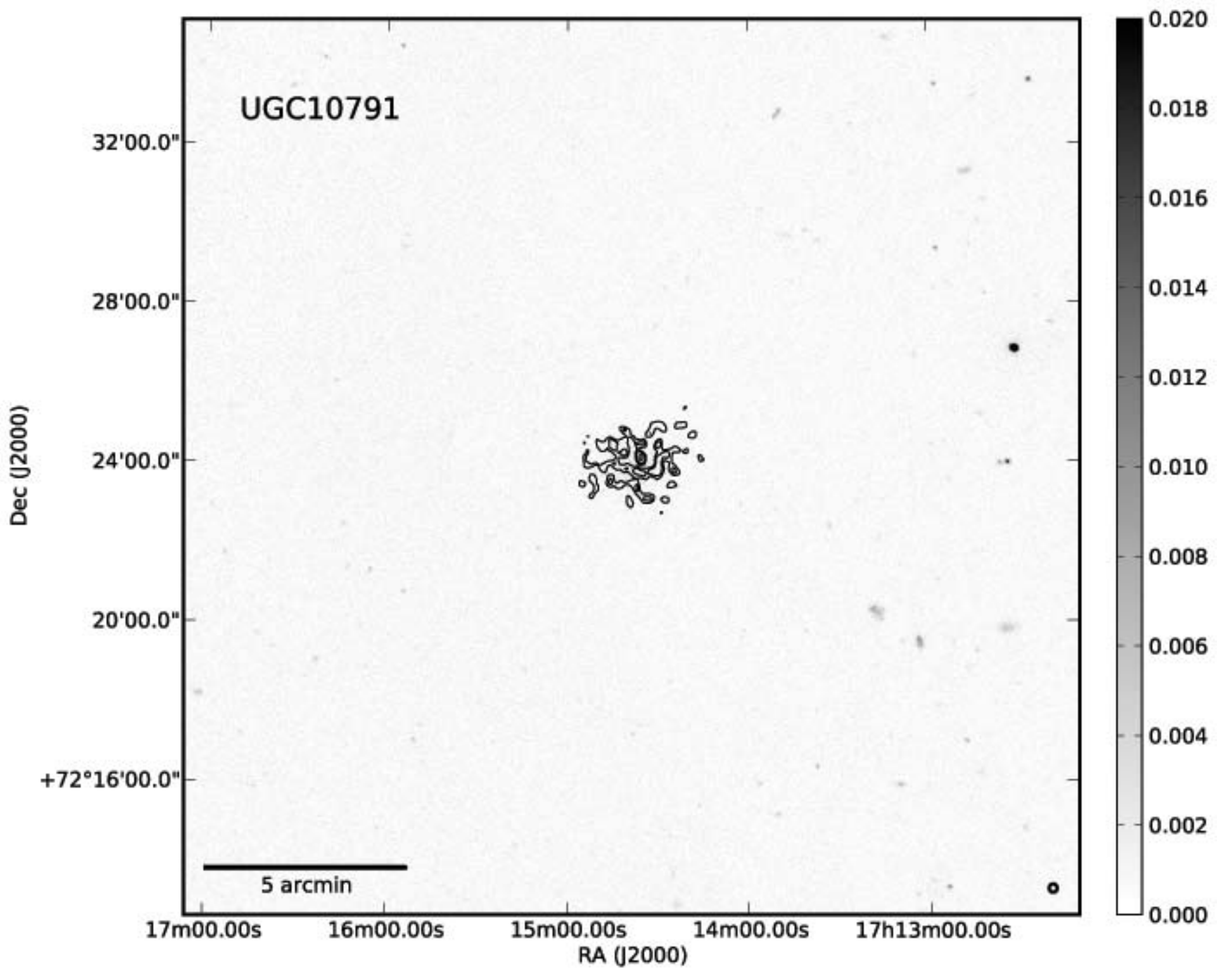}
\caption{The few examples of  \xuv\  disks, Type 1 (individual \uv\ complexes) as classified by \protect\cite{Thilker07b} with complementary \hi\ data in \whisp. The tidal feature in UGC 04862 is studied in detail by \protect\cite{Torres-Flores12}, who report high metallicities and 1-11 Myr ages for the \uv\ complexes. They speculate that their origin is either from the galaxies that formed UGC 04862 or that the current star-formation is a later generation, enriched system.  }
\label{f:xuvt1}
\end{center}
\end{figure*}

\begin{figure*}
\begin{center}
\includegraphics[width=0.49\textwidth]{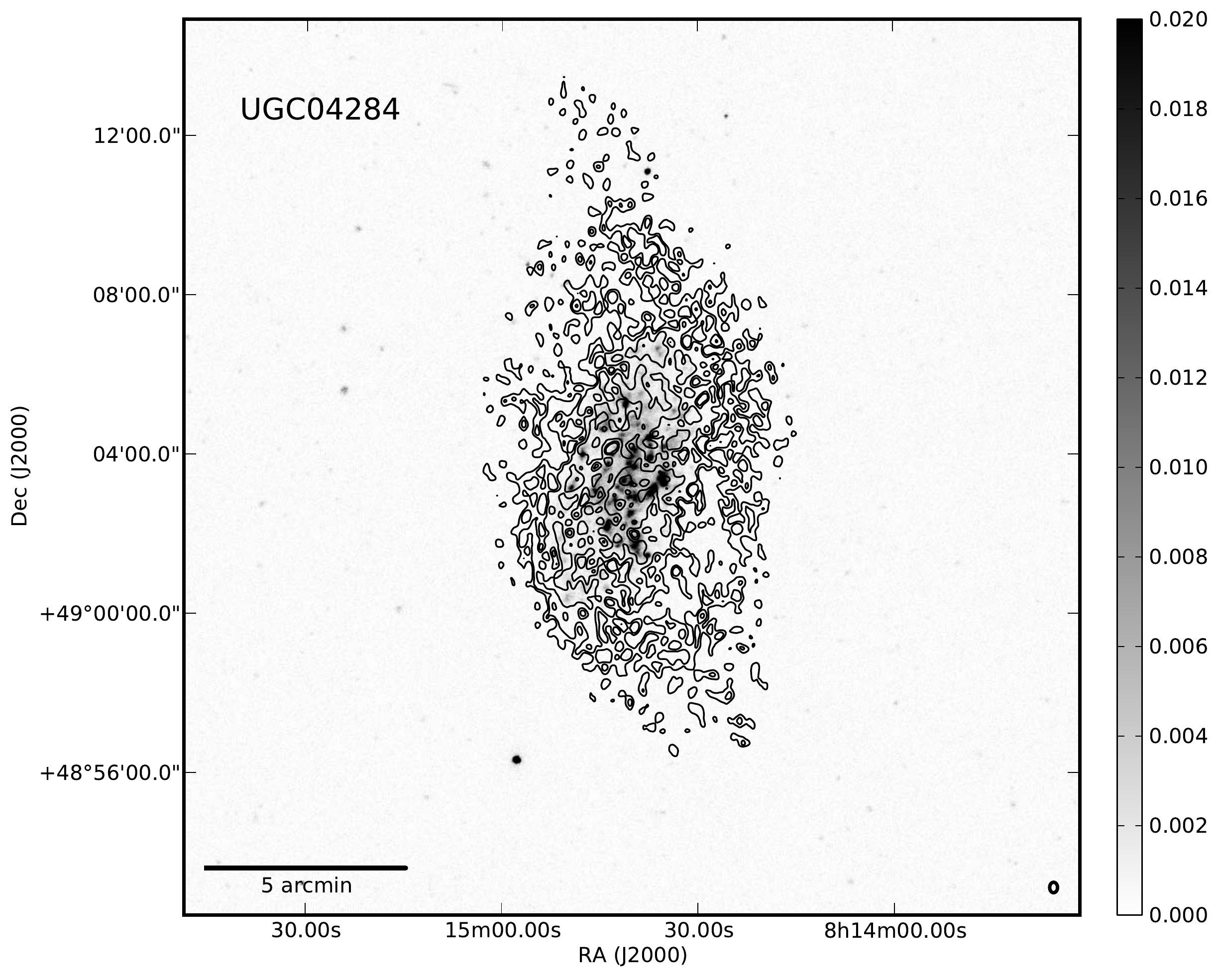}
\includegraphics[width=0.49\textwidth]{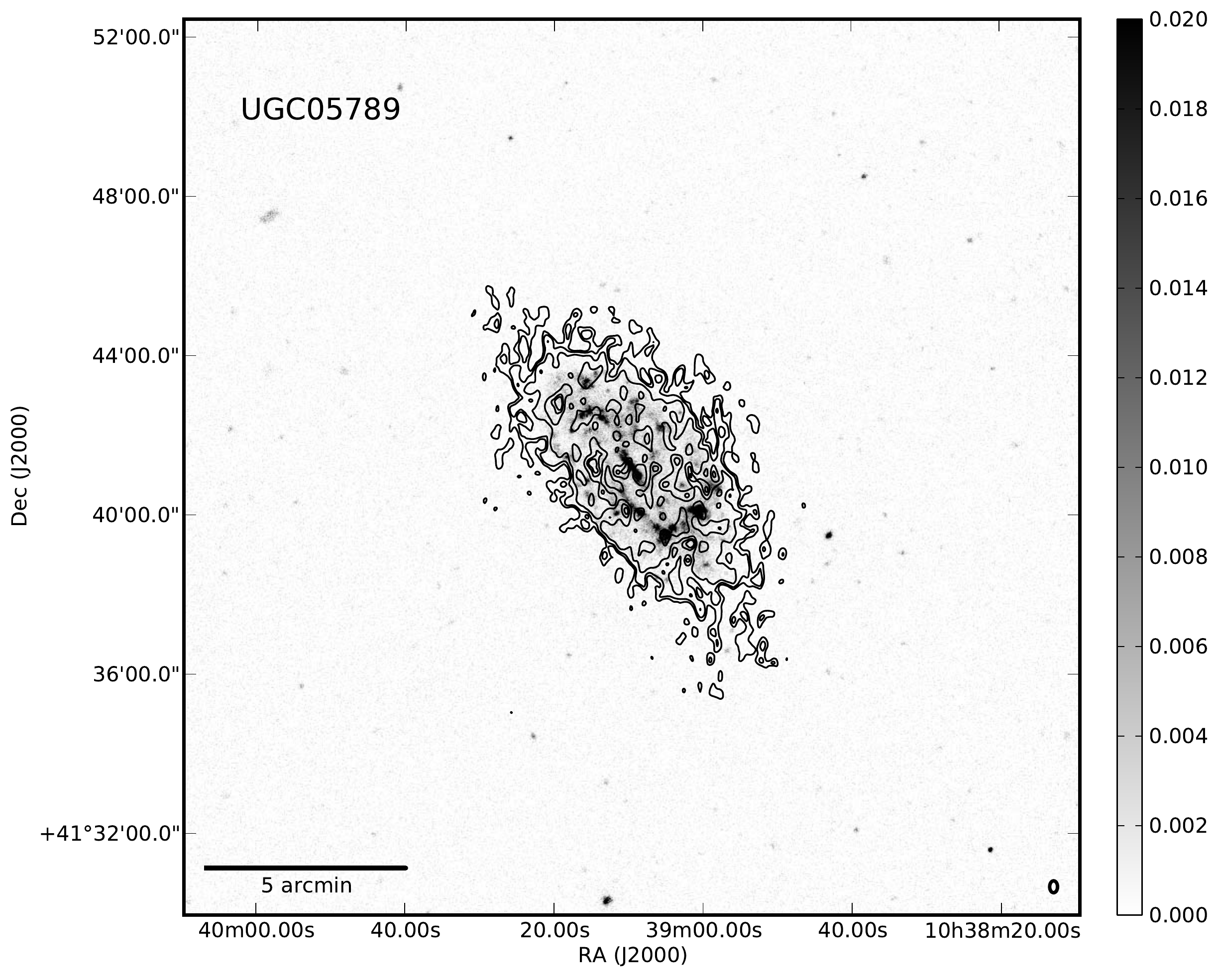}
\includegraphics[width=0.49\textwidth]{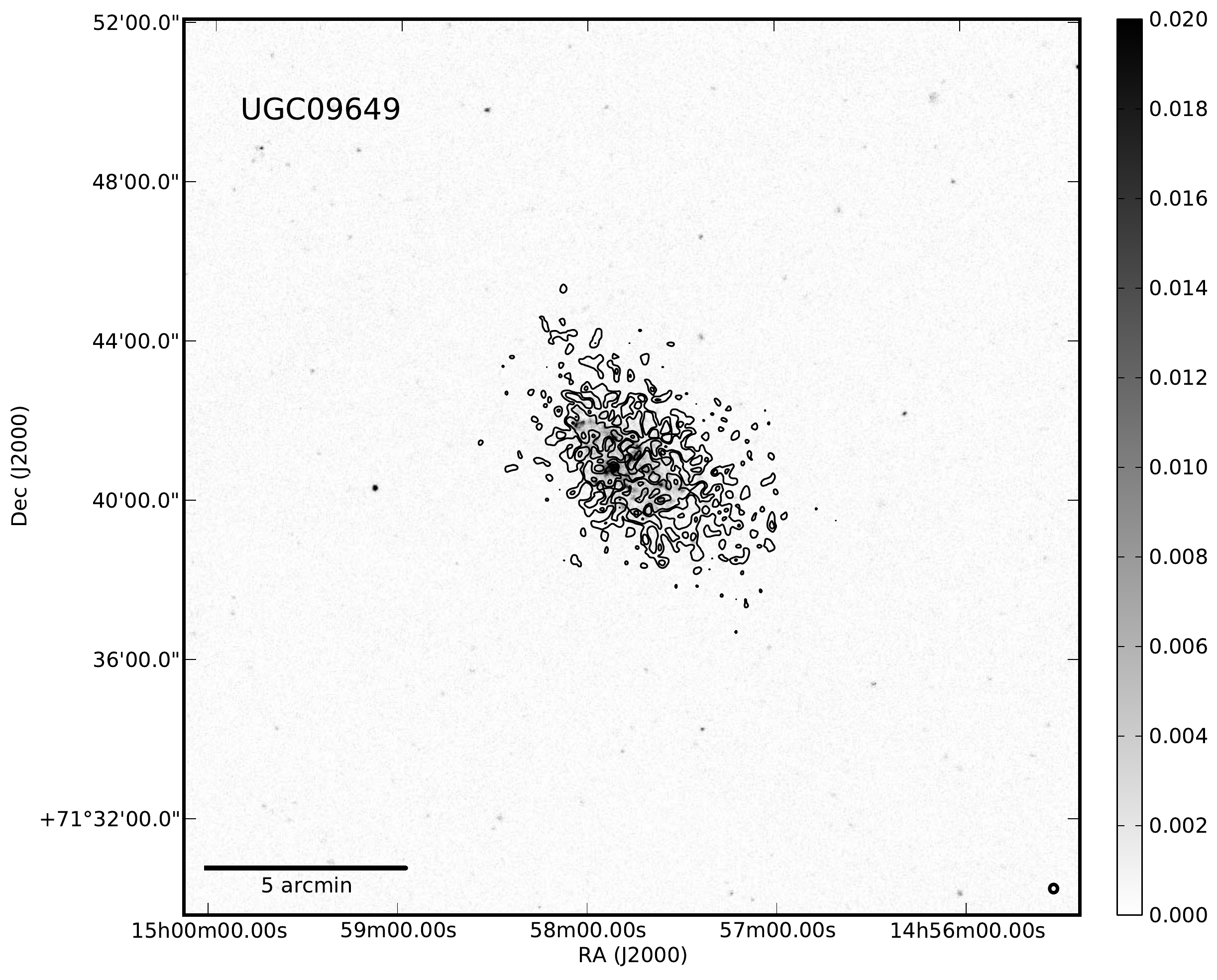}
\includegraphics[width=0.49\textwidth]{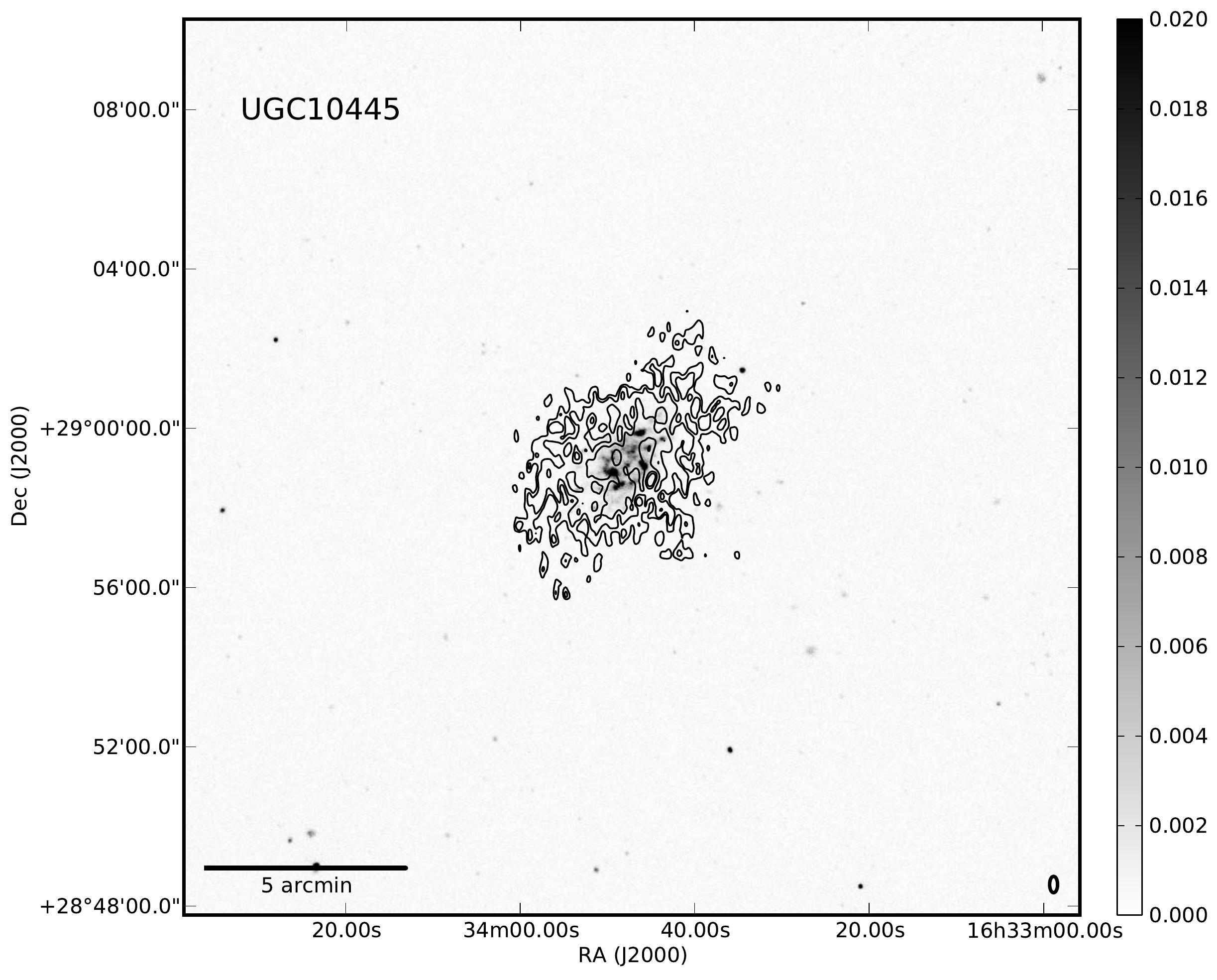}
\caption{The two known examples of Type 2 \xuv\  disks, a ring of \uv\ outside the optical disk (top) and the two
examples of Type 1/2, a combination of Types 1 and 2,  as classified by \protect\cite{Thilker07b}, which are also in the \whisp\ sample. The \hi\ morphology is remarkably flocculant and lacks grand spiral structure.}
\label{f:xuvt2}
\end{center}
\end{figure*}

\begin{figure*}
\begin{center}
\includegraphics[width=0.32\textwidth]{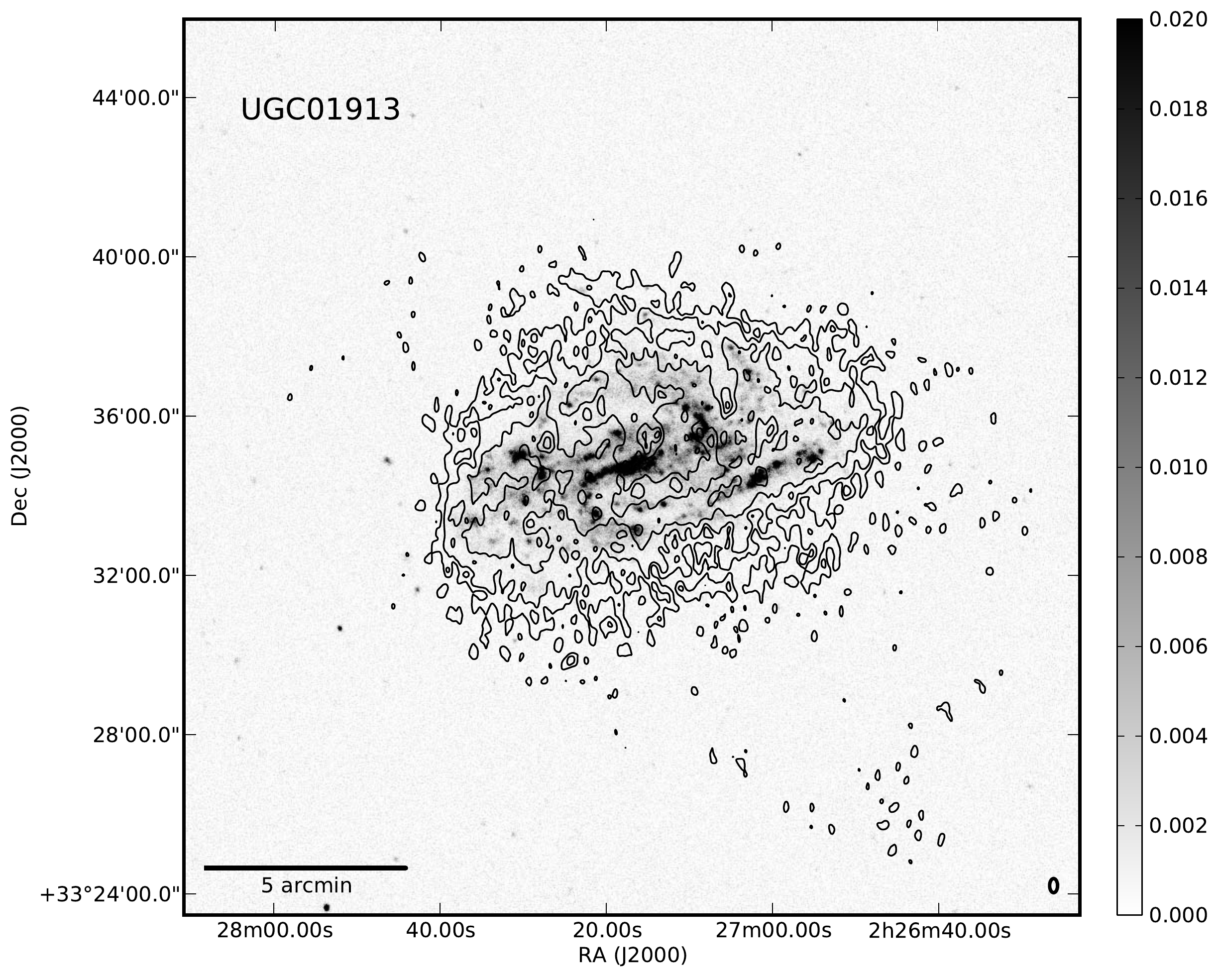}
\includegraphics[width=0.32\textwidth]{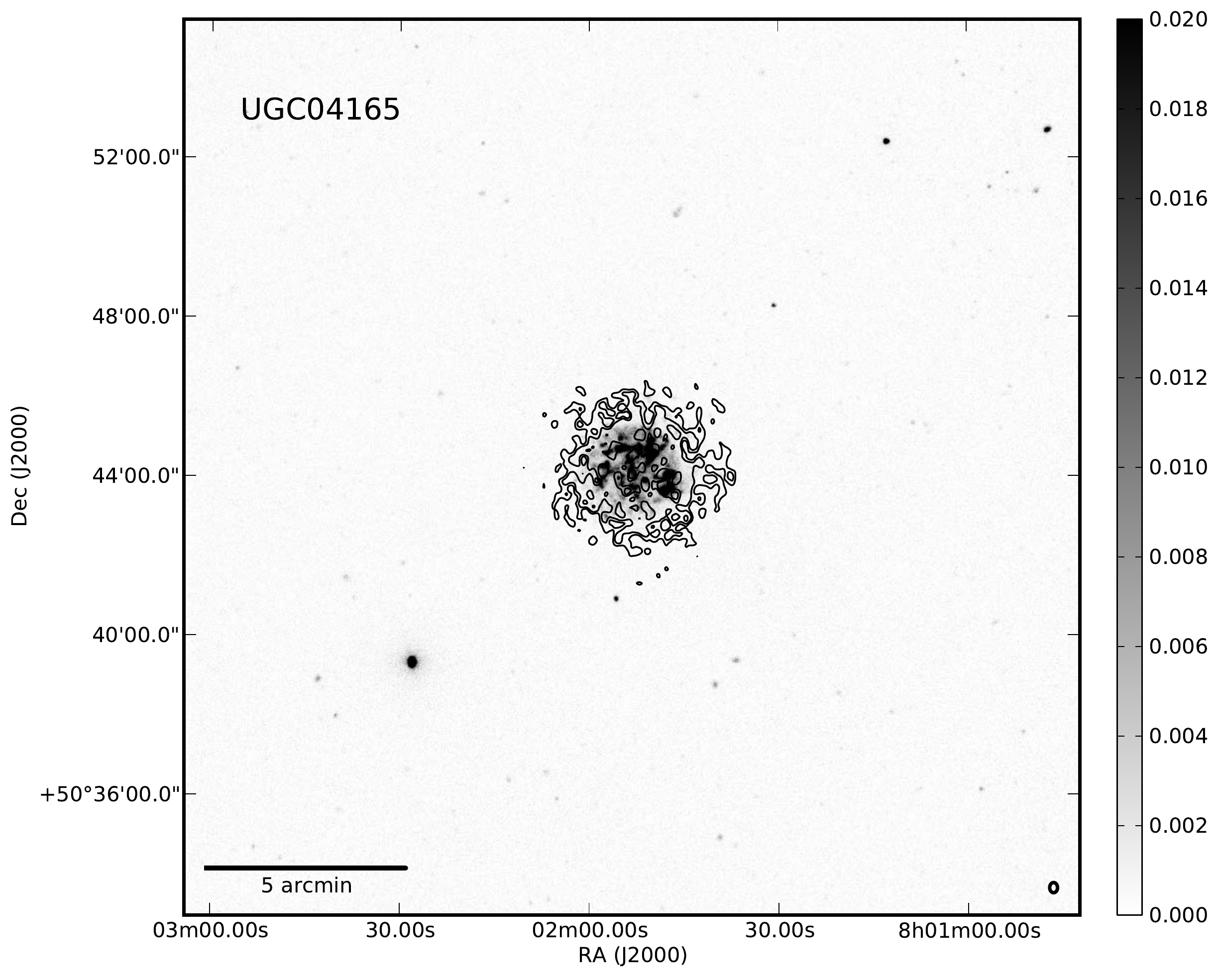}
\includegraphics[width=0.32\textwidth]{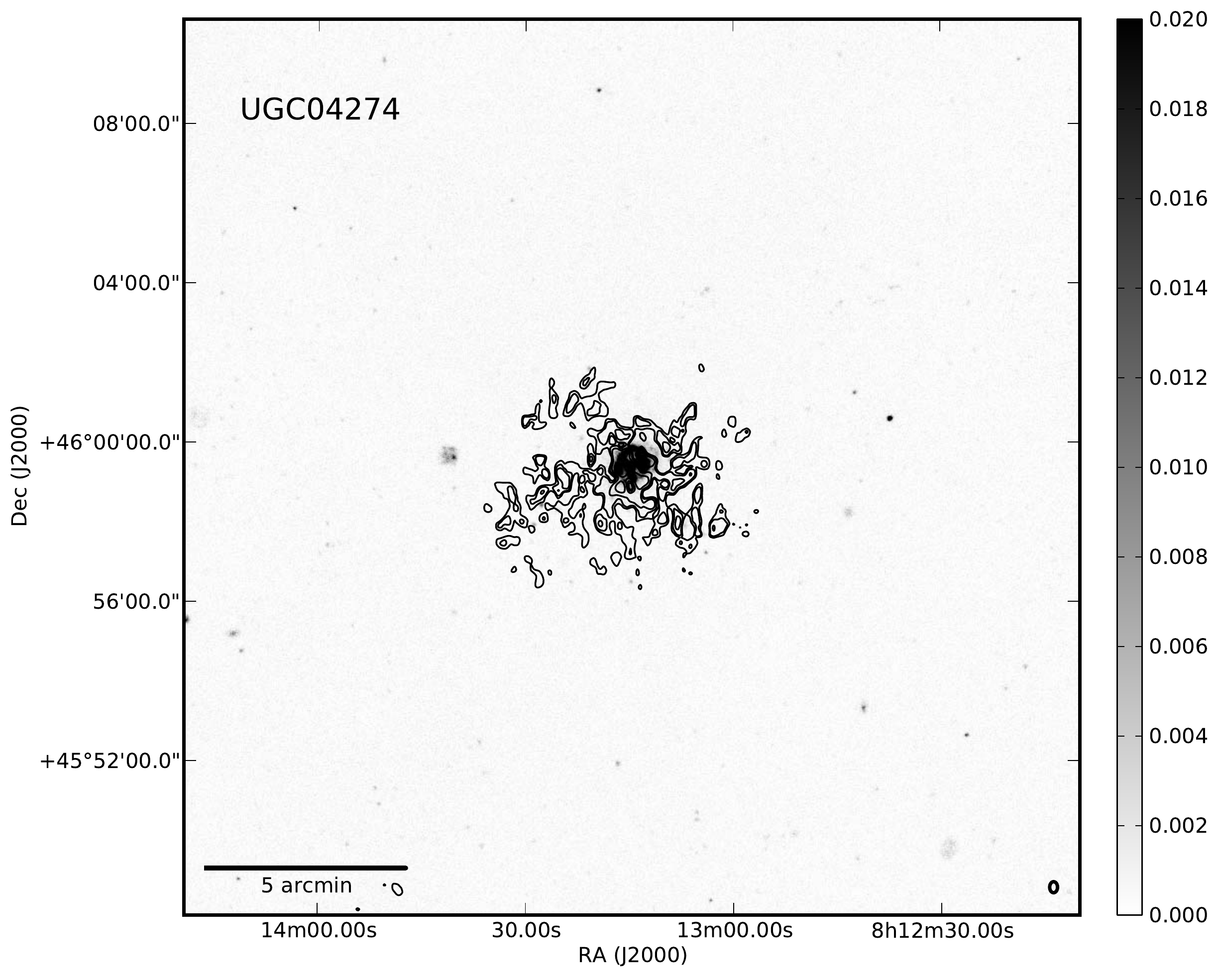}
\includegraphics[width=0.32\textwidth]{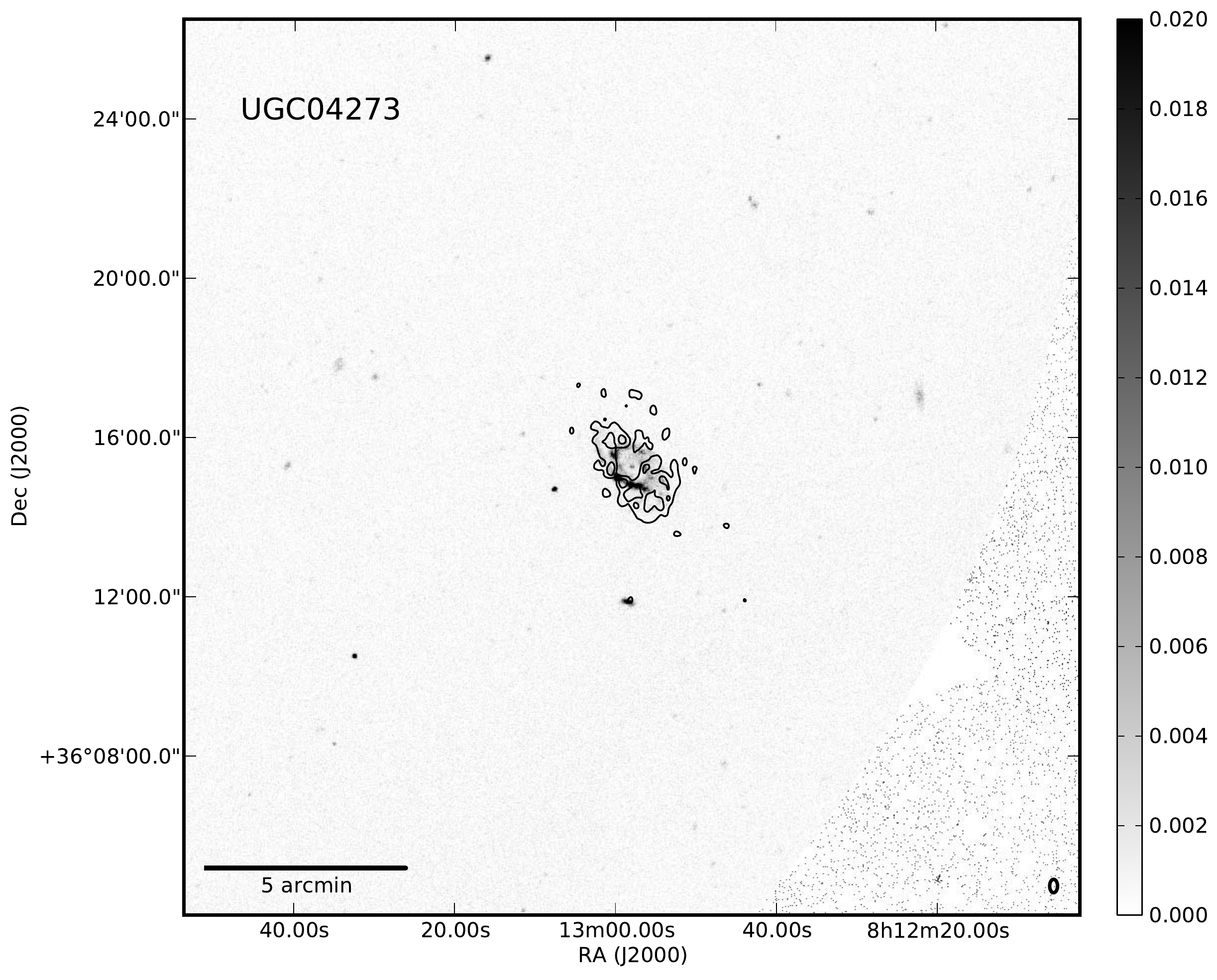}
\includegraphics[width=0.32\textwidth]{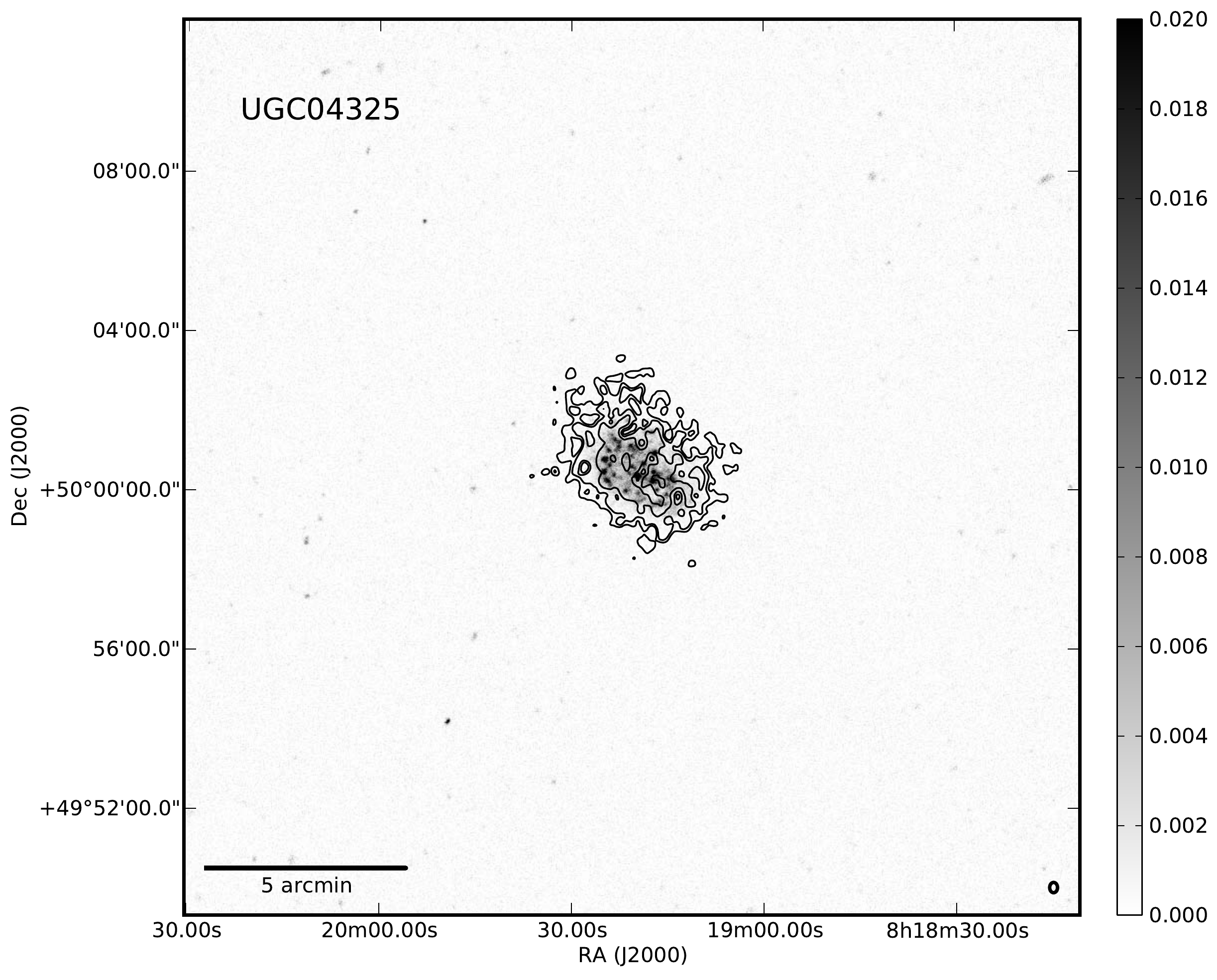}
\includegraphics[width=0.32\textwidth]{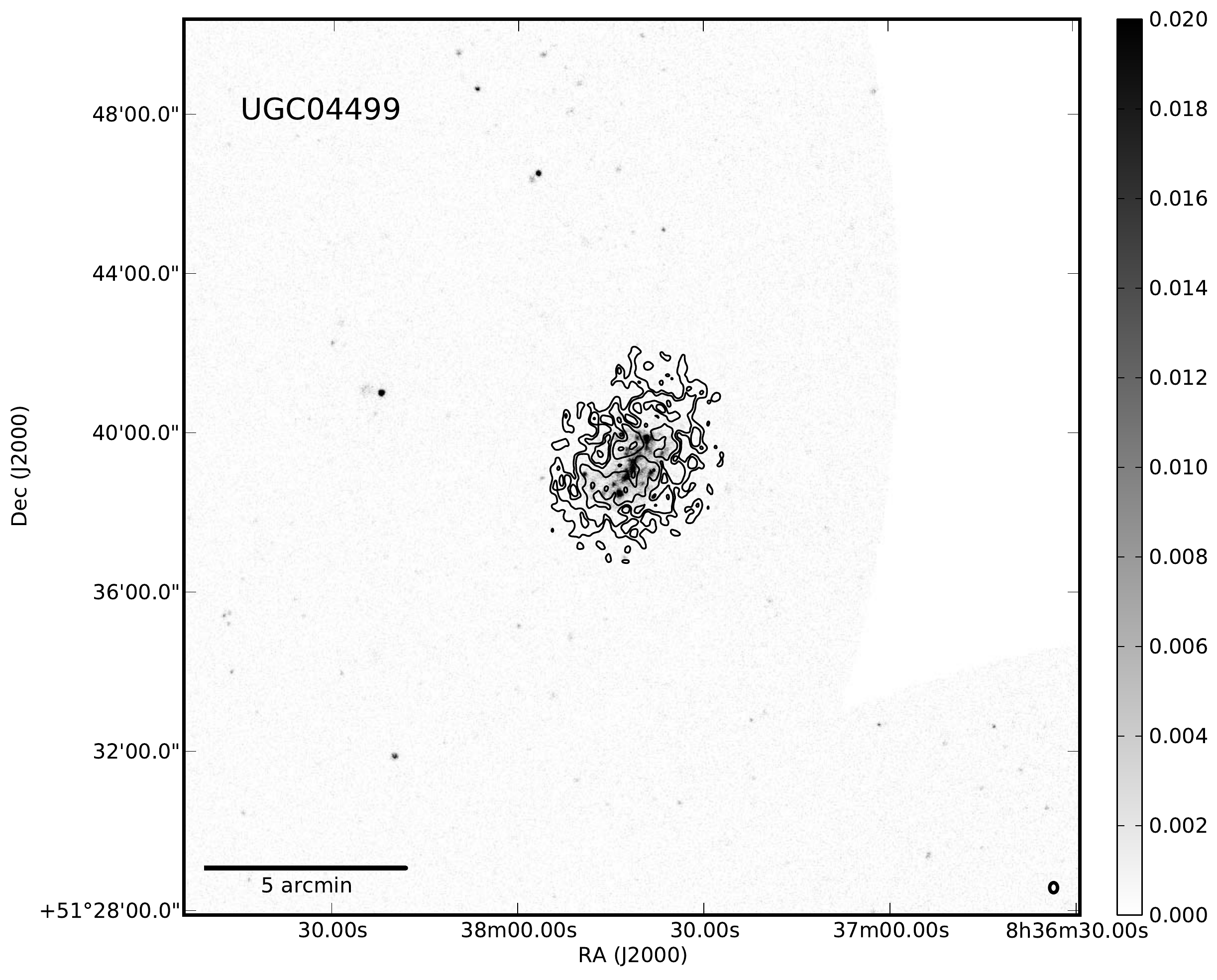}
\includegraphics[width=0.32\textwidth]{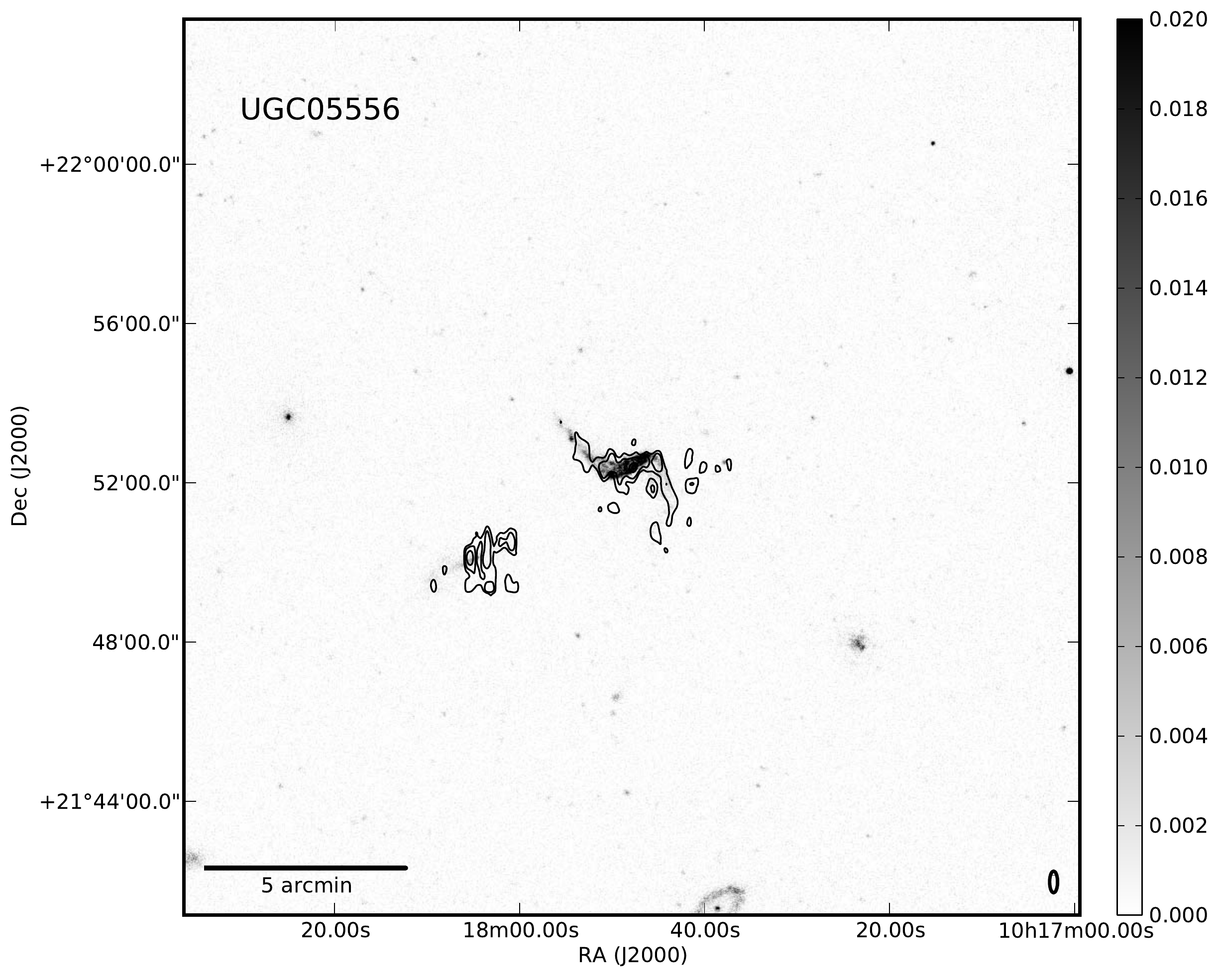}
\includegraphics[width=0.32\textwidth]{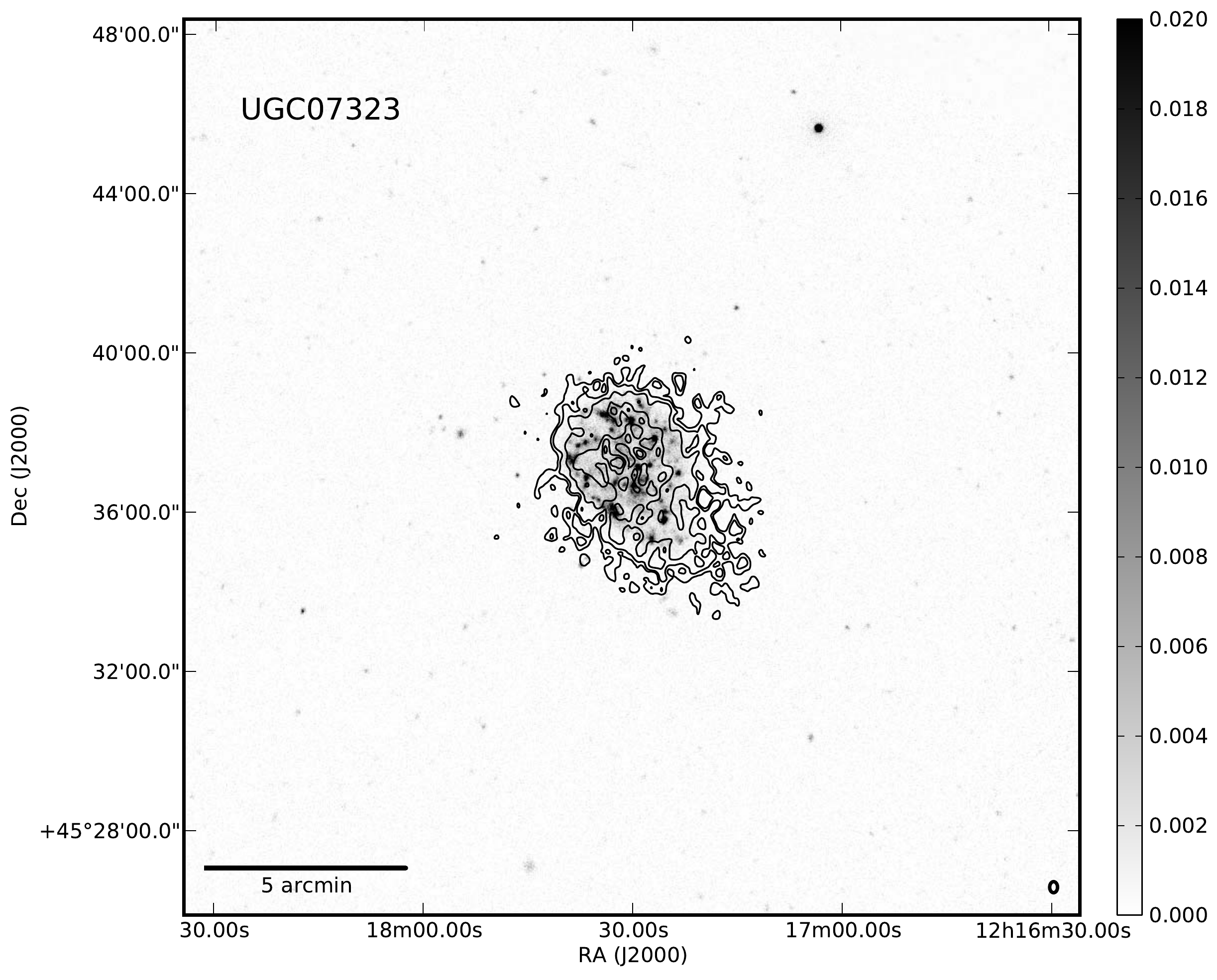}
\includegraphics[width=0.32\textwidth]{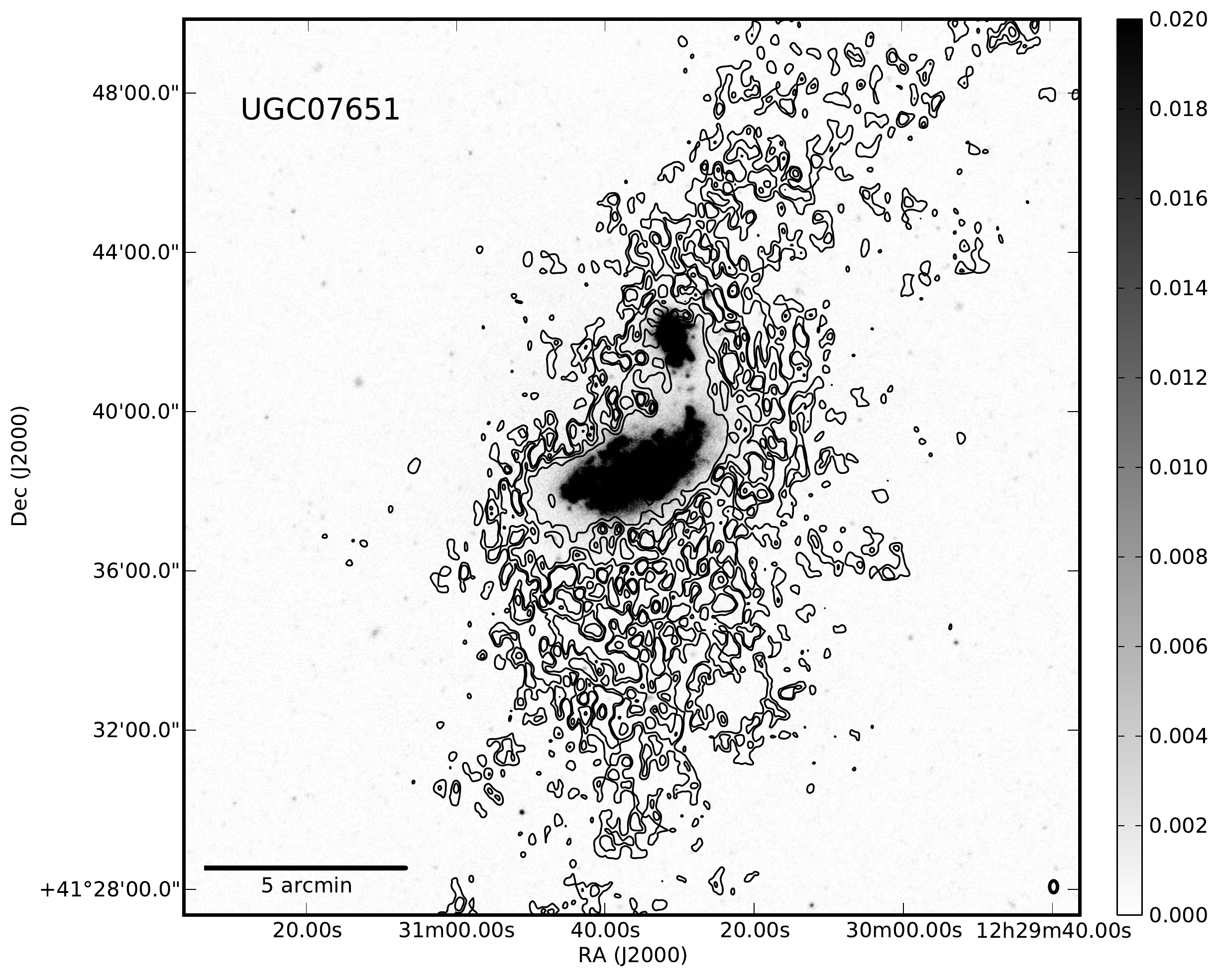}
\includegraphics[width=0.32\textwidth]{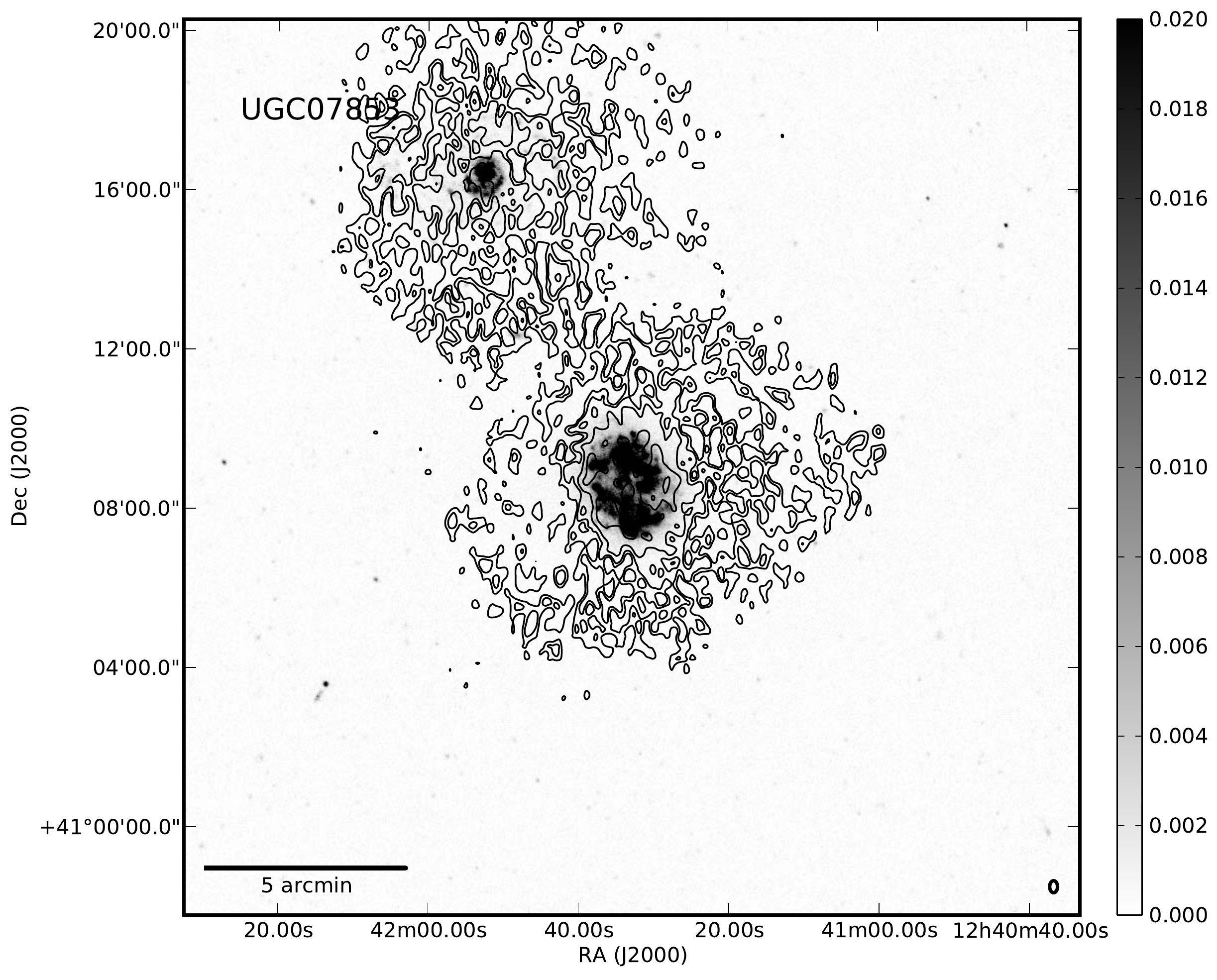}
\includegraphics[width=0.32\textwidth]{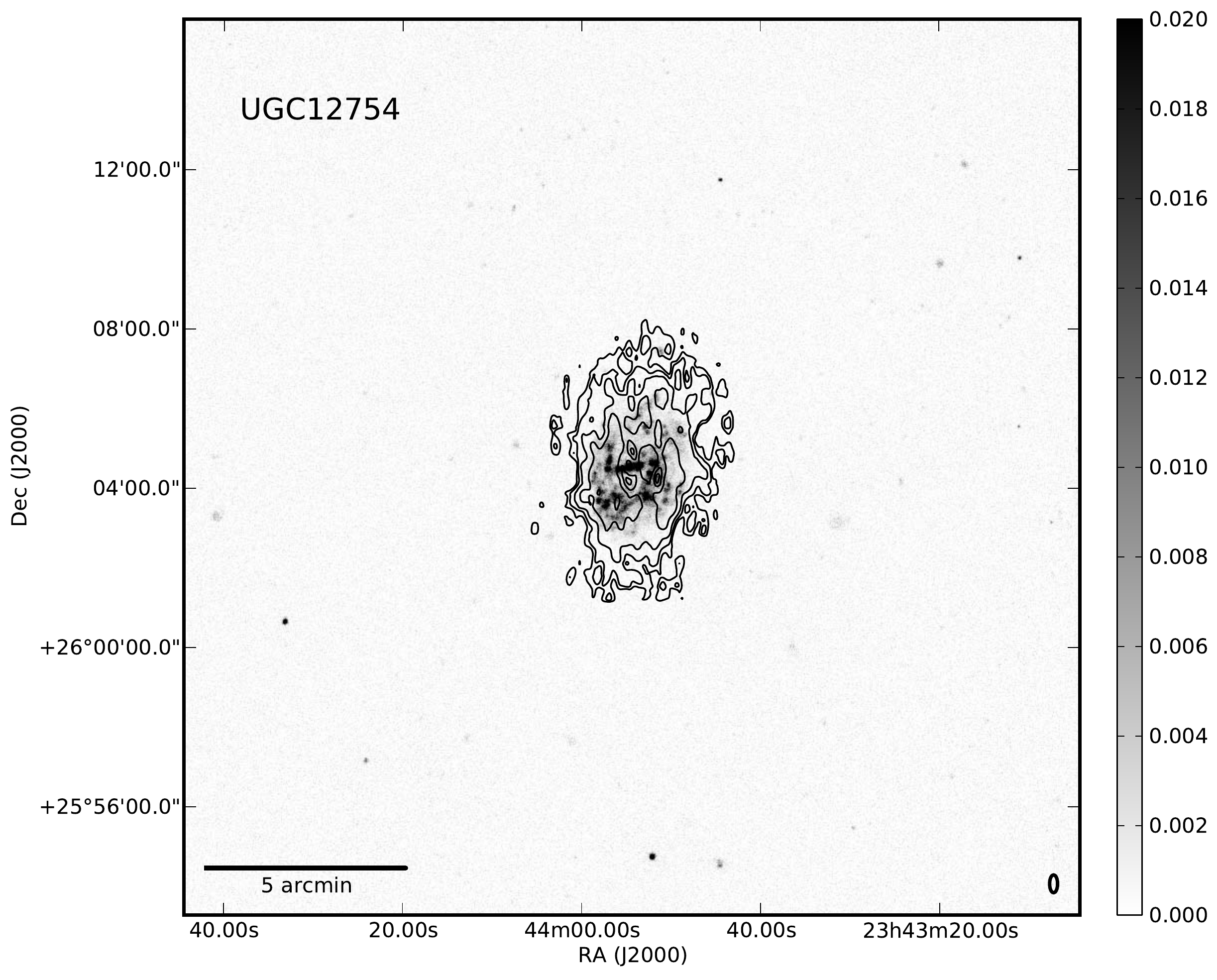}
\caption{The Type 0 \xuv\  disks --no \xuv\ disk identified by \protect\cite{Thilker07b}-- in the \whisp\ sample. There are two ongoing major mergers as well as a range of \hi\ morphologies. }
\label{f:xuvt0}
\end{center}
\end{figure*}

Figures \ref{f:xuvt1} and \ref{f:xuvt2} show the overlap between \whisp\ and the \xuv\ disks identified by Thilker. 
The Type 1 collection includes a single merger remnant (UGC 4862) and some \hi\ disks that appear slightly lopsided, tentatively supporting a tidal origin as one path to generate Type 1 disks. 
Figure  \ref{f:xuvt0} shows the \hi\ contours around those \whisp\ galaxies inspected by Thilker et al. but not found to contain an \xuv\ disk. There are two instances of clear ongoing mergers as well as a variety of \hi\ morphologies.


\subsection{\things}
\label{s:things}

Since there overlap between \whisp\ sample and the Thilker et al \xuv\  classification sample is small, we also compare the \things\ survey's \hi\ and \fuv\ morphologies to their classifications of \xuv\ disks (Table \ref{t:things}).
We use the morphological information from \cite{Holwerda11a}, with the concentration definition appropriately amended. The morphological classifications are based on the public datasets from the \hi\ Nearby Galaxy Survey \citep[THINGS,][]{Walter08}\footnote{\url{http://www.mpia-hd.mpg.de/THINGS/}}, and GALEX Nearby Galaxy Atlas \citep[NGA,][]{nga}\footnote{\url{http://galex.stsci.edu/GR4/}}, retrieved from MAST (\url{http://galex.stsci.edu}).
We use the robust-weighted (RO) \things\ column density maps as these are similar in resolution as the native resolution of \galex\ for the contour of $0.3 \times 10^{20}$ cm$^{-2}$, approximately the spatial extent of the \hi\ disk. Table \ref{t:things} lists the THINGS galaxies that have \xuv\  disk classifications from \cite{Thilker07b}. 
The benefits of the \things\ sample are that there is an \xuv\ classification for all galaxies, the \hi\ and \uv\ data are of comparable spatial resolution, and the \galex\ observations are of uniform depth. 
Drawbacks are that is is an equally small sample as the \whisp overlap and there is only a single ongoing merger in this sample as it was constructed to be quiescent.


%
%


\begin{table}
\caption{\xuv\  disks classified by \protect\cite{Thilker07b} and one from \protect\cite{Lemonias11} for the \whisp\ galaxies in their respective samples. }
\begin{center}
\begin{tabular}{l l l }
Name	 	& UGC	& Type\\ 
\hline
\hline
	-		& 1913 	& 0 \\
	-		& 4165 	& 0 \\
	-		& 4274 	& 0 \\
	-		& 4273 	& 0 \\
NGC 2541  	& 4284 	& 2	\\
	-		& 4325 	& 0 \\
	-		& 4499 	& 0 \\
NGC 2782	& 4862 	& 1 \\ 
	-		& 5556 	& 0 \\
NGC 3319	& 5789  	& 2 \\
NGC 3344	& 5840  	& 1\\
	-		& 6856 	& 0 \\
	-		& 7323 	& 0 \\
NGC 4258	& 7353  	& 1\\
	-		& 7651 	& 0 \\
NGC 4559	& 7766  	& 1\\
	-		& 7853 	& 0 \\
NGC 4625	& 7861  	& 1\\
NGC 5832	& 9649  	& 1/2 \\
	-		& 10445  	& 1/2\\
	-		& 10791  	& 1\\
	-		& 12754 	& 0 \\

\hline
\end{tabular}
\end{center}
\label{t:thilker}
\end{table}%

\begin{table}
\caption{\xuv\  disks classified by \protect\cite{Thilker07b} for those \things\ galaxies in their sample. }
\begin{center}
\begin{tabular}{l l l }
Name	 		& Type	& Reference\\ 
\hline
\hline
NGC 628			& 1		& \cite{Lelievre00}\\
				&		& \cite{Alberts11} \\
NGC 925			& 0		& \\
NGC 2403		& 1/2		& \\
Holmberg-II		& 0		& \\
M81A			& 0		& \\
DDO53			& 0		& \\
NGC 2841		& 1		& \cite{Alberts11}\\
NGC 2903		& 0		& \\
HolmbergI		& 0		& \\
NGC 2976		& 0		& \\
NGC 3031		& 1		& \\
NGC 3184		& 0		& \\
NGC 3198		& 1		& \\
IC2574			& 2		& \\
NGC 3351		& 0		& \\
NGC 3521		& 0		& \\
NGC 3621		& 1		& \cite{Alberts11}\\
NGC 3627		& 0		& \\
NGC 4736		& 0		& \cite{Zaritsky07} \\
DDO154			& 0		& \\
NGC 4826		& 0		& \\
NGC 5055		& 1		& \cite{Alberts11} \\
NGC 5194		& 0		& \\
NGC 5457		& 1		& \\
NGC 6946		& 0		& \\
NGC 7331		& 0		& \\
NGC 7793		& 0		& \\
\hline
\end{tabular}
\end{center}
\label{t:things}
\end{table}%

\section{Application of Quantified Morphology to H\,I and {\sc galex} data}
\label{s:app}

Before we compute the quantified morphology parameters described above, we preprocess the images as follows. First, we cut out a postage stamp of both the \hi\ maps and {\sc galex} data centered on the UGC position. Similar to \cite{Holwerda11d}, we use a stamp size of $22 \times 22$ arcmin around this position. We then smooth the {\sc galex} data to approximately the same spatial resolution as the \whisp\ HI data ($\sim$ 12"). We use the \hi\ map to define the part of the data over which the morphological parameters are computed over in the same manner as in \cite{Holwerda11b, Holwerda11d}; a threshold of $10^{20}$ atoms/cm$^2$.


We then compute the above quantified morphology parameters for the \hi\ map and far-, and near-ultraviolet images (summarized in Tables \ref{t:hi}, \ref{t:fuv} and \ref{t:nuv}, {\em full versions in the electronic edition}). In addition to the above parameters, we compute the effective or half-light radius ($R_{50}$) for each image as an indication of disk size on the sky.

\section{Analysis}
\label{s:analysis}

To explore in what type of \hi\ disk \xuv\ disks predominantly occur, 
we first compare the relative effective radii of the \hi\ and \uv\ disks, 
secondly compare the individual morphological parameters in both \hi\ and \uv\ against one another, 
and thirdly explore the \uv\ and \hi\ morphologies alone, to explore if \xuv\ can be identified from \uv\ morphology and to see if \xuv\ disks are predominantly in merging or interacting disks.


%
 
\begin{figure*}
\begin{center}
\includegraphics[width=0.49\textwidth]{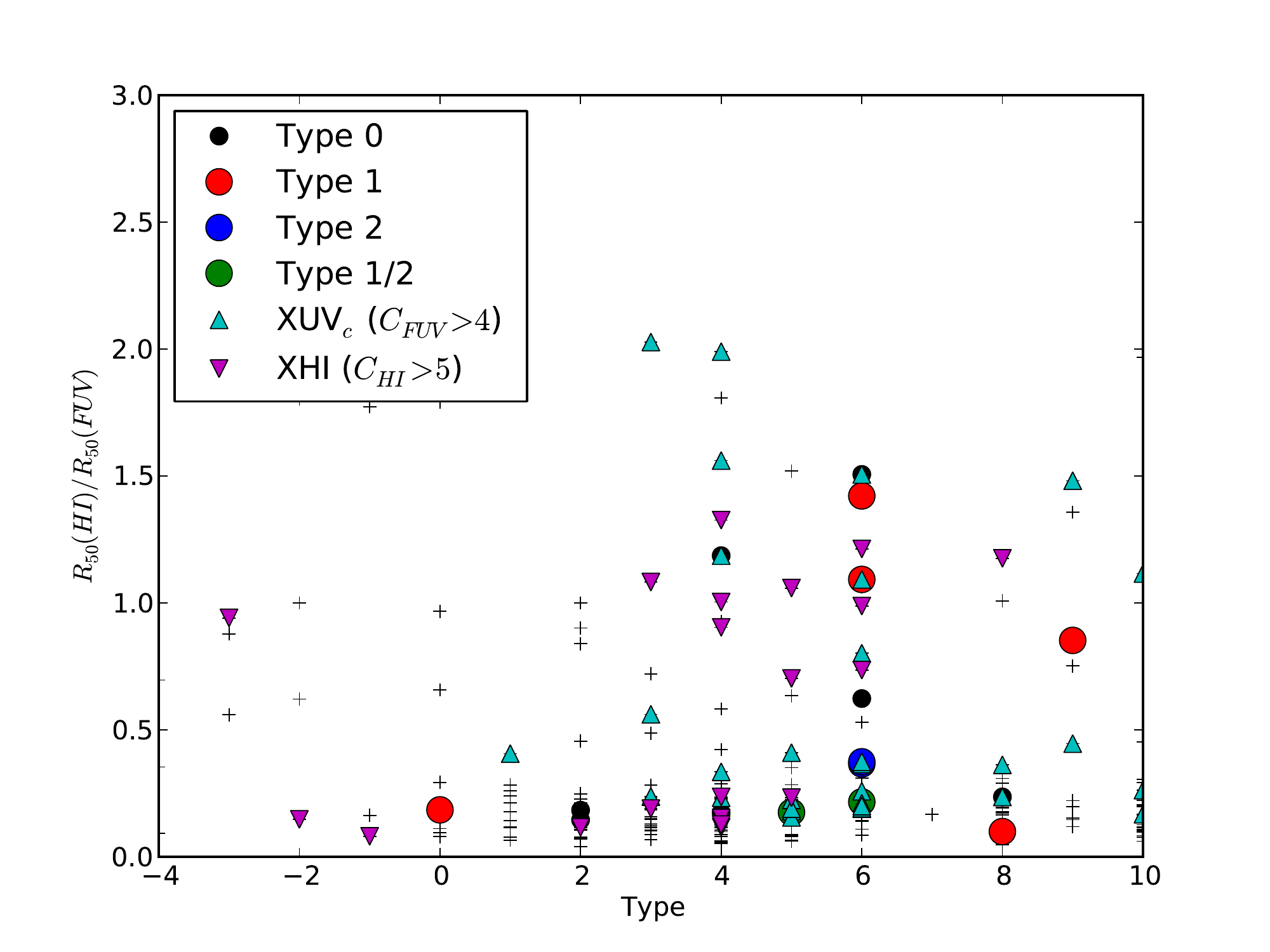}
\includegraphics[width=0.49\textwidth]{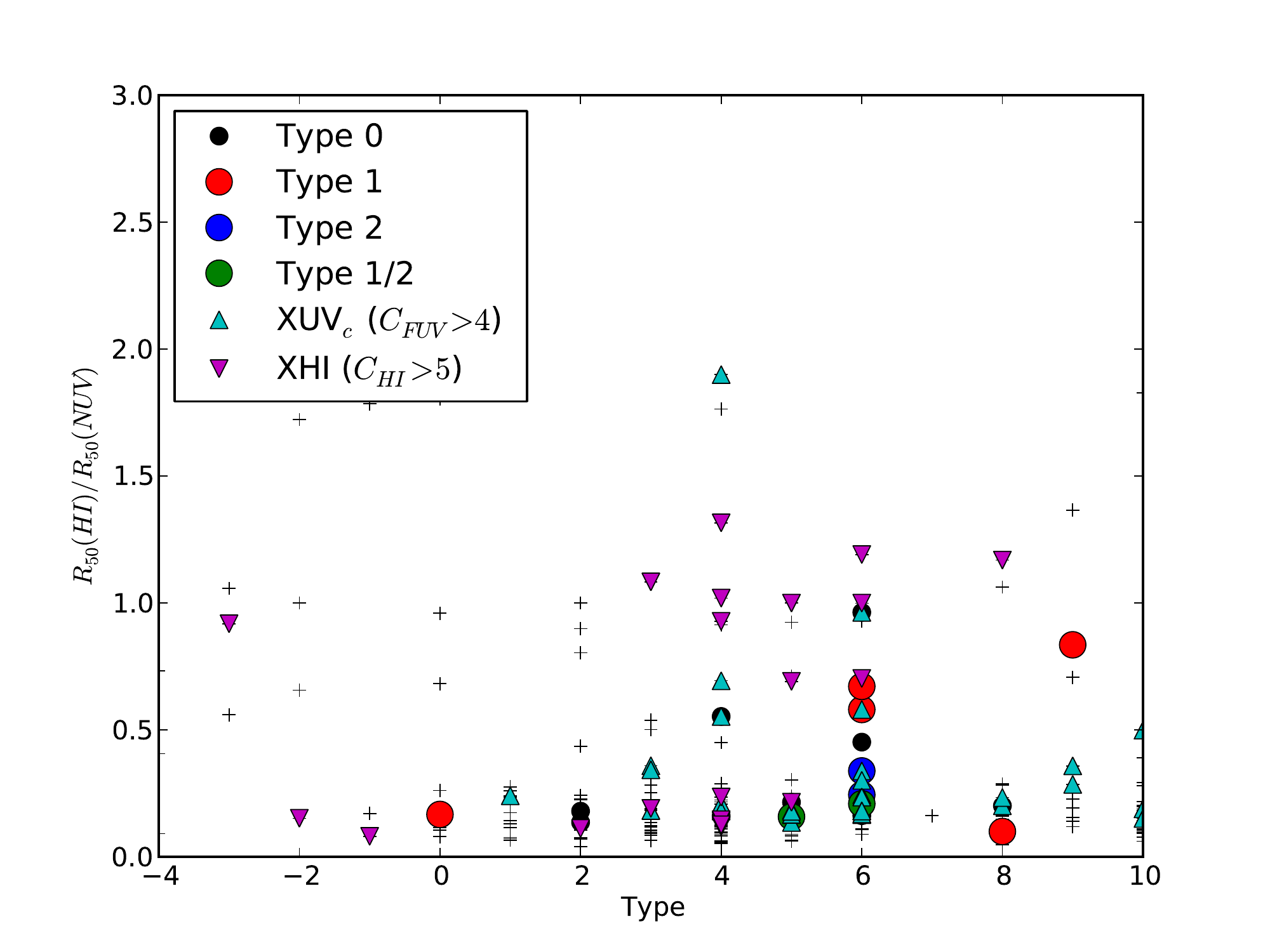}
\caption{The ratio of effective radii ($R_{50}$) of the \hi\ over \uv\ as a function of Hubble type as a function of type.}
\label{f:R50rat}
\end{center}
\end{figure*}

\subsection{Effective Radii}
\label{s:r50}

First we compare the relative sizes of the \nuv\ and \fuv\ disks and the \hi\ disk throught their effective radii, the radius which contains 50\% of the total flux.
A naive expectation would be that in the case of extended \uv\ disks, these would be similar to the \hi\ values. 
Figure \ref{f:R50rat} shows the ratio of the effective radii ($R_{eff}$) for \hi\ and \fuv\ or \nuv. Neither ratio singles out those identified by \cite{Thilker07b} as 
extended \uv\ disks. Thus the ratio of effective radii is not a good metric to automatically select \xuv\ disks. 

\begin{figure*}
\begin{center}
\includegraphics[width=0.49\textwidth]{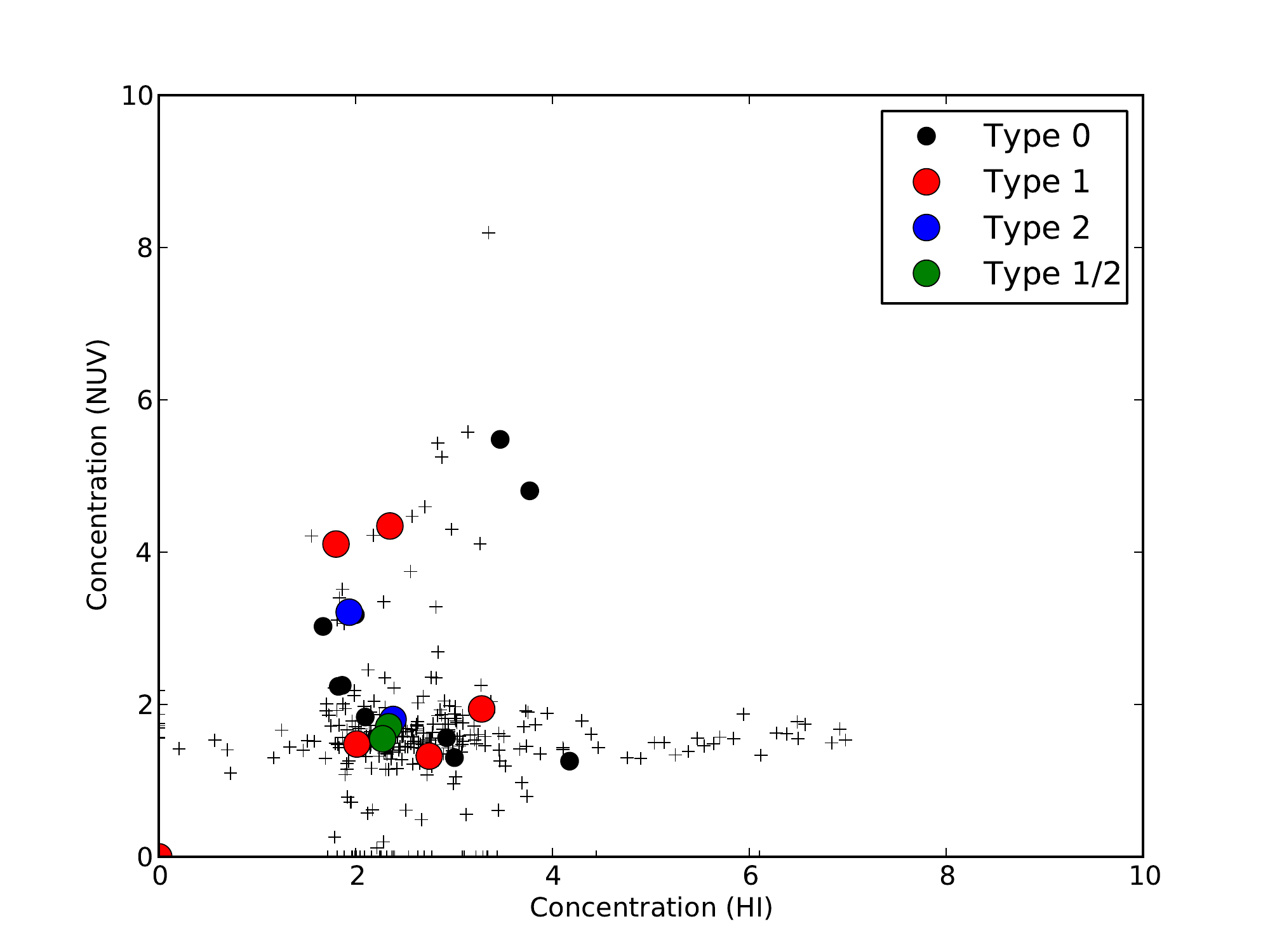}
\includegraphics[width=0.49\textwidth]{./holwerda_f5a.pdf}
\includegraphics[width=0.49\textwidth]{./holwerda_f5a.pdf}
\includegraphics[width=0.49\textwidth]{./holwerda_f5a.pdf}
\includegraphics[width=0.49\textwidth]{./holwerda_f5a.pdf}
\includegraphics[width=0.49\textwidth]{./holwerda_f5a.pdf}

\caption{The morphological parameters of the \hi\ disk and \nuv\ data; concentration (C), asymmetry (A), smoothness (S), the Gini index (G), and the second order moment of the brightest 20\% of the pixels (\m20), and the gini index of the second order moment ($G_M$).  }
\label{f:morphnuv}
\end{center}
\end{figure*}

\begin{figure*}
\begin{center}
\includegraphics[width=0.49\textwidth]{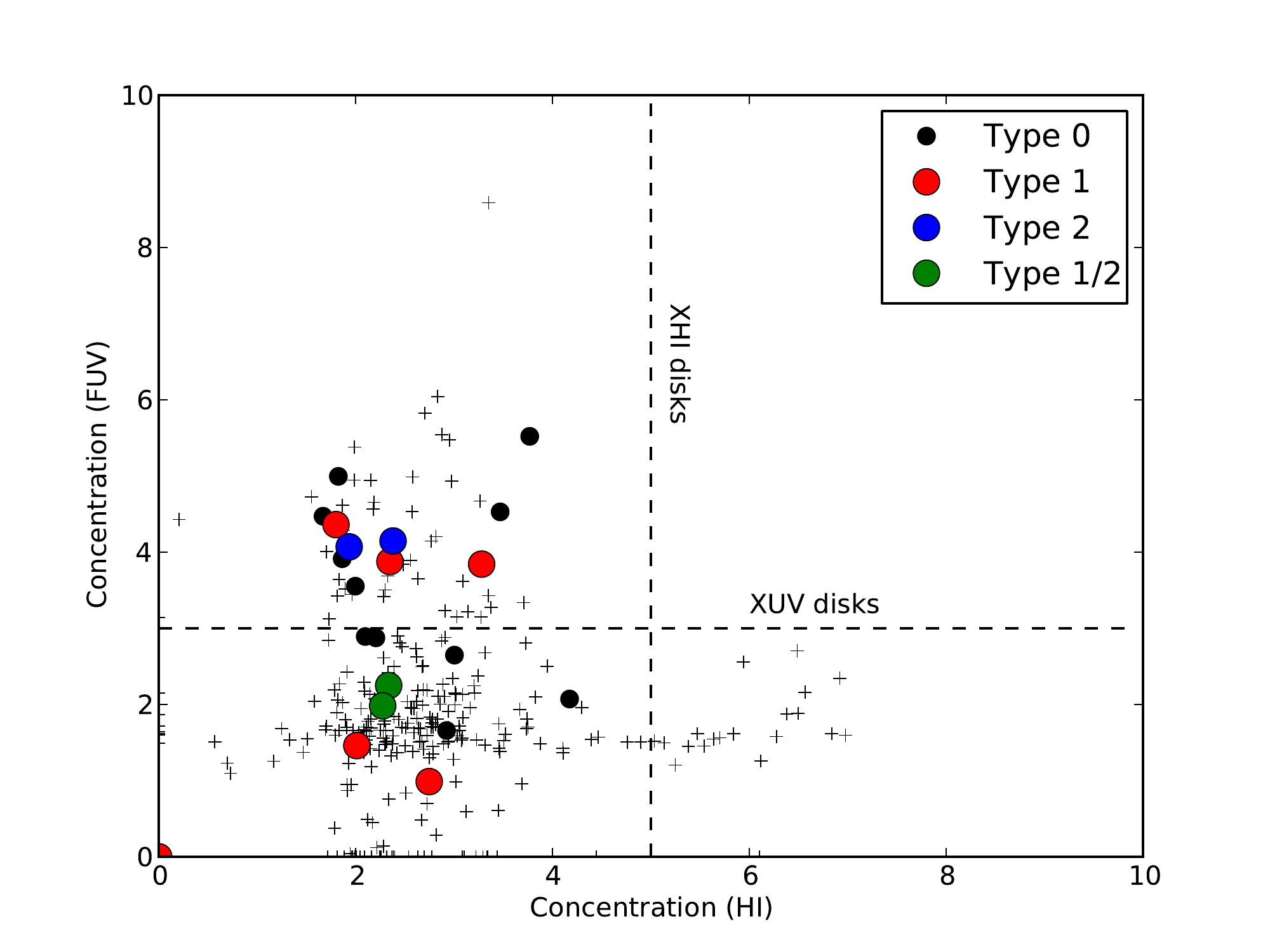}
\includegraphics[width=0.49\textwidth]{./holwerda_f6a.pdf}
\includegraphics[width=0.49\textwidth]{./holwerda_f6a.pdf}
\includegraphics[width=0.49\textwidth]{./holwerda_f6a.pdf}
\includegraphics[width=0.49\textwidth]{./holwerda_f6a.pdf}
\includegraphics[width=0.49\textwidth]{./holwerda_f6a.pdf}

\caption{The morphological parameters of the \hi\ disk and \fuv\ data; concentration (C), asymmetry (A), smoothness (S), the Gini index (G), and the second order moment of the brightest 20\% of the pixels (\m20), and the gini index of the second order moment ($G_M$).  }
\label{f:morphfuv}
\end{center}
\end{figure*}

\begin{figure*}
\begin{center}
\includegraphics[width=0.49\textwidth]{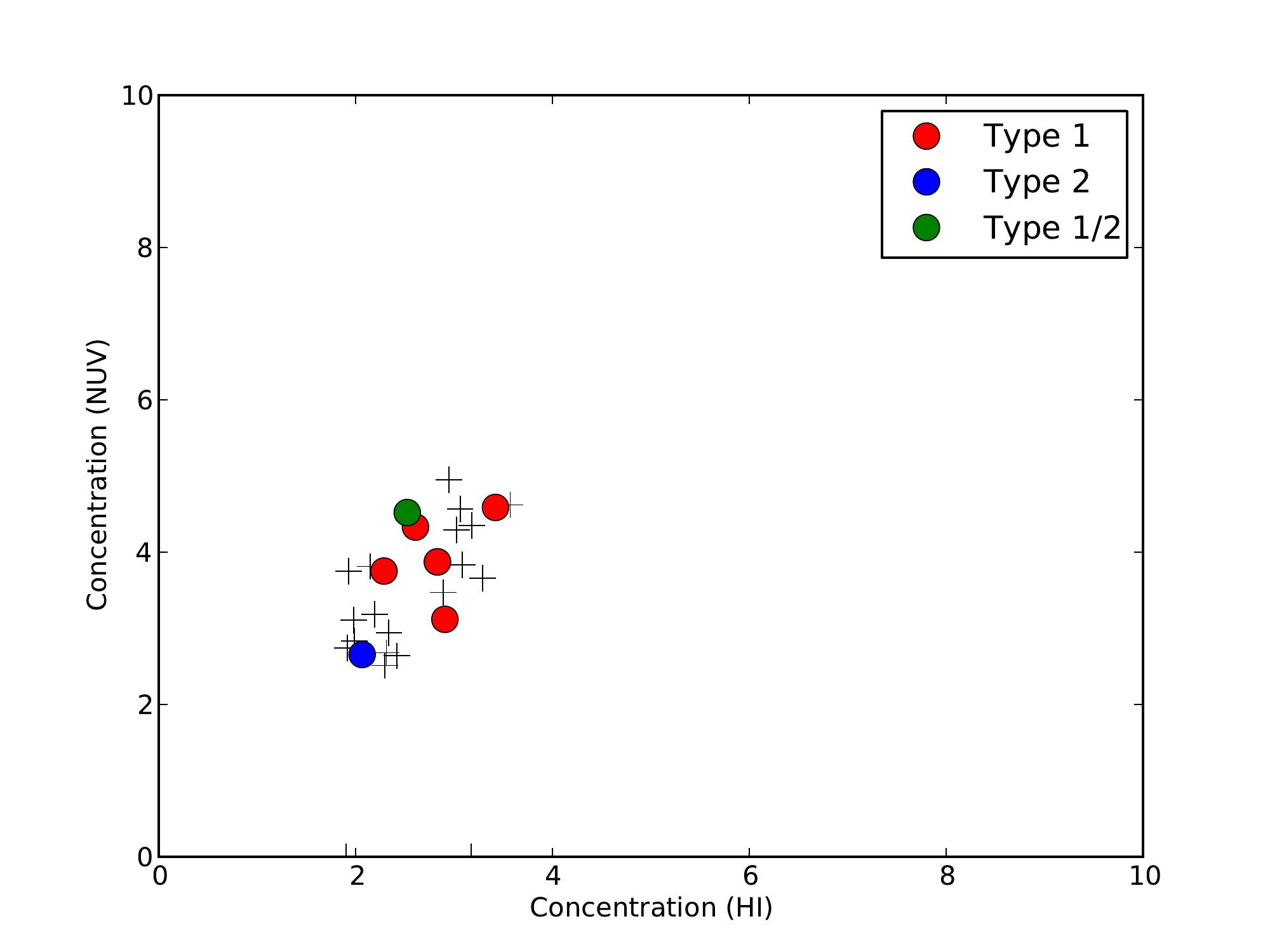}
\includegraphics[width=0.49\textwidth]{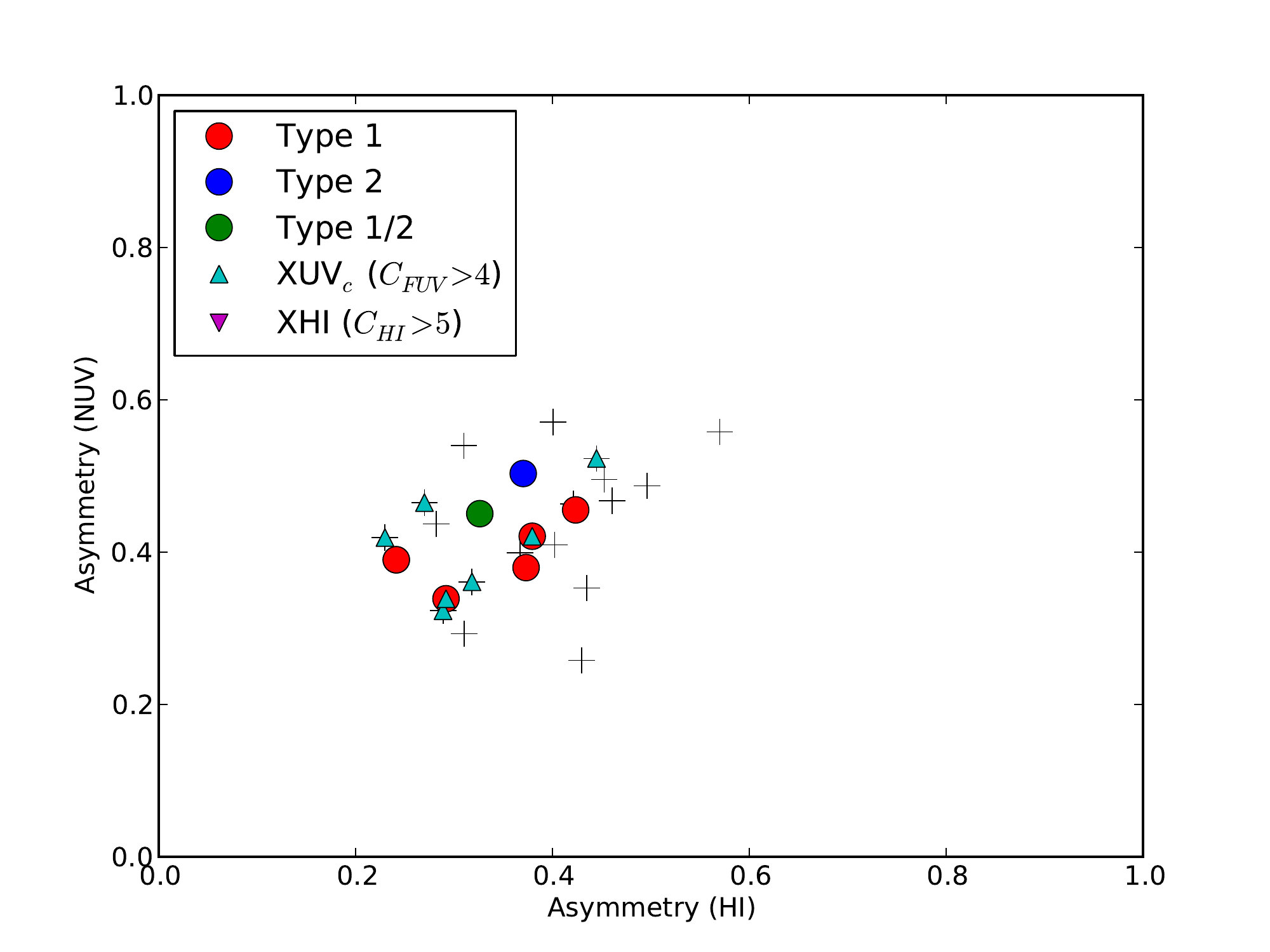}
\includegraphics[width=0.49\textwidth]{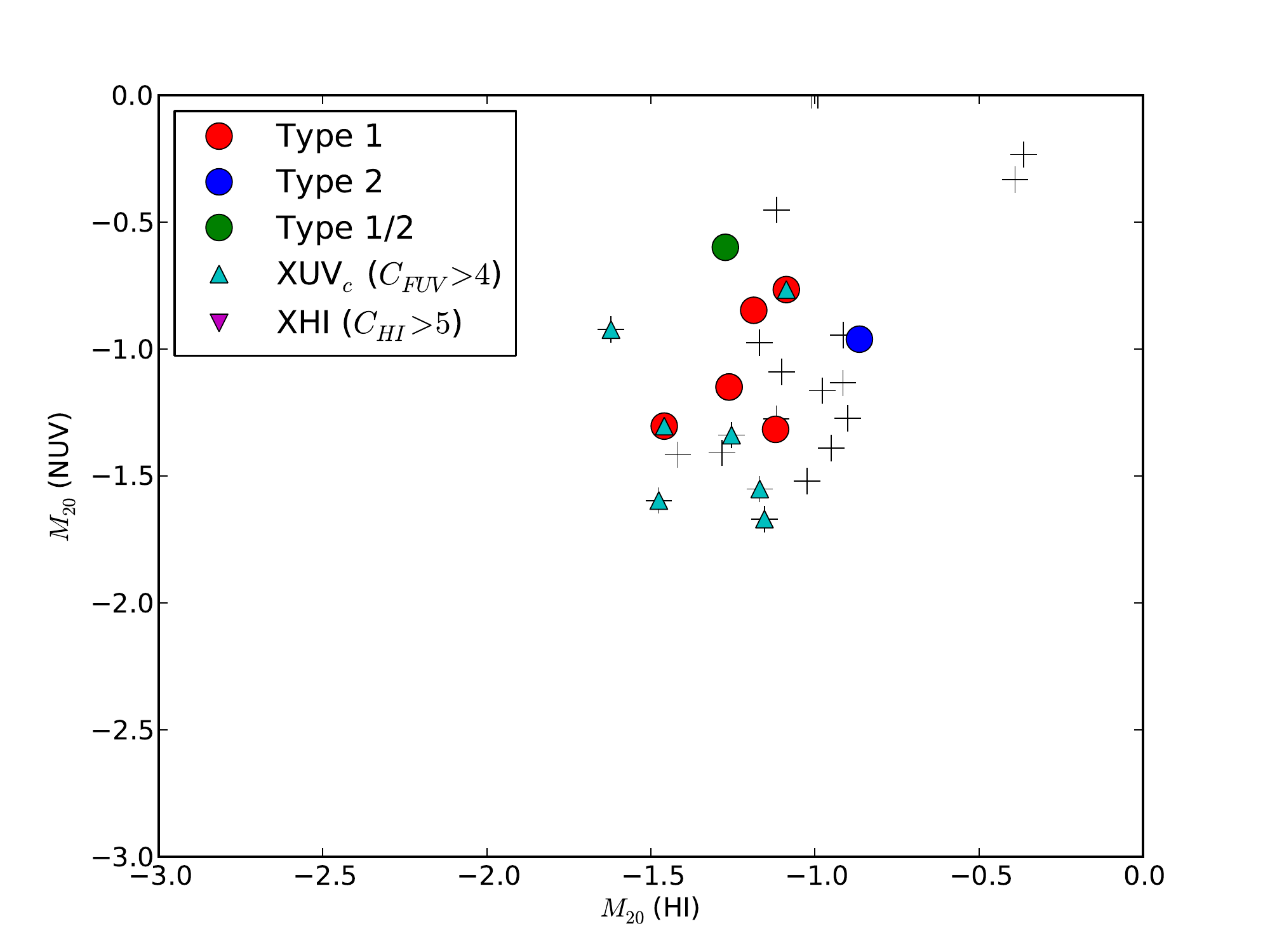}
\includegraphics[width=0.49\textwidth]{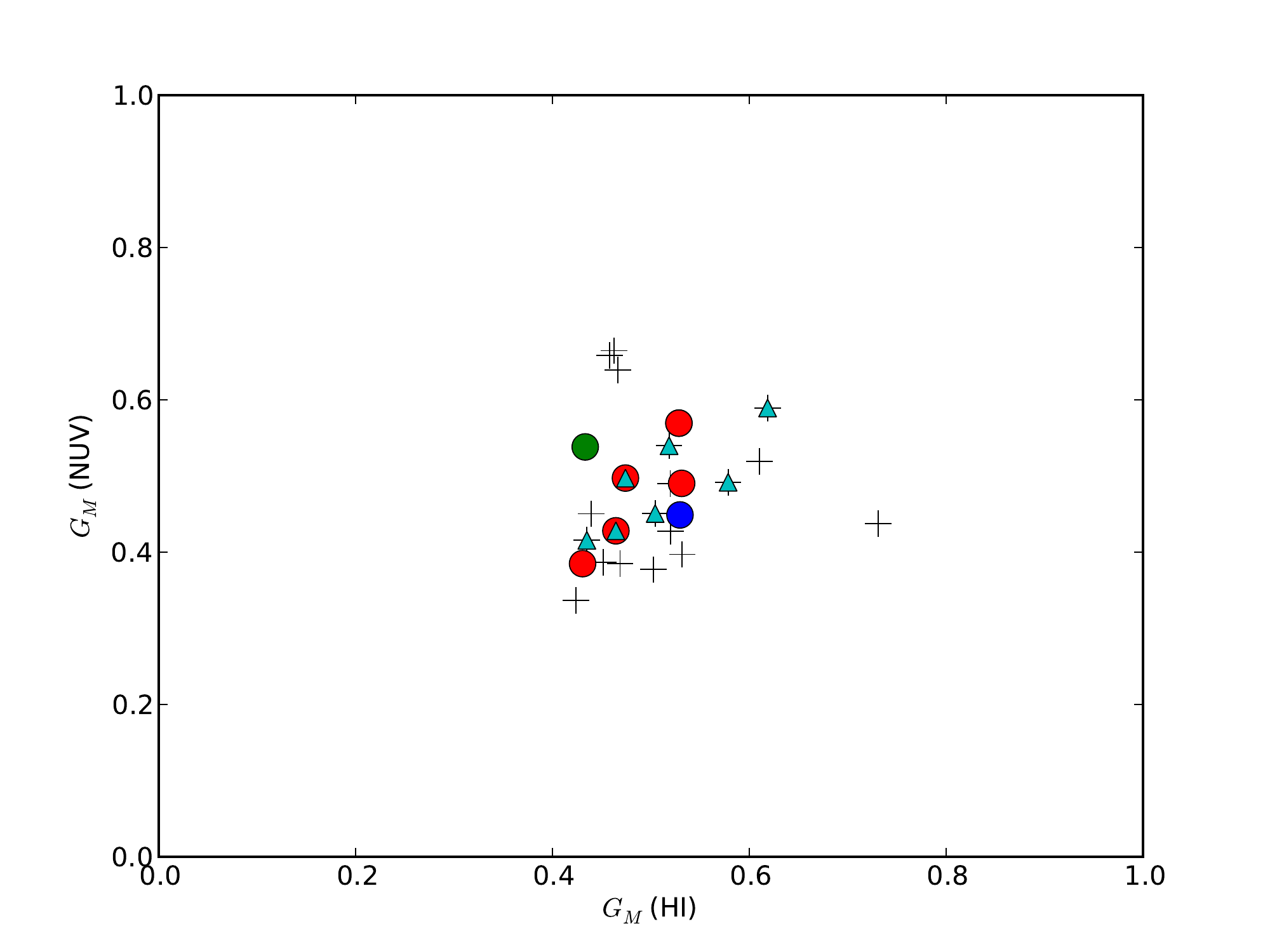}

\caption{The morphological parameters of the \hi\ disk and \nuv\ for the \things\ galaxies with the \xuv\ disk classification from \protect\cite{Thilker07b}. }
\label{f:thingsnuv}
\end{center}
\end{figure*}

\begin{figure*}
\begin{center}
\includegraphics[width=0.49\textwidth]{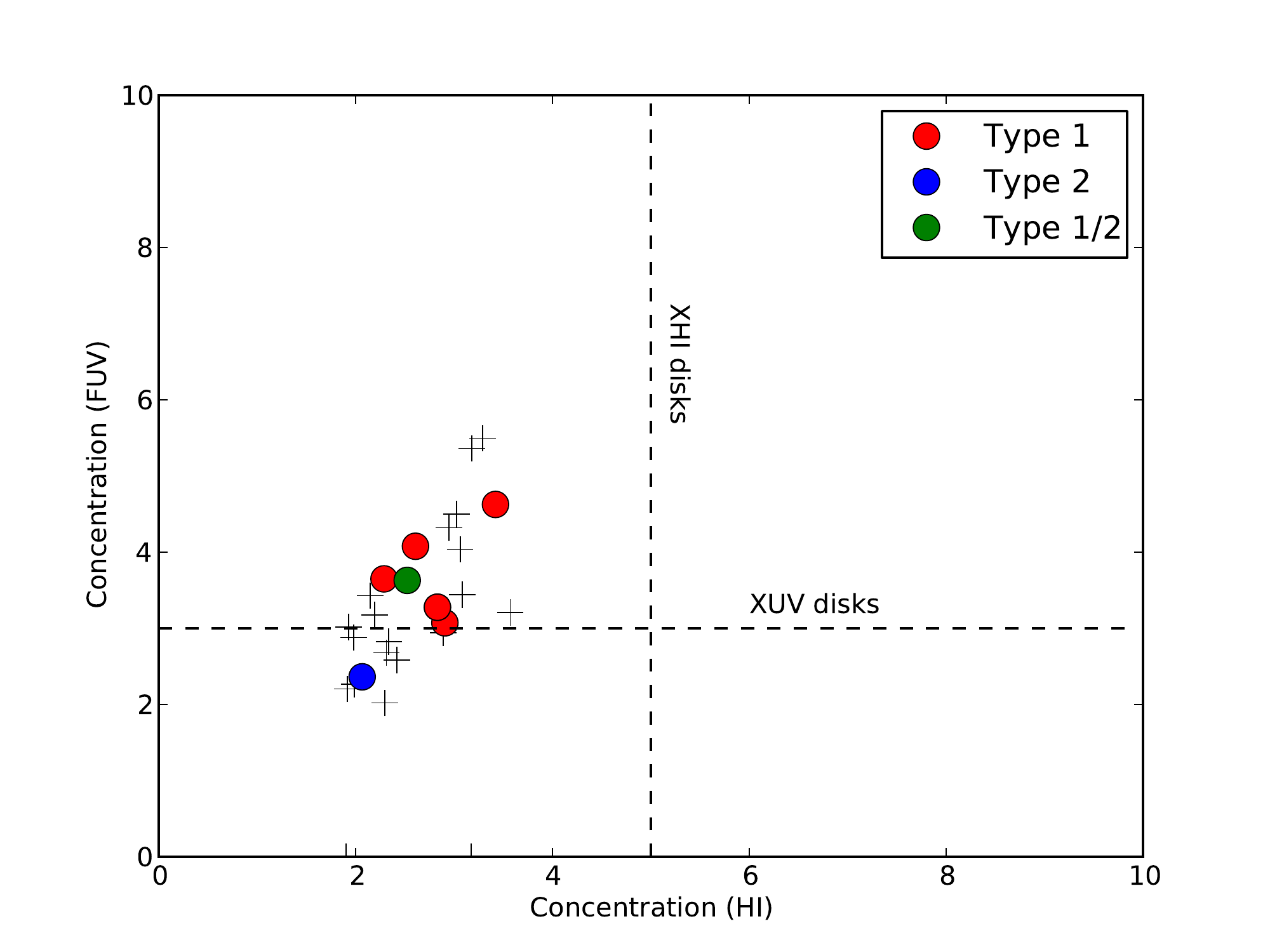}
\includegraphics[width=0.49\textwidth]{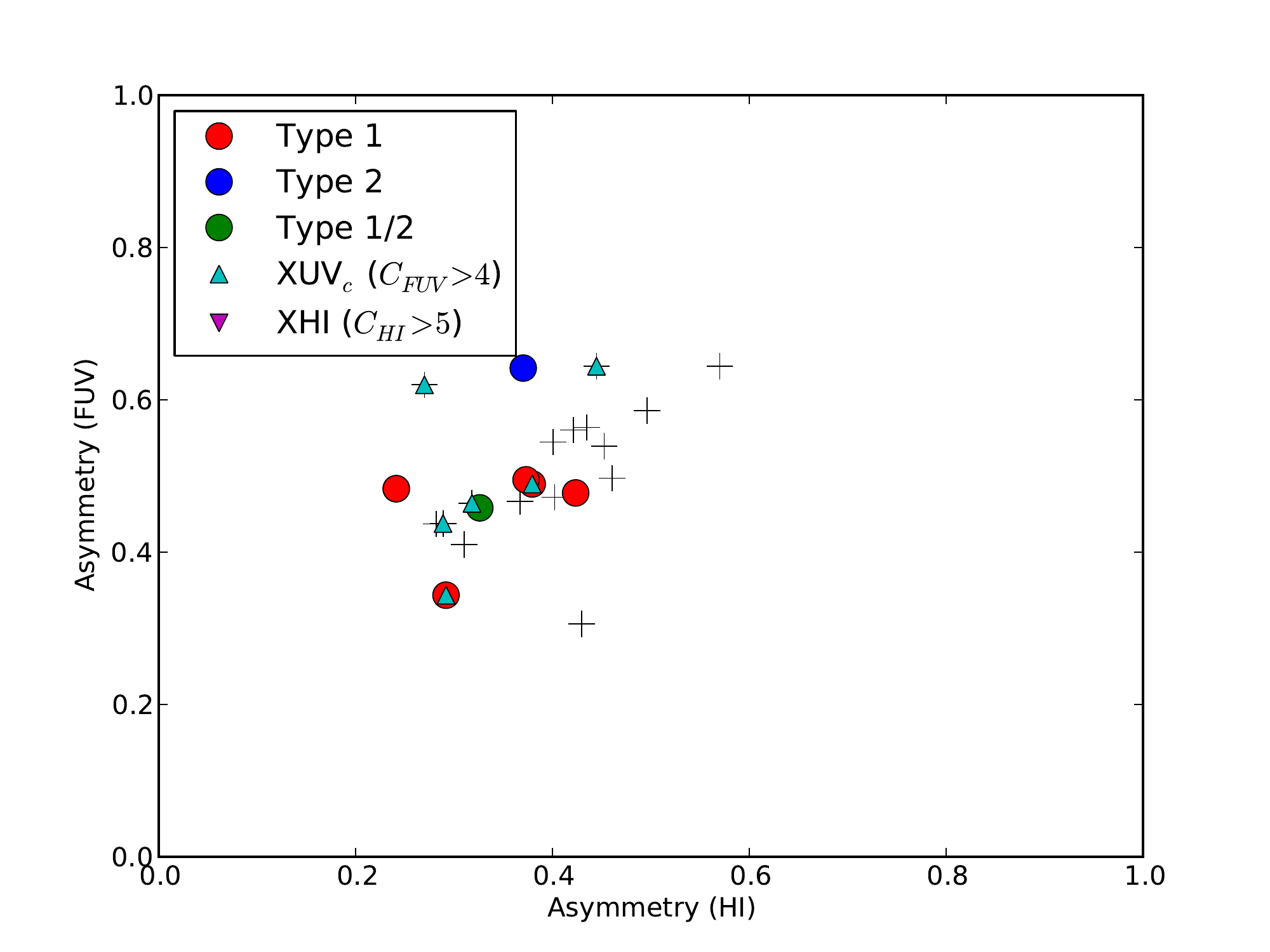}
\includegraphics[width=0.49\textwidth]{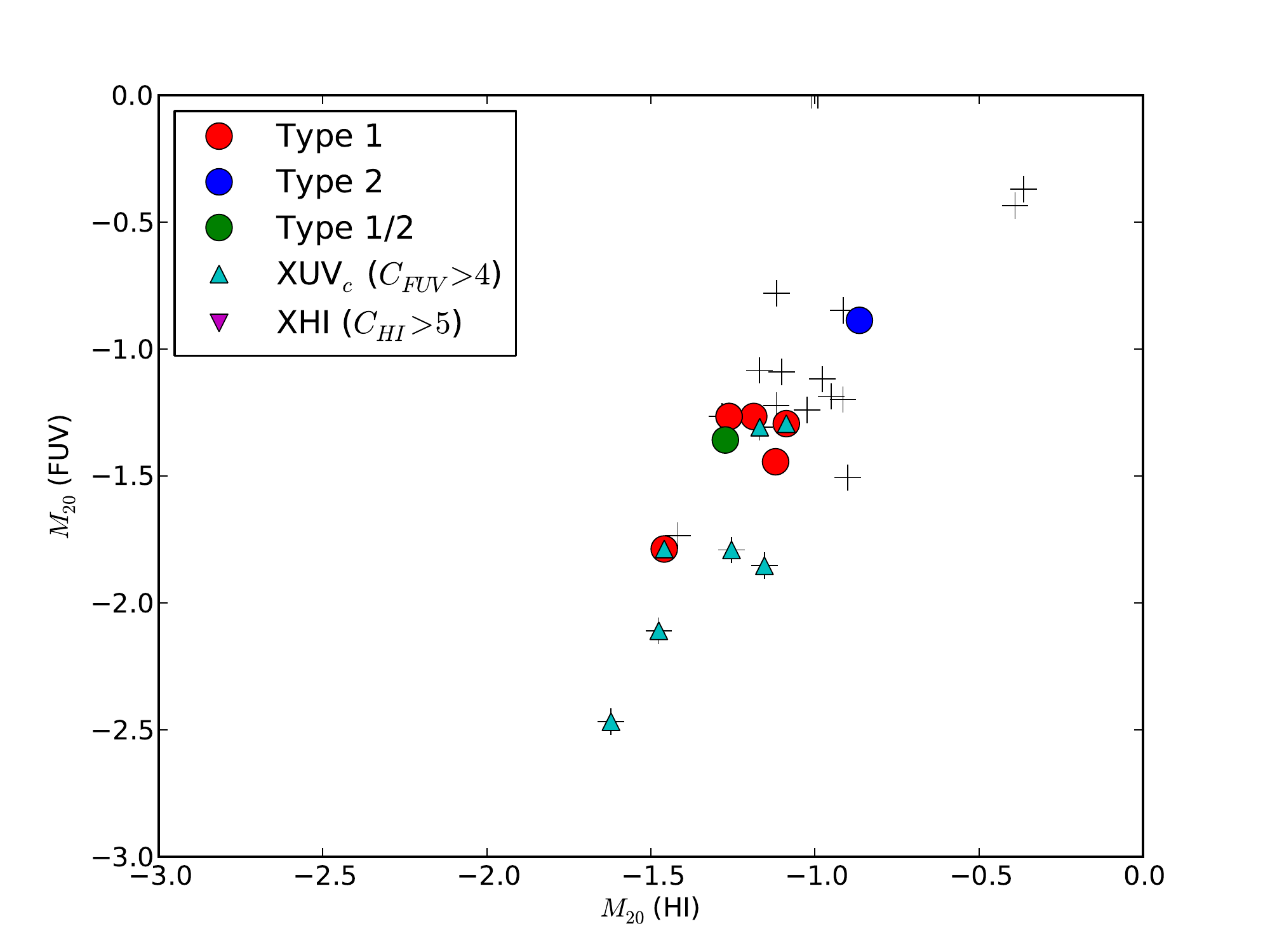}
\includegraphics[width=0.49\textwidth]{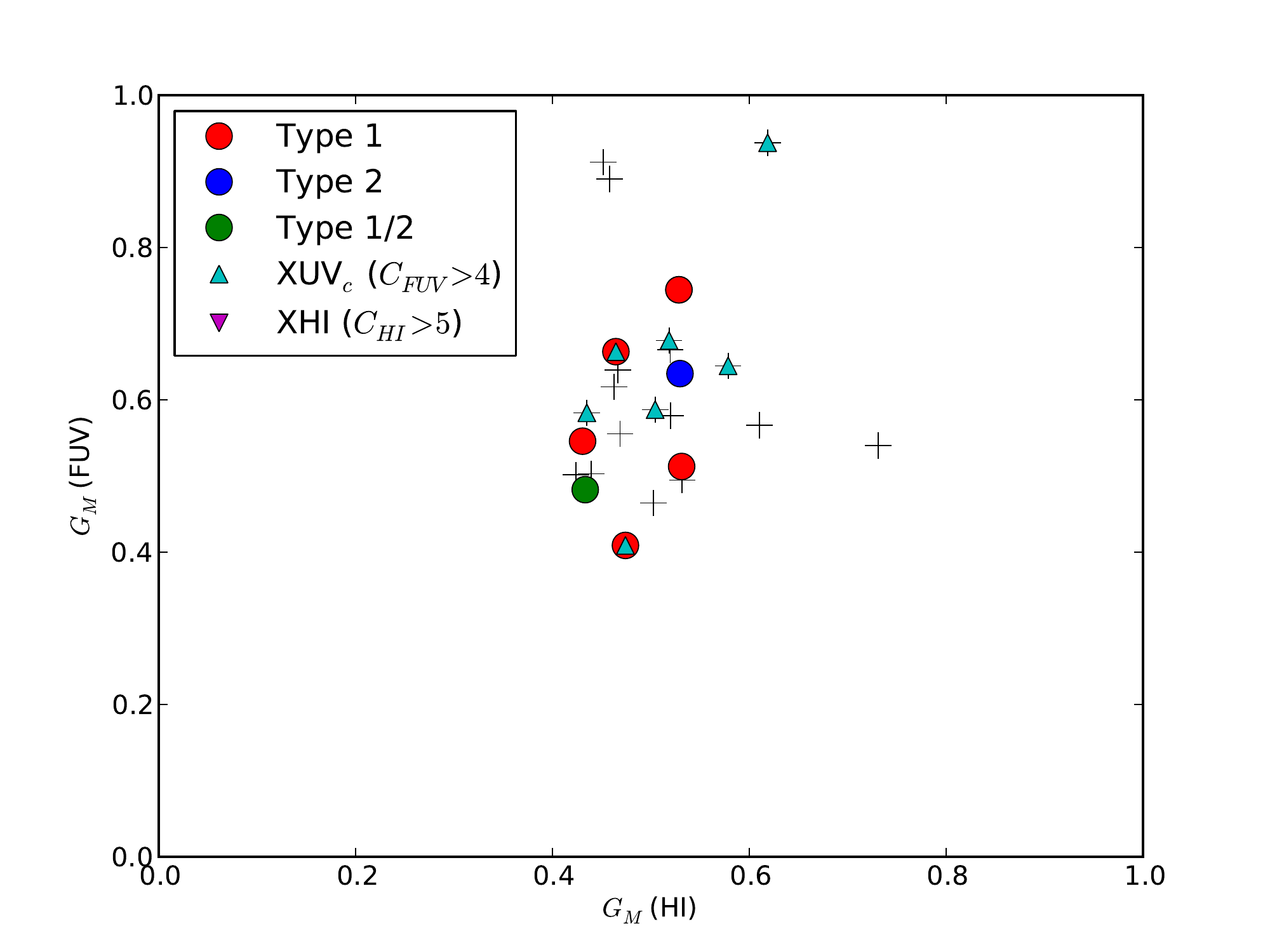}
\caption{The morphological parameters of the \hi\ disk and \fuv\ for the \things\ galaxies with the \xuv\ disk classification from \protect\cite{Thilker07b}. Dashed lines are the \hi\ and \uv\ concentration criteria from Figure \ref{f:morphfuv}.}
\label{f:thingsfuv}
\end{center}
\end{figure*}

\subsection{\uv vs \hi\ Morphologies}
\label{s:morphres}

Figures \ref{f:morphnuv} and \ref{f:morphfuv} show the Concentration, Asymmetry, Smoothness, Gini, \m20 and the $G_M$ of the \nuv\ and \fuv\ compared to their values in the \hi\ maps, as computed over the same area and convolved to the same spatial resolution. We note that our approach is different from that of \cite{Lemonias11} who use the {\em optical} concentration to identify the approximate galaxy type. 

\begin{figure}
\begin{center}
\includegraphics[width=0.49\textwidth]{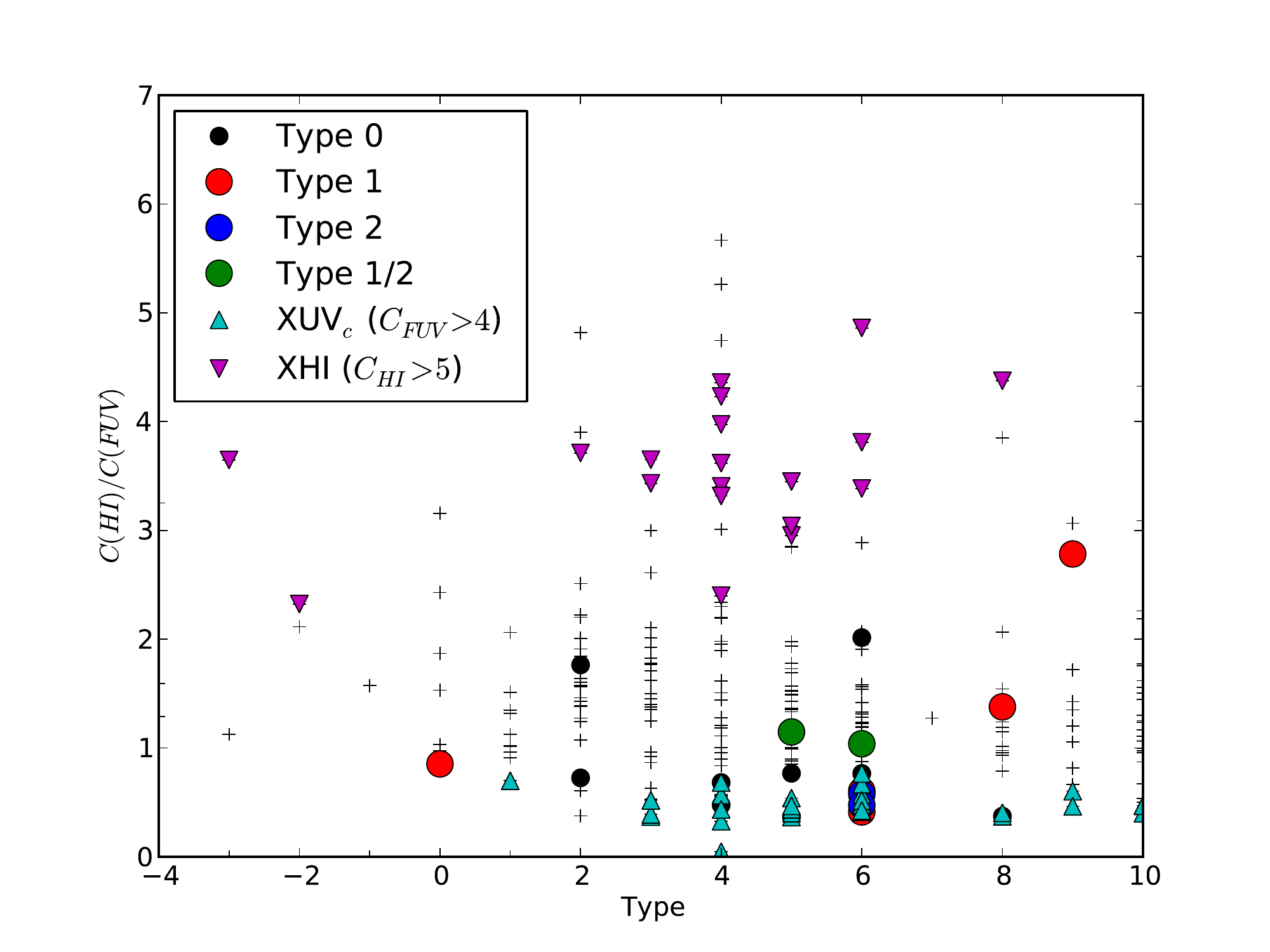}
\caption{The ratio of Concentration in \hi\ over that in \uv\ as a function of Hubble type. In this case (as expected), the XUV and XHI disks stand out. }
\label{f:Crat}
\end{center}
\end{figure}

\begin{figure}
\begin{center}
\includegraphics[width=0.49\textwidth]{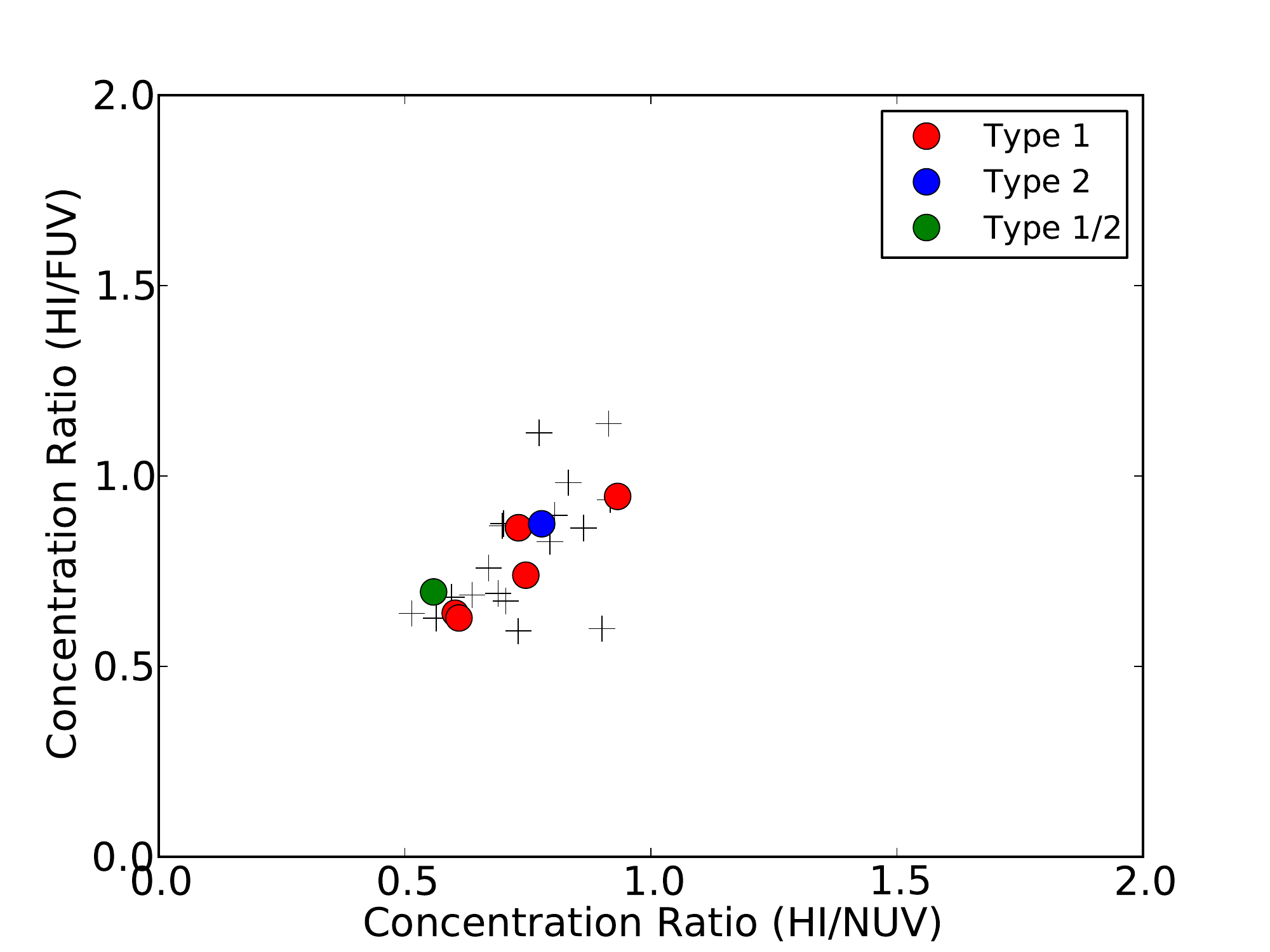}
\caption{The ratio of concentration of the \hi\ over \uv\ (\nuv\ and \fuv) for the \things\ galaxies}
\label{f:thingsCrat}
\end{center}
\end{figure}

Concentration of the \uv\ and the \hi\ shows two distinct sets of outliers. One where the \uv\ concentration is more than three; these are relatively extended disks. Some of the \xuv\  disks from \cite{Thilker07b} can be found here, but by no means all; two type 1/2 and one type 1 is not. However, six of the non-\xuv\ disks are also above the $C_{FUV} > 4$ line. Figures \ref{f:thingsnuv} and \ref{f:thingsfuv} show that most Type 1 in the \things\ sample are here.
The second set of outliers have \hi\ concentration indices exceeding $C_{HI} > 5$. However, these are not necessarily the classical \hi\ disks which extend well outside the stellar disk. High concentration values such as these often indicate recent tidal activity \citep{Holwerda11b}.
We will use the Concentration indices in \fuv\  and \hi\ to set our own supplementary definition of an extended disk; \xuvc\ disks are defined by $C_{FUV} > 4$ and extended \hi\ disks (\xhi) by $C_{HI} > 5$ in the other plots. We do so to identify these outliers in the other plots to explore their nature, not to supplant the original detailed classification by Thilker et. al. The \xuvc\ disks have \uv\ flux distributed throughout their otherwise unremarkable \hi\ disk and \xhi\ disks are extended in \hi\ but not special in \uv.
If we compare the ratio of concentration indices in Figure \ref{f:Crat} for the \whisp\ galaxies, both the \xuvc\ and \xhi\ disks --unsurprisingly-- stand out very clearly.  
This is confirmed by the ratios of \things\ concentration index in Figure \ref{f:thingsCrat}. 
We note, however, that if one applies these criteria blindly to the \things\ data (dashed lines in Figure \ref{f:thingsfuv}), one would conclude that most \things\ galaxies are \xuvc\ disks. Morphological criteria such as these need to be recalibrated when applied to different quality data.

%

\begin{figure*}
\begin{center}
\includegraphics[width=0.49\textwidth]{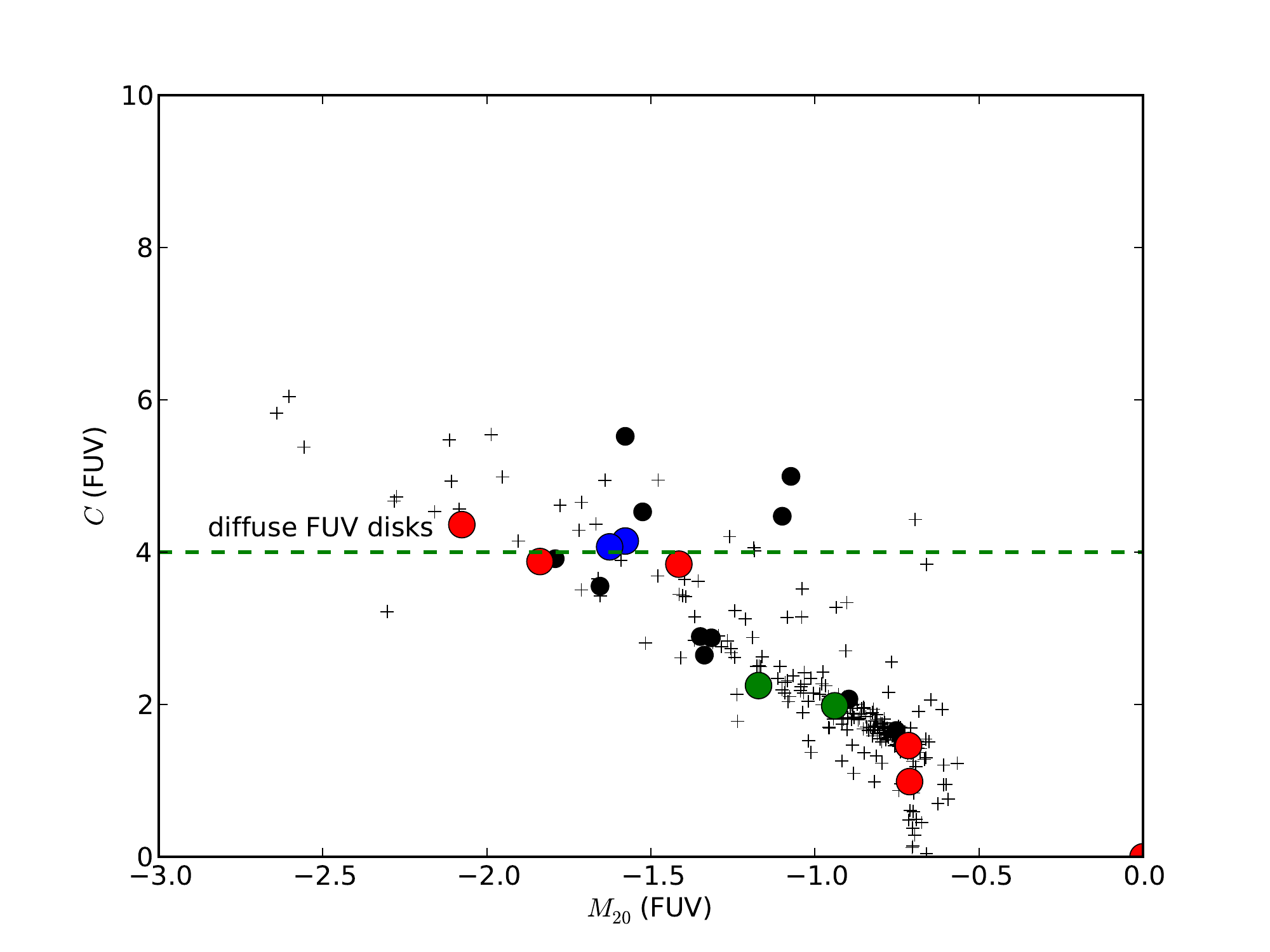}
\includegraphics[width=0.49\textwidth]{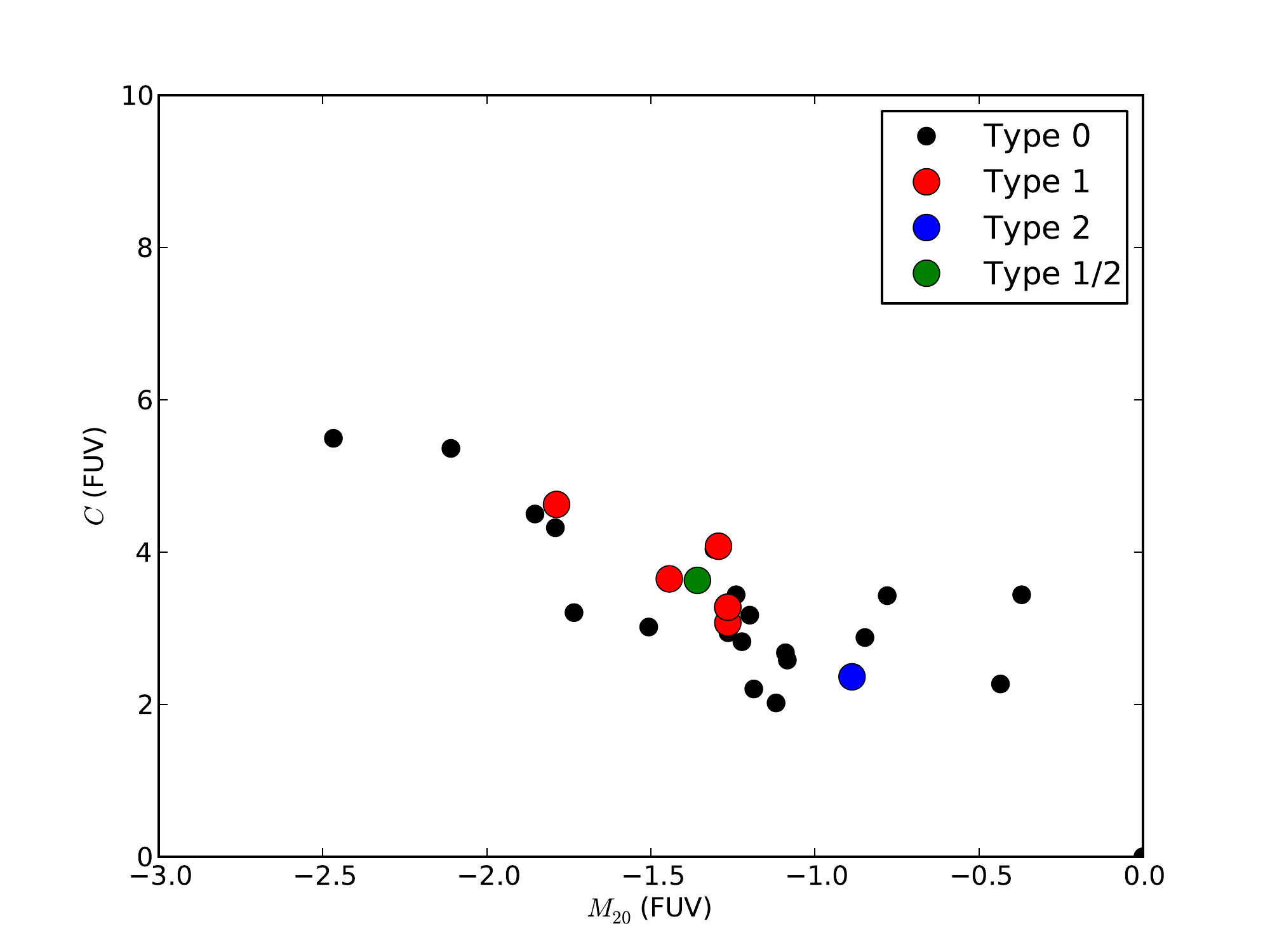}
\includegraphics[width=0.49\textwidth]{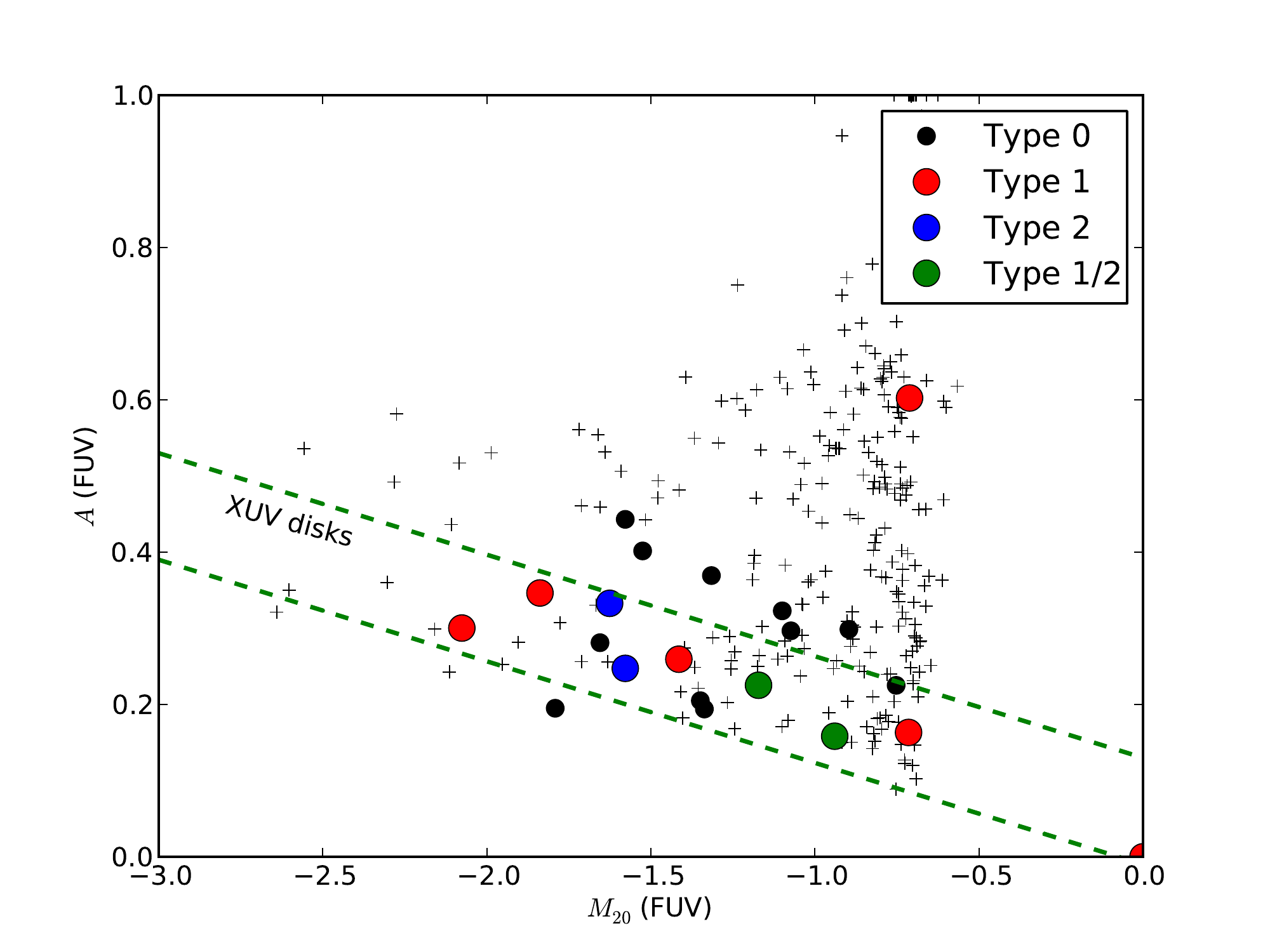}
\includegraphics[width=0.49\textwidth]{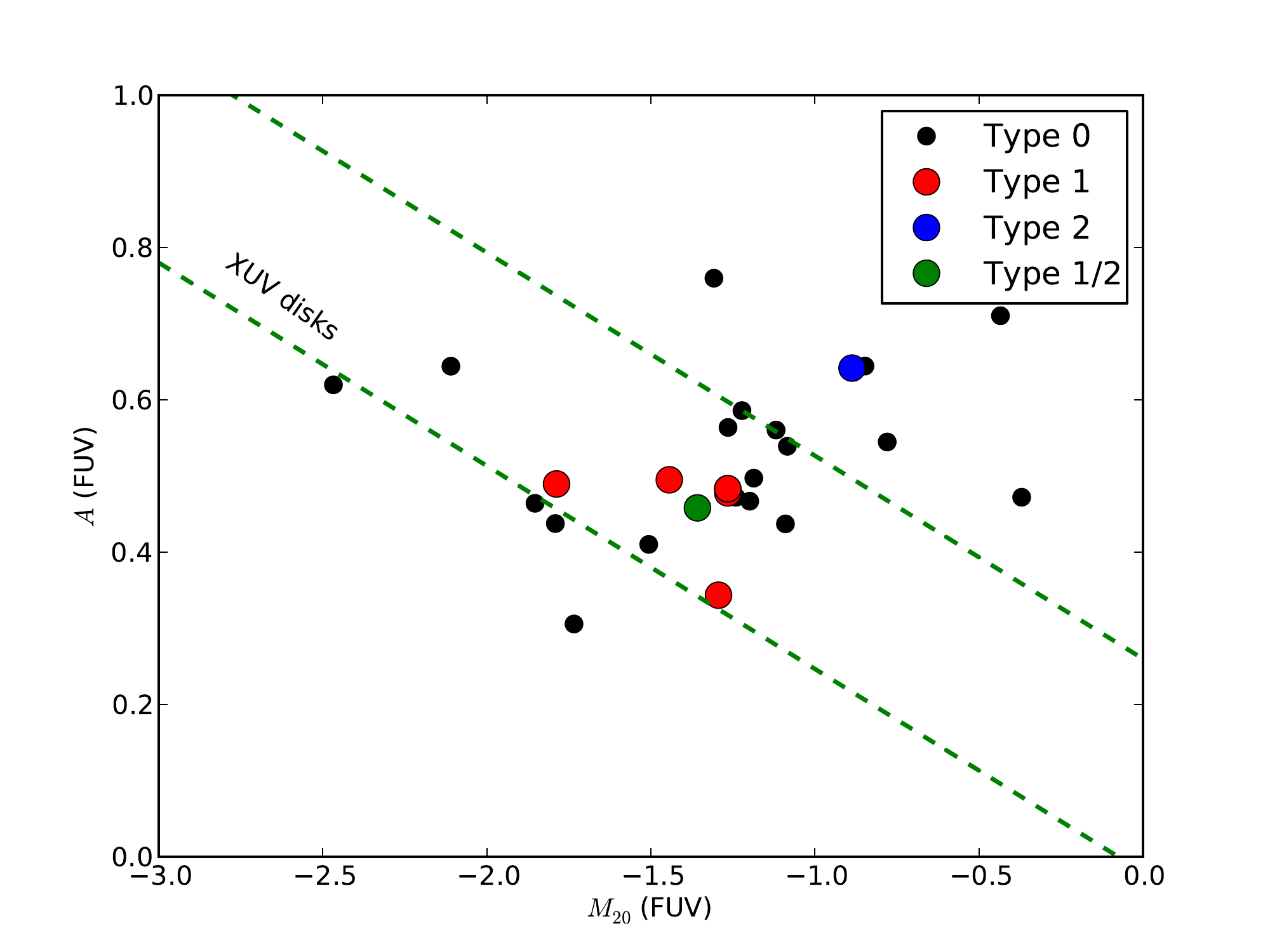}
\includegraphics[width=0.49\textwidth]{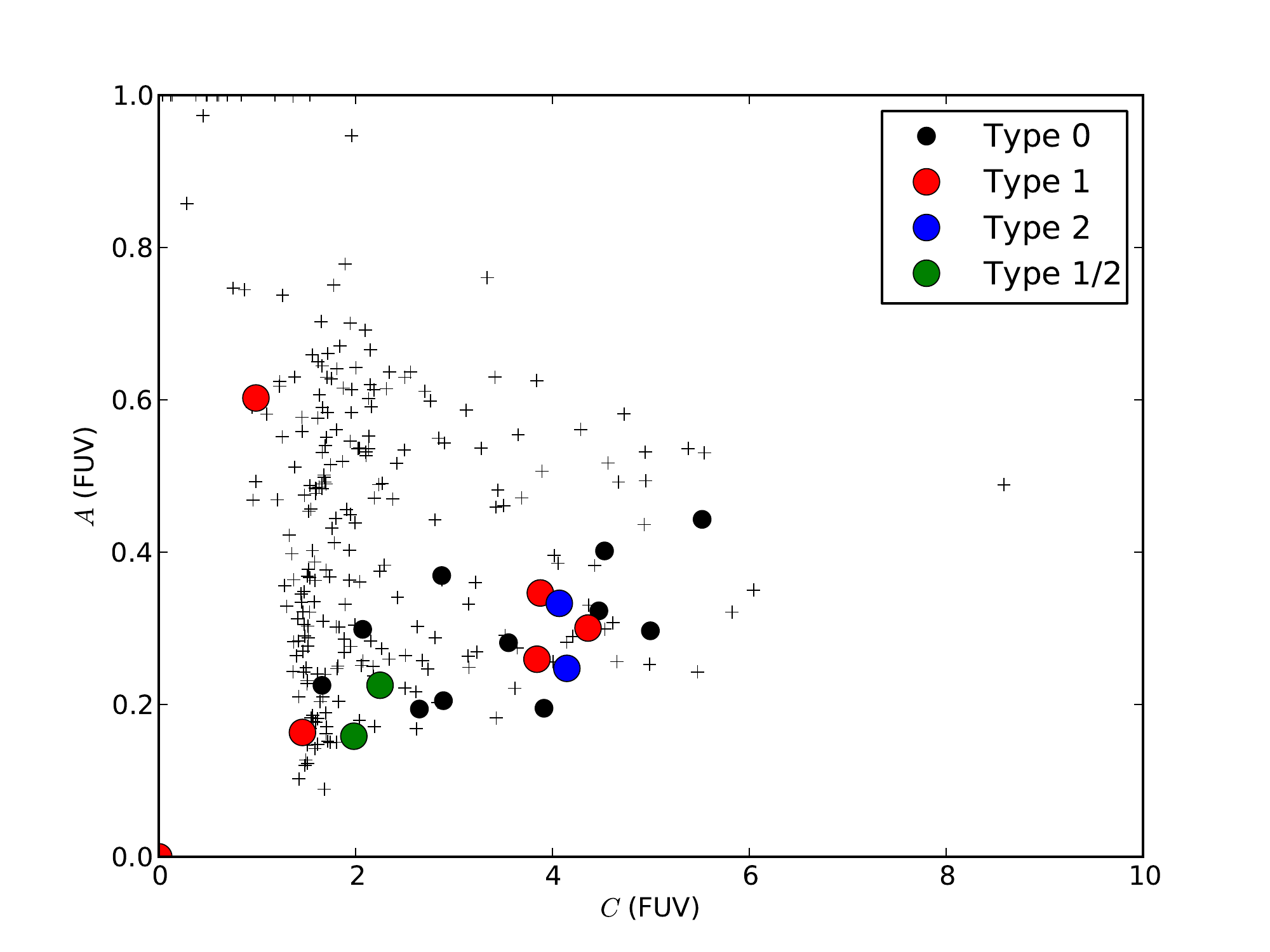}
\includegraphics[width=0.49\textwidth]{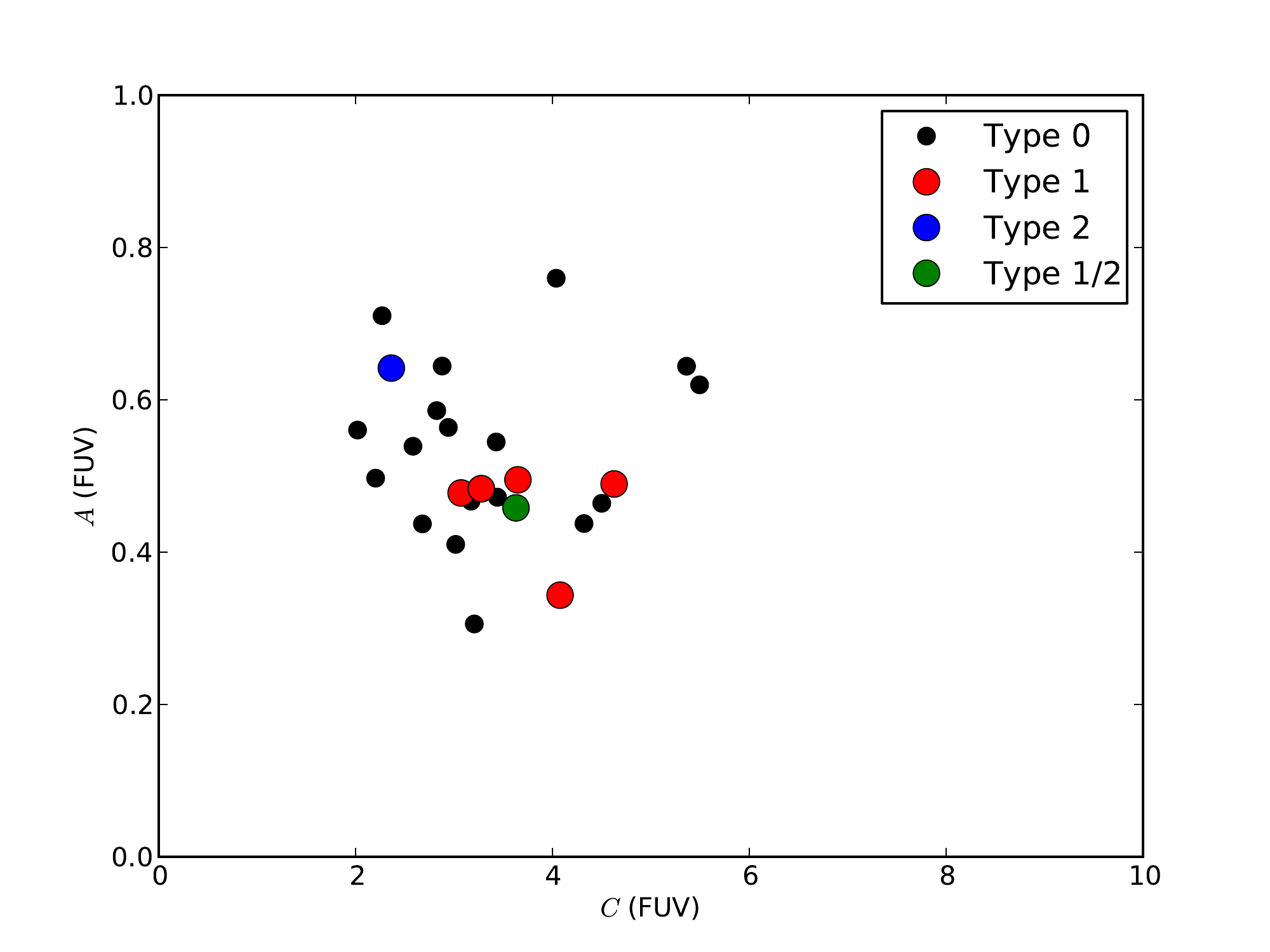}

\caption{Concentration, Asymmetry and $M_{20}$ in \fuv\ for the \whisp\ (left) and \things\ (right) samples with the Thilker et al. classifications marked. $C_{FUV}>4$ does not select cleanly the \xuv\ disks previously identified by Thilker et al. The \m20-Asymmetry criteria (green dashed lines in the middle panels), do select a relatively clean sample of \xuv\ disks from both \whisp\ and \things.  }
\label{f:fuvmorph}
\end{center}
\end{figure*}

\xuv\ and \xuvc\ disks have much lower values of the \m20\ parameter in the \uv\ than the general \whisp\ population. 
There is a general trend between \m20 and concentration \citep[specifically for the 3.6 $\mu$m Spitzer images][Holwerda et al. {\em in preparation}]{Munoz-Mateos09a}, 
so extreme values for this parameter are not surprising as such, especially if identified with a concentration criterion (see also Figure \ref{f:fuvmorph}. 
The {\em lower} value of \m20\  implies that regions that make up the \xuv\ disk do not contribute to the 
brightest 20\% of all the pixels in these \uv\ images.; the general \uv\ flux is more extended and hence the 
second order moment is less concentrated in the brightest few regions. The \hi\ \m20 values however are quite 
within the normal range of \whisp\ galaxies. Thus, perhaps the \uv\ \m20\  values offer an alternative automated 
definition of \xuv\ disks in combination with concentration.

The \xuv\ disks show a median value for Asymmetry in \fuv, but the full range in \hi. 

In \uv\ smoothness, the \xuv\ disks do not stand well apart from the general \whisp\ population. 
In the case of the G and \gm\ parameters, \xuv\ disks avoid extreme values for the \uv and generally fall below the G$_M$(\hi) =0.6 criterion for interactions.
In the case of most of the \hi\ morphological parameters, the values for the \xuv\ disks are typical for the bulk of \whisp\ galaxies. 

Based on their relative \hi\ and \uv\ morphologies, the previously identified or automatically identified \xuv\ disks mostly stand out in their \uv\ concentration and \m20 values. 
In the remaining \uv\ parameters and all \hi\ morphological values, they appear to be mixed in with the bulk of the \whisp\ (Figures \ref{f:morphnuv} and \ref{f:morphfuv}) and \things\ (Figures \ref{f:thingsnuv} and \ref{f:thingsfuv}).

\begin{table*}
\caption{\xuv\ disk selection criteria and their success (percentage of all bona-fide \xuv\ disks included) and contamination rate (percentage of included objects that are non-\xuv):}
\begin{center}
\begin{tabular}{l l l l l l}
\hline	
\hline	

Criterion								&	WHISP	&			&	& THINGS 	&	 		\\
\hline	
									& XUV (\%)	& Cont (\%)	& Nr	& XUV (\%)	& Cont (\%) 	\\
$C_{FUV} > 4$ 						& 30.			& 57			& 31 	& 25			& 71			\\

$|A - (-0.26667*M_{20})+0.12)/2 | < 0.14$ 	& 80			& 55			& 109 & - 			& -			\\

$|A - (-0.26667*M_{20})+0.12) | < 0.14$ 		& -			& -			& -	& 87			& 56			\\

\hline	
\hline	

\end{tabular}
\end{center}
\label{t:crit}
\end{table*}%

\subsection{\xuv\  disk \uv\ morphology}
\label{s:uvmorph}

In this section we look at the \uv\ morphological parameters in detail to explore if bona-fide \xuv\ disks can 
be identified from their \uv\ morphology alone --as computed over an area defined by the \hi\ disk.
For this we can use the 22 galaxies in \whisp\ and the \things\ sample, both classified by Thilker et al. 

Figure \ref{f:fuvmorph} shows the \fuv\ parameters for the \xuv\ disks identified by \cite{Thilker07b} as well 
as those disks inspected but without an extended \uv\ component for both samples.

The above Concentration-only criterion (\xuvc, $C_{FUV}>4$) pre-selects a number of the \xuv\ disks but it would miss several bona-fide 
\xuv\ disks as well as include many false positives. Using the information in two (or possibly more) parameters is how we identify mergers and may identify (candidate) \xuv\ disks
On inspection of the distributions of the \nuv\ and \fuv\ parameters (Figures \ref{f:scarlata_nuv} and  \ref{f:scarlata_fuv}), there are several combinations of parameters that offer the possibility of pre-selecting 
\xuv\ disks without too much loss: Gini-Asymmetry, \m20-Asymmetry and Gini-\gm. The goal is not to just select only the bona-fide, previously identified \xuv\ disks but also to exclude as many disks, that were previously classified as non-\xuv\ as practical. This latter criterion immediately excludes Gini-Asymmetry and Gini-\gm\ combinations. The combination of \m20\ and Asymmetry in \fuv\ is promising and we define the following criterion to identify \xuv\ disks in an \fuv\ survey (6" resolution):
\begin{equation}
\label{eq:things:M20A}
| A - (-0.27 \times M_{20} + 0.12) | < 0.14,
\end{equation}
which works well for the \things\ sample and we divide the linear relation by a factor two for the \whisp\ survey (to account for the $\sim$12" resolution):
\begin{equation}
\label{eq:whisp:M20A}
| A - (-0.27 \times M_{20} + 0.12) / 2  | < 0.14. 
\end{equation}
Both are illustrated in Figure \ref{f:fuvmorph} with dashed lines. Table \ref{t:crit} lists the two criteria the simple concentration criterion and the \m20-Asymmetry one (adjusted for each survey) 
and their success rate (percentage of all bona-fide \xuv\ disks included) and contamination rate (percentage of all classified objects that are non-\xuv). 
The  \m20-Asymmetry criterion selects most \xuv\ but with a contamination of some 55\% of non-\xuv\ disks. A future search for \xuv\ disks could therefore use this as a first cut before visual classification. 
Given the 109 \whisp\ galaxies selected by this criterion (Table \ref{t:xuvauto}), 61 of these should be bona-fide \xuv\ disks, 23\% of our sample of 266.

\subsection{Are \xuv\  disks in interacting \hi\ disks?}
\label{s:merger}

\begin{figure*}
\begin{center}
\includegraphics[width=0.49\textwidth]{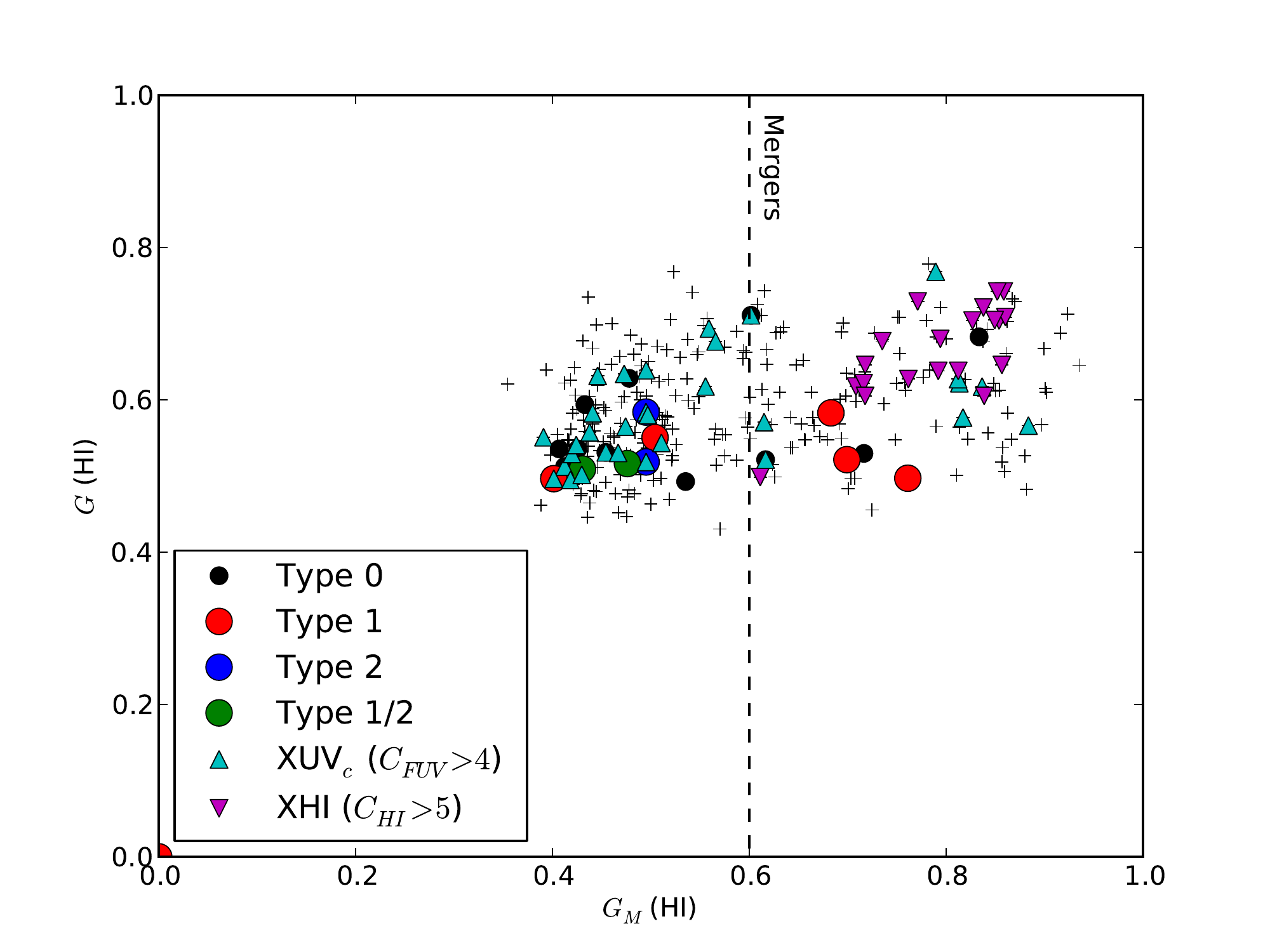}
\includegraphics[width=0.49\textwidth]{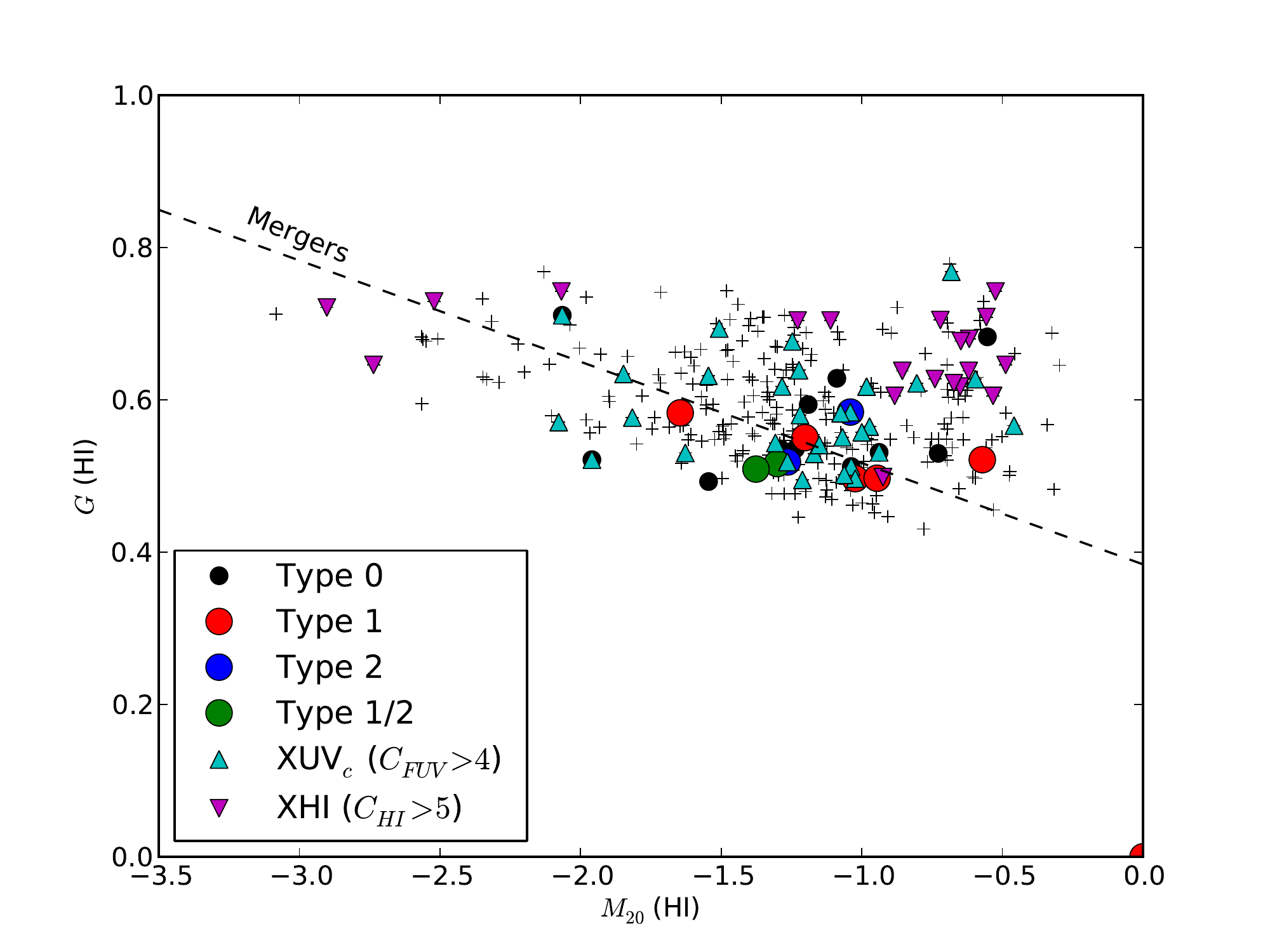}
\includegraphics[width=0.49\textwidth]{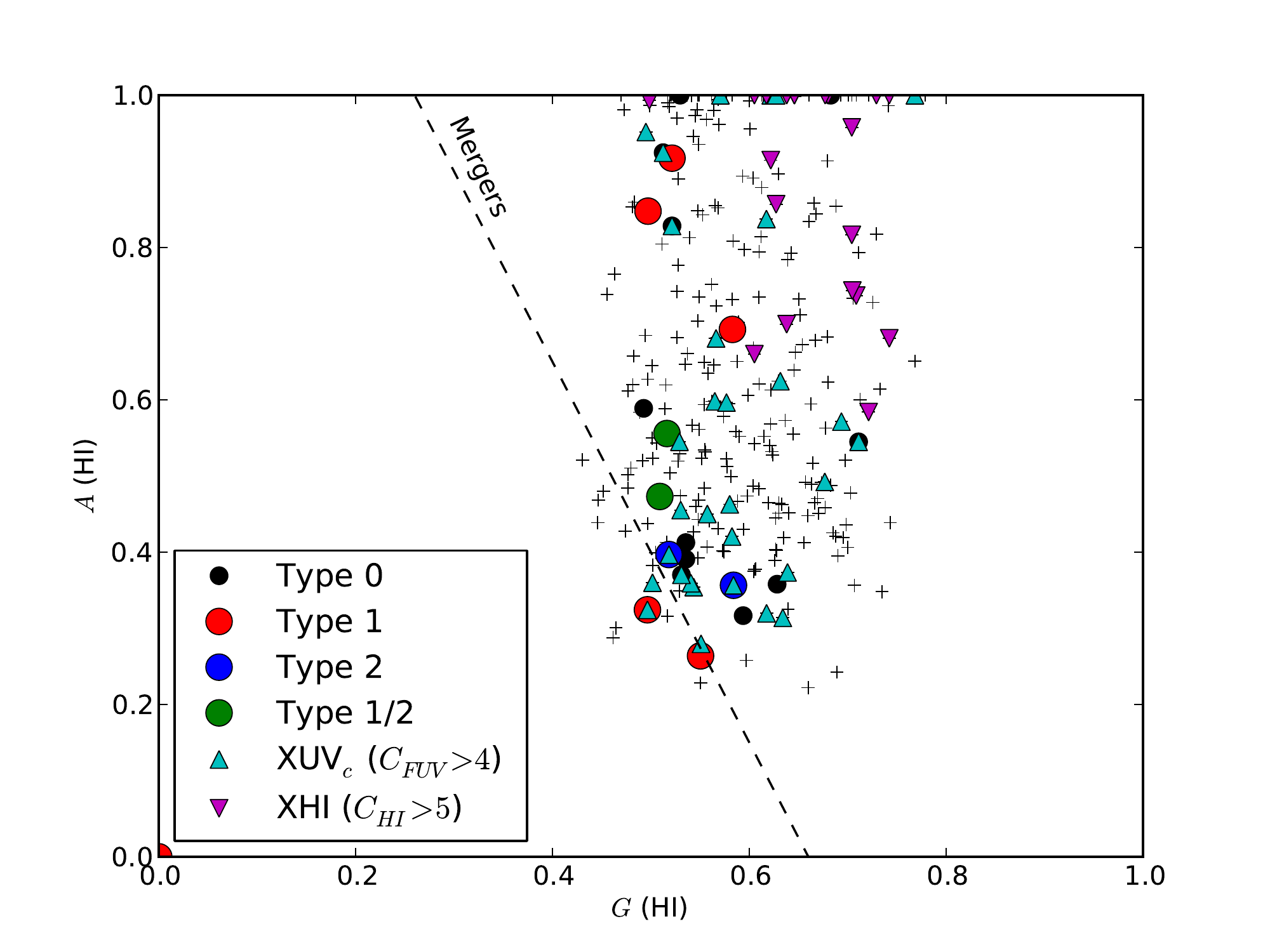}
\includegraphics[width=0.49\textwidth]{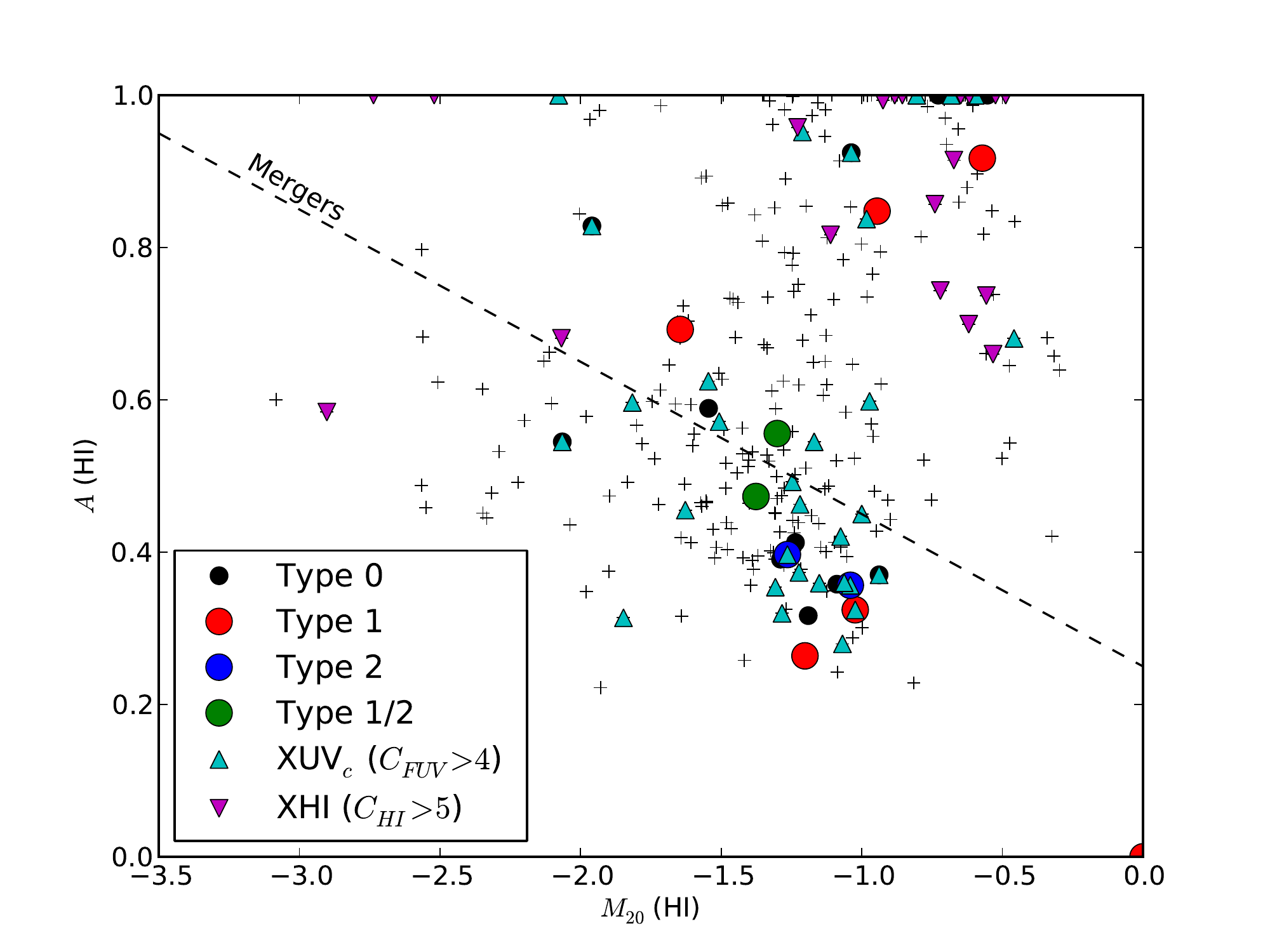}
\includegraphics[width=0.49\textwidth]{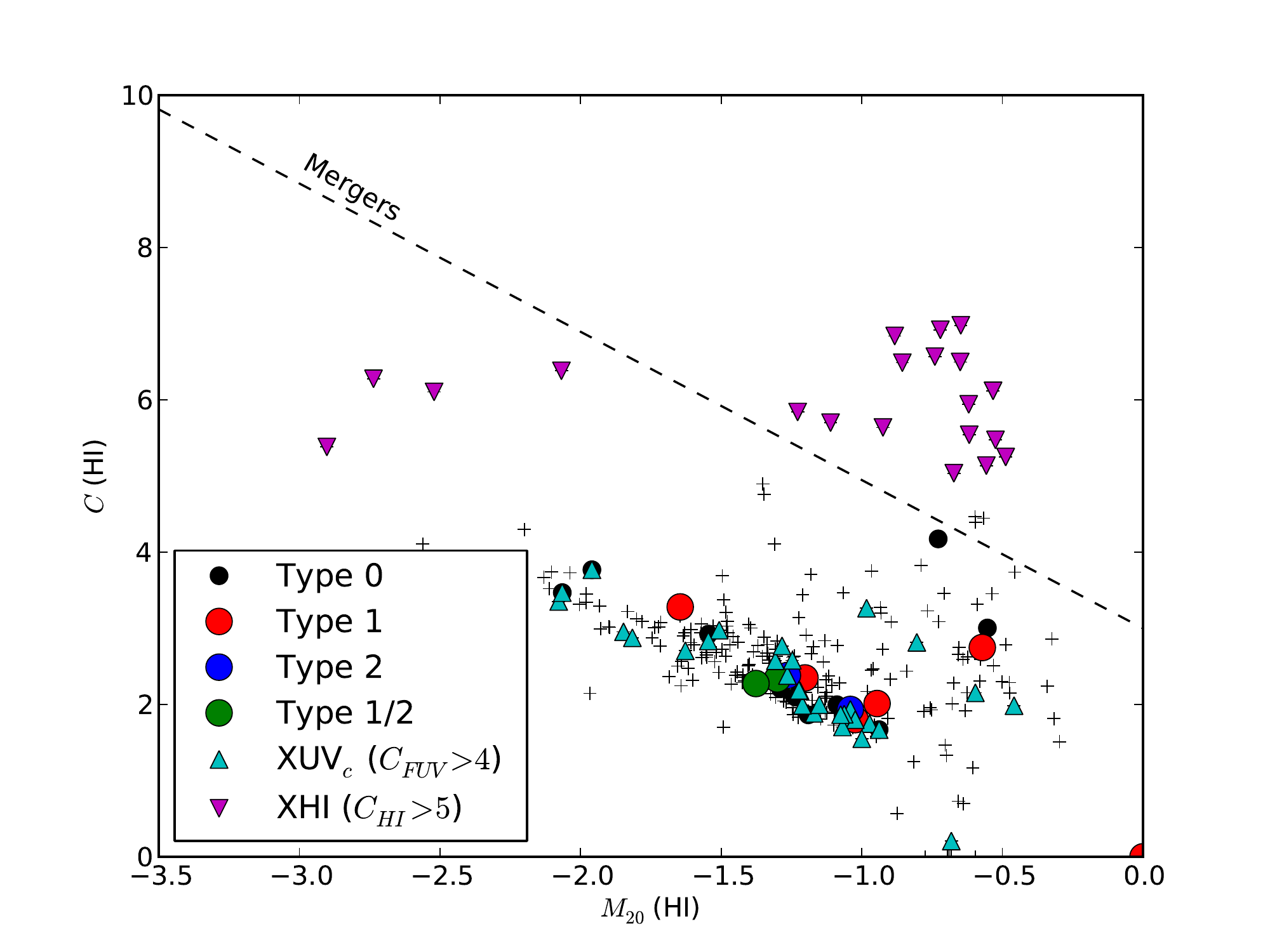}
\caption{The morphological selection criteria (dashed lines) for \hi\  applied to the \whisp\ sample. 
Disks are interactions above and to the right of the dashed lines.
\xuv\  disks identified either by Thilker et al. or high \fuv\ or \hi\ concentration are marked.}
\label{f:himorph}
\end{center}
\end{figure*}

Figure \ref{f:himorph} shows five of the morphological criteria we used or defined in \cite{Holwerda11c} to identify those disks that are currently undergoing an interaction (see Figure \ref{f:scarlata_hi2} for a similar plot to Figure 5 in \cite{Holwerda11b} for a direct comparison).
We mark those galaxies identified by \cite{Thilker07b} and those identified as extended UV or \hi\ disks by the concentration parameters. 

Based on the \hi\  morphologies, the $G_M$ and $C-M_{20}$ criteria, those \hi\ disks that host an \xuv\ disk are not interacting in the majority of cases. 
In the case of the other morphological criteria ($A-G$, $A-M_{20}$, $G-M_{20}$), this distinction is not as clear-cut but \xuv\ disks --either selected by 
Thilker et al or by their concentration-- appear to reside in morphologically very typical \hi\ disks.

At the same time, extended \hi\  disks are often interacting. This is in part because of our definition of an interaction is based on concentration but even in the 
$G_M$ criterion, these are conspicuously separated from the bulk of the \hi\ disks. Typically, the bona-fide \xuv\ disks selected by an \hi\ morphological selection of an ongoing merger appear to be Type 1, consistent with the first impression from Figures \ref{f:xuvt0} - \ref{f:xuvt2}. Thus, mergers appear to be one avenue to generate a Type 1 \xuv\ disk.

If we plot these --as a sanity check-- for the \things\ sample, the \xuv\ disks identified by \cite{Thilker07b} equally appear not to be interacting based on the
\gm criterion, but several are based on G-A or \m20-G criteria. Several of the non-\xuv\ disks are interactions in A-\m20. 
These criteria would need to be adjusted for the higher resolution of the \things\ sample \citep[difficult to do without many mergers in the \things\ sample, as we noted in][]{Holwerda11a}.

\begin{figure*}
\begin{center}

\includegraphics[width=0.49\textwidth]{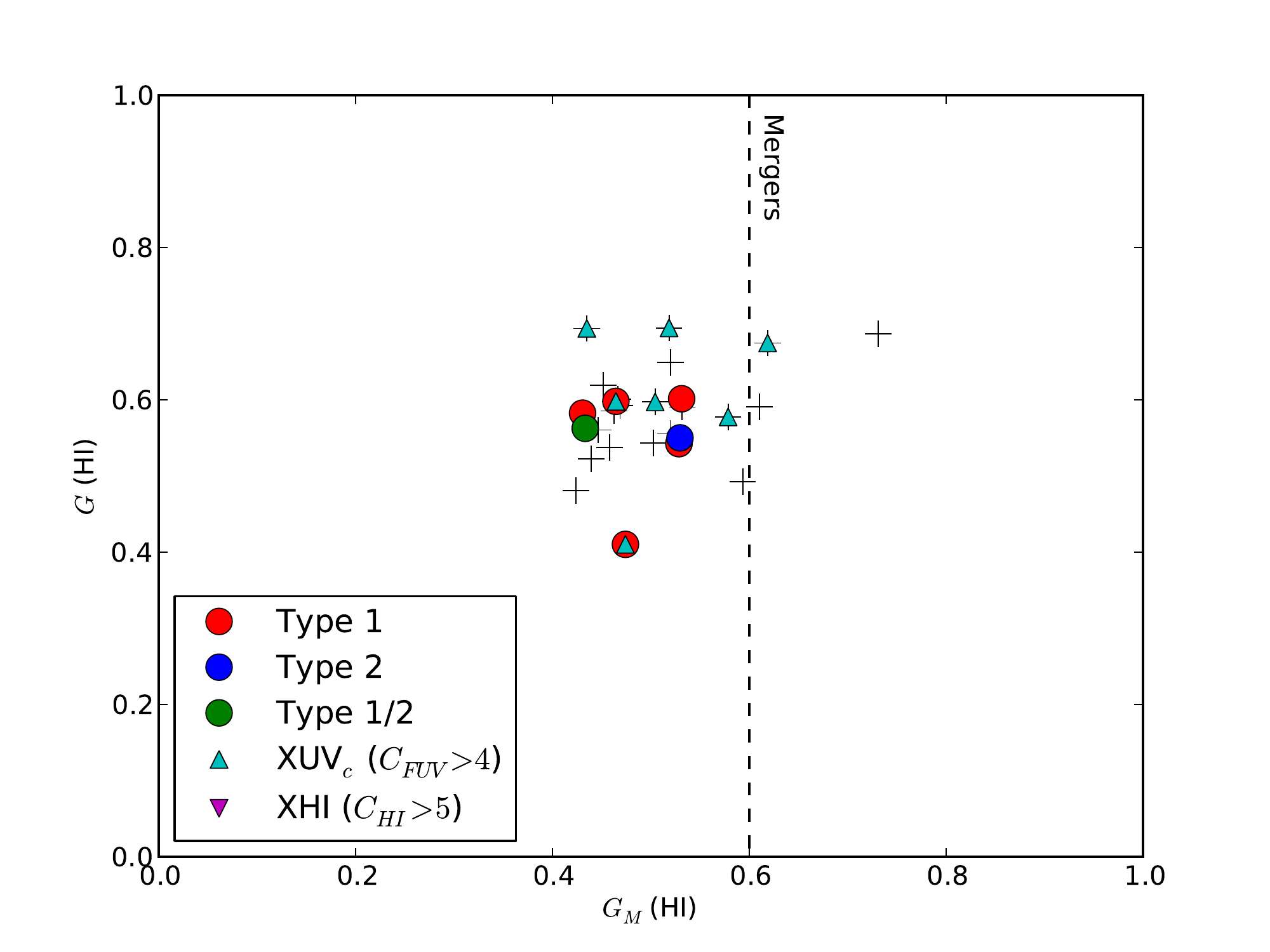}
\includegraphics[width=0.49\textwidth]{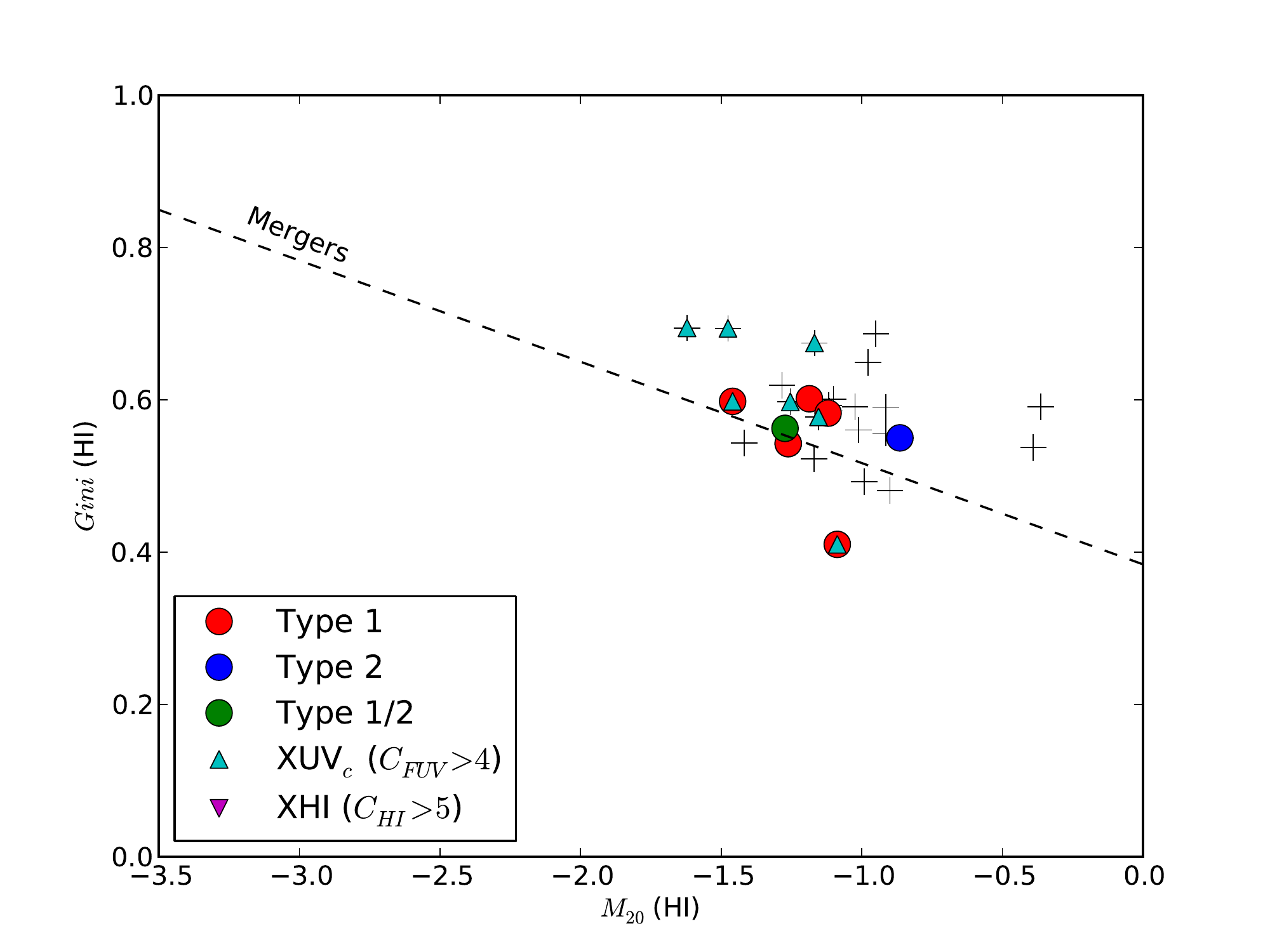}
\includegraphics[width=0.49\textwidth]{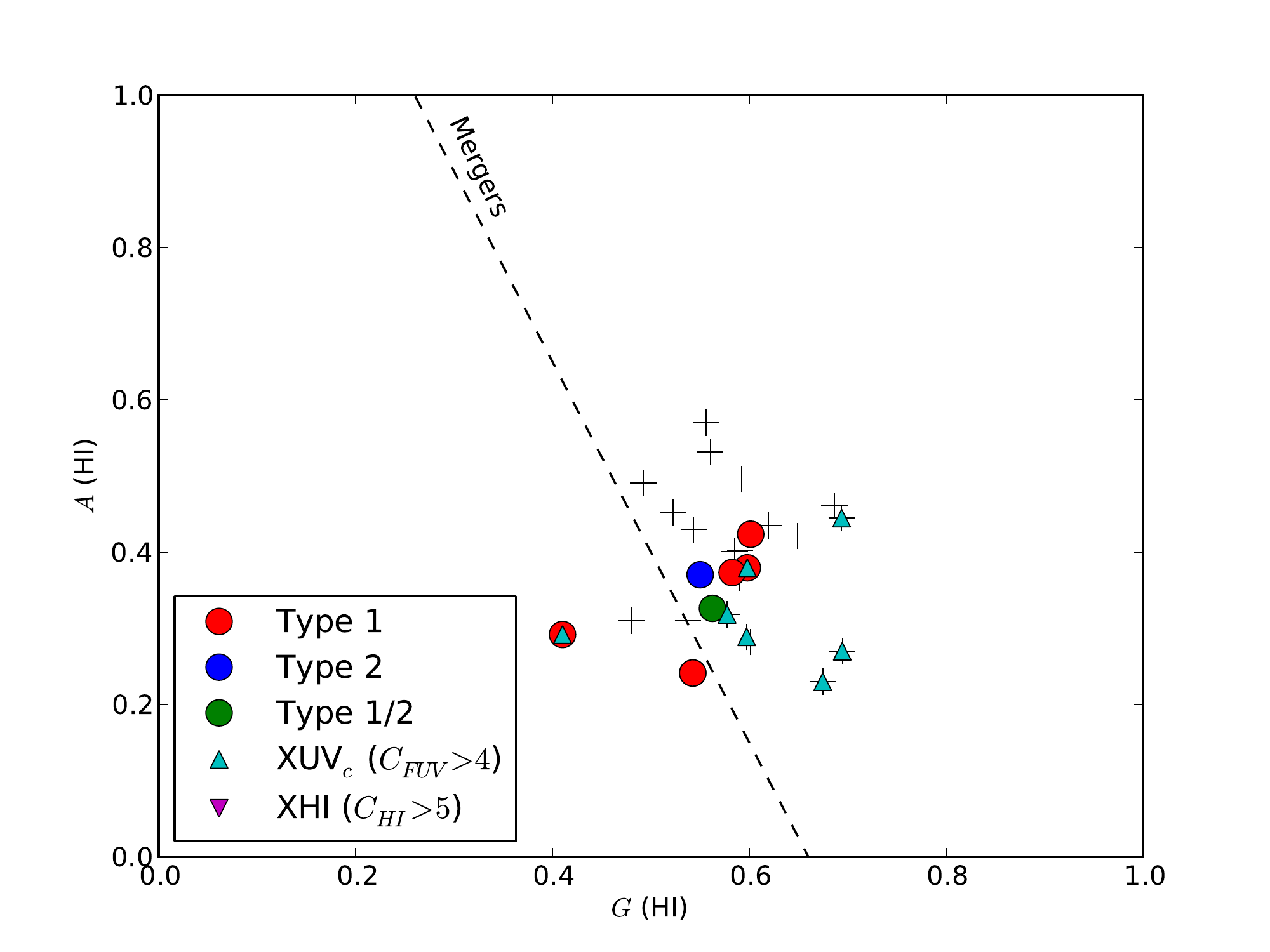}
\includegraphics[width=0.49\textwidth]{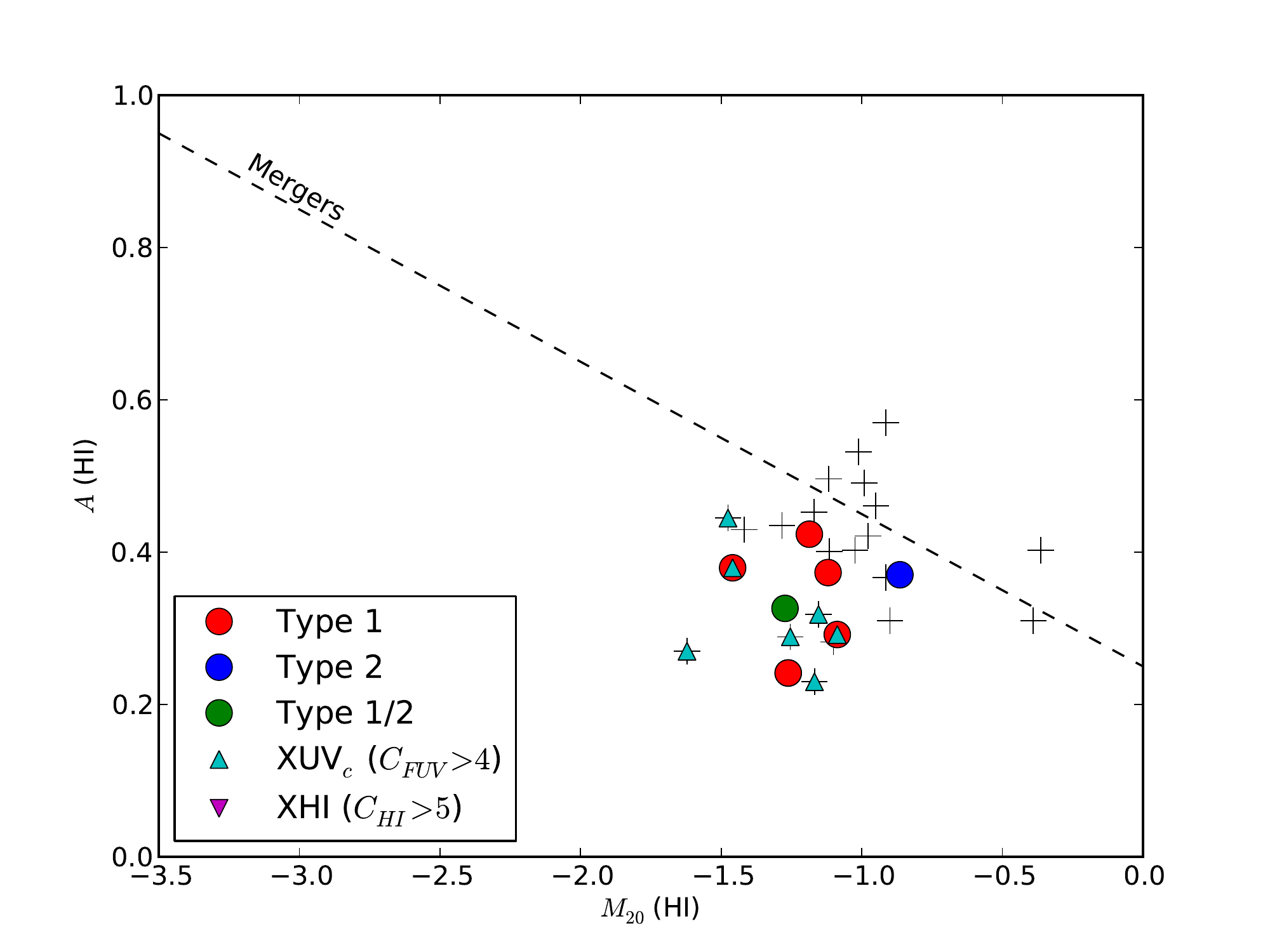}
\includegraphics[width=0.49\textwidth]{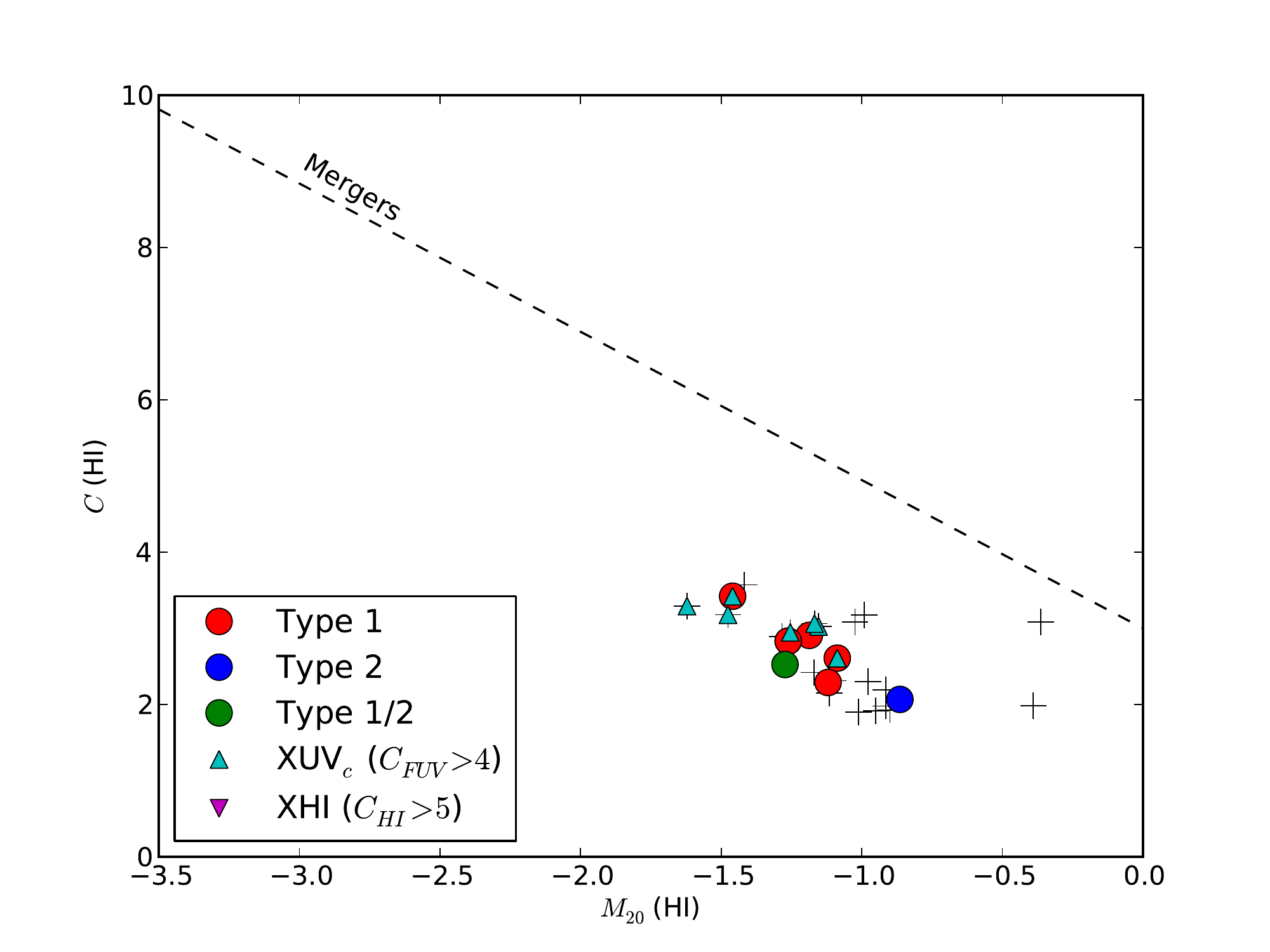}

\caption{The morphological selection criteria (dashed lines) for \hi\  applied to the \things\ sample. 
\xuv\  disks identified either by Thilker et al. or via their concentration are marked.}
\label{f:hithings}
\end{center}
\end{figure*}

%
%
%

\section{Discussion}
\label{s:disc}

Our first motivation for this study was to explore the possibility if \xuv\ disks could be reliably found using the morphological parameters in \uv\ and \hi, especially when computed over a wide aperture such as the \hi\ disk.
Based on their morphology parameters in \uv\ and \hi, \xuv\ disks identified in \cite{Thilker07b} rarely stand out from the bulk of the \whisp\ sample, with the exception of \m20-A and concentration in \uv.
In the \fuv, the \m20-A relation appears reasonably successful in identifying \xuv\ disks and the rate in the \whisp\ sample (23\%) is consistent with the \cite{Thilker07b} and \cite{Lemonias11} result, for a survey of late-types such as \whisp.
The \fuv\ values of Asymmetry and \m20\ could therefore be used to identify candidate \xuv\ disks in the local Universe with \galex\ or similar quality \uv\ data or in deep rest-frame \uv\ observations with {\em HST}.
However, these will have to be computed over a much larger area than commonly used (e.g., the de Vaucouleur diameter $D_{25}$ or the Petrosian radius), such as the one defined by the outer \hi\ contour.
%
The quantified \hi\ and \uv\ morphologies do not correlate as closely as they did for the \things\ sample in \cite{Holwerda11a} (their Figure 6). 
In our opinion, this is due to the wider range of environments probed by the \whisp\ survey as well as its lower resolution compared to \things. 
A closer relation may return for the inner stellar disk and possibly at higher spatial resolution.

Our second motivation for this work on \xuv\ disks was to explore their origin. 
Are they the (by)product of (a) cold flow accretion, (b) a gravitational interaction or mergers with another galaxy or (c) is photo-dissociation of molecular clouds by the \uv\ regions of the \xuv\ disks responsible for the \hi\ disks?

The fact that the majority of the \xuv\ disks do not reside in morphologically distinct \hi\ disks (Figure \ref{f:morphfuv} and \ref{f:thingsfuv}), 
precludes the notion that most \xuv\ disks originate from major interaction between galaxies. Few are selected by the previously defined interaction criteria for \hi\ disks.
There are, however, some obvious exceptions (Figure \ref{f:xuvt1}). So our na\"ive interpretation for the origin of Type 1 disks in Figure \ref{f:xuvt1} 
is only partly supported by \hi\ morphology and only holds for Type 1.

Since extended \uv\ (\xuvc) and extended \hi\ disk (\xhi) populations --as defined by their concentration parameter computed over the {\em same} area-- do not coincide (Figure \ref{f:morphfuv},  top left panel), a direct link, such as the photo-dissociation origin of the \hi\ disk, appears less likely.
A closer \uv--\hi\ relation in morphology may well still hold within the inner stellar disk, especially since the photo-dissociation scenario was developed to explain the higher column densities of \hi\ in this environment \citep{Allen97, Allen04, Heiner08a, Heiner08b, Heiner09, Heiner10}. 
However, the existence of both the $C_{FUV} > 4$ disks without complementary extended \hi\ disks as well as extended ($C_{HI}>4$) \hi\ disks without a complementary \xuv\ disk appears to be contradictory to the photo-dissociation origin of \xhi\ disks.
Two considerations are still in its favor however, \cite{Gogarten09} showed that UV emission does not fully trace the lower-lever star-formation in the low-column density environments of the outer disk. Thus, the ionizing stars may be lower-mass Main Sequence stars and simply remain undetected in \galex\ imaging due to their extreme low surface-brightness. 
And with higher resolution and more sensitive \hi\ and \uv\ imaging, a closer relation may still return in their quantified morphology (witness the \things\ sample).
Thus, we cannot completely rule out a photo-dissociation scenario explaining the relation between \hi\ and \uv\ disks. Similarly, \cite{Gil-de-Paz07} find their time-scale argument for a photo-dissociation origin of \xuv\ disks to be inconclusive. 

This leaves cold flows as a likely origin for most \xuv\ disks. A number of these Type 1 \xuv\ disks may be the result of a later-stage full merger such as UGC 04862 but judging from the \hi\ morphologies, this is small fraction of the full \xuv\ or \xuvc\ sample. Cold flows occur as both a steady trickle along a filament of the Cosmic Web as well as discrete accretion in the form of the cold gas originally associated with a small satellite.

There are indirect signs that cold flows are indeed the most common cause for \xuv\ disks. 
For example, at least three of the nine \things\ galaxies with an \xuv\ disk (M83, NGC 4736 and NGC 5055) are currently cannibalizing a small companion \citep{Martinez-Delgado10}. Satellites embedded in the gas filament are one of two fueling mechanisms of cold flow; discrete and trickle cold flows \citep[][]{Keres05}.
The acquired cold gas accompanying the cannibalized companion is a likely trigger for the formation \uv\ complexes.
Similarly, those disks selected by their extended \uv\ emission (Figure \ref{f:xuv}), often appear to have flocculent, low-column density \hi\ morphologies, which {\em may} point to recent cold accretion. 
A few cases (UGC 7081, 7651, and 7853) are obviously interacting major mergers with accompanying \hi/\uv\ tidal features. 
In contrast, the \xhi\ disks are nearly always small galaxies with an obvious close companion (Figure \ref{f:xhi}), which makes a tidal origin likely for their very extended \hi\ disk. 

The quantified morphologies of the \whisp\ \hi\ and \uv\ disks are not conclusive evidence that cold flows are responsible for the \xuv\ complexes but certainly 
suggest this is the most likely origin for most, with a few exception due to major interactions.


\section{Conclusions}
\label{s:concl}

Based on the quantified morphology of the \hi\ and \uv\ maps of the \whisp\ sample of galaxies, we conclude the following:
\begin{itemize}
\item[1.] There are distinct galaxy populations that stand out by their high concentration values in \fuv\ or \hi\ (\xuvc\ and \xhi\ thoughout the paper). These population do not overlap. Some known \xuv\ disks are in the \xuvc\ sample (Figures \ref{f:morphfuv}-a \ref{f:morphfuv}-e, \ref{f:thingsfuv}-a and \ref{f:thingsfuv}-c).To compute this concentration index, the outer \hi\ contour is needed to delineate the extent of the disk.
\item [2.] The fact that relative dilute \uv\ disks ($C_{FUV} > 4$) and \hi\ disks ($C_{HI} > 5$) population do not overlap at all and the lack of close morphological relations in any other parameters together suggest that a common small-scale origin for the \uv\ and \hi\ disks, such as a photodissociation of molecular hydrogen scenario, is less likely for \xuv\ disks (but may still very much hold for the inner disks).
\item [3.] Asymmetry and \m20\ can be used in combination to select \xuv\ disks with reasonable reliability (80\% included) but substantial contamination (55\%), when properly calibrated for the survey's spatial resolution (equation \ref{eq:things:M20A} and \ref{eq:whisp:M20A}) and the morphologies computed over a large enough aperture. With this selection, one can find the number of \xuv\ disks in a survey or candidates for visual classification (Figure \ref{f:fuvmorph}).
\item [4.] Based on the morphology of the \hi\ disk in which they occur, \xuv\ disks appear not to occur often in tidally disturbed gas disks (Figures \ref{f:himorph} and \ref{f:hithings}).
\item [5.] In a few cases, the \xuv\ disk {\em is} the product of a major merger; for example UGC 04862, identified by \cite{Thilker07b} and UGC 7081, 7651, and 7853, identified by their large \fuv\ concentration. This appears to be an avenue to form {\em some} of the Type 1 \xuv\ disks.
\item [6.] The \hi\ morphology and anecdotal evidence for small satellite cannibalism all point to a third mechanism for the origin of most \xuv\ disks; cold flow accretion. 

\end{itemize}

With the emergence of new and refurbished radio observatories in preparation for the future Square Kilometre Array \citep[SKA;][]{ska}, a new window on the 21 cm emission line of atomic hydrogen gas (\hi) is opening. The two SKA precursors, the South African Karoo Array Telescope \citep[MeerKAT;][]{MeerKAT,meerkat1,meerkat2}, and the Australian SKA Pathfinder \citep[ASKAP;][]{askap2, askap1, ASKAP, askap3,askap4} stand poised to observe a large number of Southern Hemisphere galaxies in \hi\ in the nearby Universe (z$<$0.2). In addition, the Extended Very Large Array \citep[EVLA;][]{evla} and the APERture Tile In Focus instrument \citep[APERTIF;][]{apertif,apertif2} on the Westerbork Synthesis Radio Telescope (WSRT) will do the same for the Northern Hemisphere. The surveys conducted with these new and refurbished facilities will add new, high-resolution, \hi\ observations on many thousands of galaxies. In the case of WALLABY (Koribalski et al. {\em in preparation}), these will be of similar quality spatial resolution as the \whisp\ survey.

Combining these with existing \uv\ observations from \galex, we can explore the relation between the morphology of the atomic hydrogen and ultraviolet light for much greater samples and in much greater detail. 
An application of the Asymmetry-\m20\ identification of \xuv\ disks in the \galex\ Nearby Galaxy Atlas \citep{nga} or a sample similar to \cite{Lemonias11} could reveal additional examples but the combination with deep \hi\ observations can prove the link with cold flows for the majority of \xuv\ disks.

\section*{Acknowledgments}

The authors would like to thank D. Thilker for discussions at the Cloister See-On conference, the audience at Leiden Observatory for encouragement for this work and the anonymous referee for his or her comments that  improved the science and writing of this paper.
The authors acknowledge the us of the HyperLeda database \url{http://leda.univ-lyon1.fr},
the NASA/IPAC Extragalactic Database (NED) which is operated by the Jet Propulsion Laboratory, California Institute of Technology, under contract with the National Aeronautics and Space Administration, 
and the Westerborg on the Web (WoW) project (\url{http://www.astron.nl/wow/}).

The Westerbork Synthesis Radio Telescope is operated by the Netherlands Institute for Radio Astronomy ASTRON, with support of NWO.
Based on observations made with the NASA Galaxy Evolution Explorer. GALEX is operated for NASA by the California Institute of Technology under NASA contract NAS5-98034.
%

\clearpage
\end{document}